\documentclass[twocolumn]{jpsj2}

\usepackage{graphicx}

\title{Theory of Interacting Bloch Electrons in a Magnetic Field}

\author{Takafumi {\sc Kita} and Masao {\sc Arai}$^\dagger$}
\inst{Division of Physics, Hokkaido University, Sapporo 060-0810\\
$^\dagger$National Institute for Materials Science, 
Namiki 1-1, Tsukuba, Ibaraki 305-0044}

\recdate{\today}

\abst{%
We study interacting electrons in a periodic potential 
and a uniform magnetic field ${\mib B}$
taking the spin-orbit interaction into account. 
We first establish a perturbation expansion
for those electrons with respect to the Bloch states in zero field.
It is shown that the expansion
can be performed with the zero-field Feynman diagrams of
satisfying the momentum and energy conservation laws.
We thereby clarify the structures of the self-energy and the thermodynamic
potential in a finite magnetic field.
We also provide a prescription of 
calculating the electronic structure in a finite magnetic field
within the density functional theory starting from
the zero-field energy-band structure.
On the basis of these formulations, we derive explicit expressions 
for the magnetic susceptibility
of ${\mib B}\!\rightarrow\!{\mib 0}$ at various approximation levels
on the interaction, 
particularly within the density functional theory, which
include the result of Roth [J.\ Phys.\ Chem.\ Solids {\bf 23} (1962) 433]
as the non-interacting limit.
We finally study the de Haas-van Alphen oscillation in metals
to show that quasiparticles at the Fermi level
with the many-body effective mass are directly relevant to the phenomenon.
The present argument may be more transparent than that by
Luttinger [Phys.\ Rev.\ {\bf 121} (1961) 1251]
of using the gauge invariance and has an advantage
that the change of the band structure with the field may be incorporated.
}

\kword{%
Bloch electrons, magnetic field, correlation effects,
magnetic susceptibility, de Haas-van Alphen effect,
density functional theory}

\begin{document}
\sloppy
\maketitle

\section{Introduction}
\label{sec:intro}

Extensive theoretical studies have been carried out on 
Bloch electrons in a magnetic field ${\mib B}$
which display many exciting phenomena.
Among them are investigations on fundamental quantities and
phenomena such as the effective Hamiltonian,\cite{Peierls33,Luttinger51,Kohn59,
Blount62,Roth62,Brown64,Brown68,Fischbeck70,Nenciu91}
the static magnetic susceptibility,\cite{Peierls33,Herring66,KO56,HS60,
Blount62,Roth62,Wannier64,HS64,Roth66,
MR69,HM70,FK70,Fukuyama71,MK72,PM72,Buot72,FM74,
VP75,MV76,Gunnarsson76,Oh76,Buot76,Janak77,YS79,MMM82,BW83,YS85,
MSS90,MSS91} 
and the de Haas-van Alphen (dHvA)
oscillation.\cite{Onsager52,LK55,Luttinger61,Gorkov62,Roth66,ES70,WS96}
Since the relevant energy scale is $\mu_{\rm B}B\!\lesssim\! 1$meV
with $\mu_{\rm B}$ the Bohr magneton,
it seems natural to start from the
energy-band structure in zero field and try to include 
the field effects perturbatively.
However, a uniform magnetic field in quantum mechanics 
necessarily accompanies a vector potential
with a non-periodic linear spatial dependence.
Hence the procedure is not at all an easy task to perform
as it apparently looks,
particularly when the electron-electron interaction is taken into account.
For example, we still do not have a satisfactory calculation
of the total magnetic susceptibility of metals, even for
Li and Na.

The purposes of the present paper are threefold.
We first establish a definite
theoretical prescription to calculate properties of 
interacting Bloch electrons
in a uniform magnetic field 
on the basis of the energy-band structure
in zero field.
This includes an extension of the density functional 
theory\cite{HK64,Mermin65,KS65,SM71,BH72,RC73,GL76,
MV79,PW86,VR88,JG89,GH94,Kohn99}
to a finite magnetic field.
We then derive explicit expressions of the total
magnetic susceptibility
$\chi({\mib B}\!\rightarrow\!{\mib 0})$
at various approximation levels on 
the electron-electron interaction.
Finally, the formulation is used to study many-body effects
on the dHvA oscillation in metals.

Let us briefly summarize the relevant
works on the effective Hamiltonian, the density functional theory, 
the magnetic susceptibility and the dHvA oscillation
together with the extensions considered here.

We now have a well-established effective Hamiltonian in a uniform magnetic
field at the one-particle level;\cite{Peierls33,Luttinger51,Kohn59,
Blount62,Roth62,Fischbeck70} see also Ref.\ \citen{Nenciu91}
on mathematical aspects.
Let $t^{(0)}({\mib R}\!-\!{\mib R}')$ denote the transfer integral
in zero field between two unit cells specified by
${\mib R}$ and ${\mib R}'$; it completely 
determines the energy-band structure.\cite{comment1}
Then the transfer integral in a finite field is obtained by
the replacement:
\begin{subequations}
\begin{equation}
t^{(0)}({\mib R}\!-\!{\mib R}')
\longrightarrow
{\rm e}^{iI_{{\mib R}{\mib R}'}}\,t({\mib R}\!-\!{\mib R}',{\mib B}) \, ,
\label{intro-t}
\end{equation}
where $I_{{\mib R}{\mib R}'}$ is the Peierls phase\cite{Peierls33} 
and $t({\mib R}\!-\!{\mib R}',{\mib B}\!\rightarrow\!{\mib 0})\!\rightarrow\!
t^{(0)}({\mib R}\!-\!{\mib R}')$.
The relation was first obtained by Peierls\cite{Peierls33} for the tight-binding model
without the ${\mib B}$ dependence in $t$. 
The extension beyond the tight-binding model is due to Luttinger.\cite{Luttinger51}
The ${\mib B}$ dependence in $t$ was taken into account by
Kohn,\cite{Kohn59} Blount,\cite{Blount62} and Roth.\cite{Roth62}
As shown by Luttinger,\cite{Luttinger51}
eq.\ (\ref{intro-t}) provides a microscopic justification for the
procedure:
${\cal E}^{(0)}({\mib k})\!\rightarrow\!
{\cal E}\!\left(-i{\mib \nabla}\!-\! \frac{e}{\hbar c}{\mib A}\right)$ with 
${\cal E}^{(0)}({\mib k})$
an energy eigenvalue in zero field, 
${\mib k}$ a wave vector in the first Brillouin zone,
and ${\mib A}$ the vector potential,
which was a key assumption in the Onsager-Lifshitz-Kosevich theory 
for the dHvA oscillation.\cite{Onsager52,LK55}
Here, we also consider interaction effects 
on the basis of the treatments by Luttinger\cite{Luttinger51}
and Roth.\cite{Roth62}
It is thereby shown that the self-energy also experiences
the change:
\begin{equation}
\Sigma^{(0)}(\varepsilon_{n},{\mib R}\!-\!{\mib R}')
\longrightarrow
{\rm e}^{iI_{{\mib R}{\mib R}'}}\,
\Sigma(\varepsilon_{n},{\mib R}\!-\!{\mib R}',{\mib B}) \, ,
\label{intro}
\end{equation}
\end{subequations}
with $\varepsilon_{n}$ the Matsubara frequency.
We will provide a definite prescription of how to calculate 
$\Sigma(\varepsilon_{n},{\mib R}\!-\!{\mib R}',{\mib B})$
for an arbitrary finite field ${\mib B}$.
Thus, our approach is more powerful than
those by the expansion in ${\mib A}$\cite{Fukuyama71,FM74} that is
effective only in the zero-field limit.
With an application to superconductors in mind, 
the whole formulation will be carried out without assuming a specific gauge
for the vector potential in such a way that
an extension to a non-uniform magnetic field may be performed easily.

The density functional theory (DFT) is 
regarded now as one of the most efficient and reliable
methods for the quantitative understanding of solids.\cite{JG89,Kohn99}
Its extension to a finite magnetic field has been performed by 
Vignale and Rasolt\cite{VR88} using the current density ${\mib j}({\mib r})$ 
as the relevant extra variable, and also by Grayce and Harris\cite{GH94}
choosing ${\mib B}$ instead of ${\mib j}({\mib r})$.
In these formulations, however, one has to calculate
the electronic structure in ${\mib B}$ from the beginning
by treating the field and the periodic potential on an equal footing.
Thus, practical calculations for solids have never been carried out.
Here, we will provide a prescription of obtaining 
the electronic structure in ${\mib B}$
within DFT based on the known zero-field energy-band structure.
This two-step procedure may be regarded as one of the main advantages
of the present formulation over the previous ones.\cite{VR88,GH94}
We choose the average flux density
${\mib B}\!\equiv\!{\mib\nabla}\!\times\!{\mib A}$ 
as the relevant variable.
The total moment may be obtained
by ${\mib M}\!=\!-(\partial \Omega/\partial {\mib B})$, and
the external field is found by
${\mib H}\!=\!{\mib B}\!-\!4\pi{\mib M}/V$ with $V$ the volume.

The static magnetic susceptibility $\chi({\mib B}\!\rightarrow\! {\mib 0})$
has been a matter of extensive theoretical 
investigations,\cite{Peierls33,Herring66,KO56,HS60,
Blount62,Roth62,Wannier64,HS64,Roth66,
MR69,HM70,FK70,Fukuyama71,MK72,PM72,Buot72,FM74,
VP75,MV76,Gunnarsson76,Oh76,Buot76,Janak77,YS79,MMM82,BW83,YS85,
MSS90,MSS91}  and
we now have several apparently different but essentially equivalent
expressions of $\chi$ for non-interacting Bloch electrons.\cite{Roth62,Wannier64,HS64,MK72}
Among them, Roth\cite{Roth62} gave a complete expression 
including the spin-orbit interaction
with a clear derivation process.
We closely follow her procedure
to extend the consideration into interacting Bloch electrons. 
It should be noted that $\chi$ of interacting Bloch electrons were already calculated
for the orbital part by
by Fukuyama,\cite{Fukuyama71} Phillipas and McClure,\cite{PM72} and
Fukuyama and McClure,\cite{FM74} 
and for both the orbital and spin parts by Buot\cite{Buot76} and
Misra {\em et al}.\cite{MMM82}
However, most of them provided only approximate treatments of $\chi$:
either the vertex corrections were neglected\cite{Fukuyama71,
FM74,MMM82} or the frequency dependence of the self-energy were 
not considered.\cite{PM72,MMM82}
Also, the treatment by Buot\cite{Buot76} 
fails to incorporate vertex corrections explicitly in
the general expression.
We here present a complete framework to calculate $\chi$ 
at various approximation levels on the interaction.
Particularly, we derive an explicit expression of $\chi$
within DFT.
Calculations of the susceptibility by DFT have been performed 
only for the spin part,\cite{VP75,MV76,Gunnarsson76,Janak77,MSS90,MSS91}
for the orbital part neglecting vertex corrections,\cite{Oh76,YS79,BW83}
and for the both parts but neglecting vertex corrections.\cite{YS85}
Thus, there is no available expression within DFT for the total susceptibility
with vertex corrections.
The formula obtained here, which incorporates the effects of
the spin-orbit interaction, core polarizations, interband transitions,
and vertex corrections, 
is expected to enhance 
our understanding on the total magnetic susceptibility of solids.
It will be shown that the vertex corrections of the spin part 
in our formula naturally include the Stoner
enhancement factor.\cite{Herring66,Stoner36}

The dHvA oscillation in metals provides unique information
on the Fermi-surface structures and interaction effects. 
The classic theory at the one-particle
level is due to Lifshitz and Kosevich,\cite{LK55} 
who applied Onsager's semiclassical quantization scheme\cite{Onsager52}
to non-interacting band electrons.
The interaction effects were considered by Luttinger\cite{Luttinger61} and 
Bychkov and Gor'kov.\cite{Gorkov62}
However, Bychkov and Gor'kov considered only an isotropic Fermi liquid.
Also, the work by Luttinger\cite{Luttinger61} is based on a gauge-invariance argument
${\cal E}^{(0)}({\mib k})\!+\!\Sigma^{(0)}(\varepsilon_{n},{\mib k})\!\rightarrow\!
{\cal E}\!\left(-i{\mib \nabla}\!-\! \frac{e}{\hbar c}{\mib A}\right)\!
+\!\Sigma\!\left(\varepsilon_{n},-i{\mib \nabla}\!-\! \frac{e}{\hbar c}{\mib A}\right)$
without clarifying the structure 
of the self-energy explicitly.
Hence it is not clear from Luttinger's argument 
whether it is the momentum ${\mib p}$
or the  crystal momentum $\hbar{\mib k}$ in the first Brillouin zone that
is really relevant. 
The replacement procedure for the self-energy
needs to be established microscopically
in the same sense as eq.\ (\ref{intro-t}) had to.
His argument also has an ambiguity as to the ``non-oscillatory part'' of the
self-energy, which later caused an interpretation\cite{ES70,WS96}
that the self-energy does not participate in making up the
quantized energy levels, i.e., one only has to replace
the one-particle part as ${\cal E}^{(0)}({\mib k})\!\rightarrow\!
{\cal E}\!\left(-i{\mib \nabla}\!-\! \frac{e}{\hbar c}{\mib A}\right)$.
To remove the confusion and also to analyze experiments unambiguously,
it will be well worth placing the theory on a firm ground.
On the basis of the formulation to derive eq.\ (\ref{intro}), we will present
a hopefully clearer argument for the many-body effects on
the dHvA oscillation.
This argument also has an advantage that the change of the 
energy-band structure with the field can be taken into account.

This paper is organized as follows. 
Section 2 provides an alternative derivation of 
the effective one-particle Hamiltonian with the spin-orbit interaction.
We combine the advantages in the treatments 
of Luttinger\cite{Luttinger51} and Roth\cite{Roth62}
to formulate the problem so that extensions
(i) to interacting systems and (ii) to a non-uniform 
magnetic field may be performed easily.
Section 3 takes the electron-electron interaction into account.
We establish a perturbation expansion for the thermodynamic potential and the self-energy
in terms of the energy-band structure in zero field.
It is shown that only the zero-field Feynman diagrams are necessary.
We also construct a DFT in a finite magnetic field.
Section 4 derives expressions of the magnetic susceptibility 
at various approximation levels,
including that of DFT.
Section 5 studies many-body effects on the dHvA oscillation in metals.
Section 6 summarizes the paper.
We put $k_{\rm B}\!=\! 1$ throughout.

\section{Effective Hamiltonian}

\subsection{Hamiltonian and basis functions}

We consider Bloch electrons in a uniform magnetic field described by a
Hamiltonian with the spin-orbit interaction and Pauli paramagnetism:
\begin{equation}
\hat{\cal H}=\frac{\hat{\mib P}^{2}}{2m}+{\cal V}({\mib r})
-g \frac{e\hbar}{2m^{2}c^{2}}(\hat{\mib s}\times{\mib \nabla}{\cal V})
\cdot \hat{\mib P}-g \frac{e\hbar}{2mc}\hat{\mib s}\cdot{\mib B}-\mu \, .
\label{Hamiltonian}
\end{equation}
Here $m$ and $e(<\!\! 0)$ are the electron mass and charge, respectively, 
$\hat{\mib P}$ is defined by
\begin{equation}
\hat{\mib P}\equiv\hat{\mib p}-\frac{e}{c}{\mib A}({\mib r}) \, ,
\end{equation}
with $\hat{\mib p}$ the momentum operator and ${\mib A}({\mib r})$ the vector potential,
${\cal V}({\mib r})$ denotes the periodic lattice potential,
$g$ is the electron $g$ factor,
$\hat{\mib s}$ is the dimensionless spin operator,
${\mib B}\!\equiv \!{\mib \nabla}\!\times\!{\mib A}({\mib r})$
is the magnetic field,
and $\mu$ is the chemical potential.

The eigenfunctions of eq.\ (\ref{Hamiltonian}) at ${\mib A}\!=\!{\mib 0}$
are the Bloch spinors:
\begin{equation}
\vec{\psi}_{b{\mib k}}({\mib r})={\rm e}^{i{\mib k}\cdot{\mib r}}\,
\vec{u}_{b{\mib k}}({\mib r}) \, ,
\label{Bloch}
\end{equation}
where ${\mib k}$ is a wave vector in the first Brillouin zone and
$b$ denotes a set of quantum numbers for the band and spin.
They are orthonormal as 
$\langle \vec{\psi}_{b{\mib k}}|\vec{\psi}_{b'{\mib k}'} \rangle 
\!=\! \delta_{bb'}\delta_{{\mib k}{\mib k}'}$ and form a complete set.
For our consideration, however, 
the wave functions (\ref{Bloch}) may not necessarily be the
eigenfunctions of eq.\ (\ref{Hamiltonian}) at ${\mib A}\!=\!{\mib 0}$.
Thus, we will proceed by assuming only the completeness and
orthonormality of eq.\ (\ref{Bloch}).
Note that, relaxing the conditions in this way,
$\vec{\psi}_{b{\mib k}}({\mib r})$ may be chosen 
analytic in ${\mib k}$ throughout the Brillouin zone, 
as shown by des Cloizeaux\cite{desCloizeaux64a,desCloizeaux64b} 
and Nenciu.\cite{Nenciu83,Nenciu91}
Those basis functions,
which are analytic in ${\mib k}$ but not 
necessarily the eigenfunctions of $\hat{\cal H}$,
were named quasi-Bloch functions
by des Cloizeaux.\cite{desCloizeaux64a,desCloizeaux64b}

A set of alternative basis functions was
introduced by Wannier,\cite{Wannier37} which are more suitable for the 
present purpose.
They are defined as a Fourier transform of 
$\vec{\psi}_{b{\mib k}}({\mib r})$ by
\begin{equation}
\vec{w}_{b{\mib R}}({\mib r})\equiv\frac{1}{\sqrt{N_{\rm c}}}
\sum_{\mib k}{\rm e}^{-i{\mib k}\cdot{\mib R}}\,
\vec{\psi}_{b{\mib k}}({\mib r}) \, ,
\label{Wannier}
\end{equation}
where ${\mib R}$ specifies a unit cell and
$N_{\rm c}$ denotes the number of
unit cells in the system.
It hence follows that 
the Wannier functions are also complete 
$\sum_{b{\mib R}}|\vec{w}_{b{\mib R}} \rangle\langle \vec{w}_{b{\mib R}}|
\!=\! 1$ and orthonormal as
$\langle \vec{w}_{b{\mib R}}|\vec{w}_{b'{\mib R}'} \rangle
\!=\! \delta_{bb'}\delta_{{\mib R}{\mib R}'}$
with
$\vec{w}_{b{\mib R}}({\mib r})\!=\!\vec{w}_{b {\mib 0}}({\mib r}\!-\!{\mib R})$.

To describe Bloch electrons in a finite magnetic field,
we adopt the basis functions introduced by Luttinger.\cite{Luttinger51}
They differ from eq.\ (\ref{Wannier}) by only a phase factor as
\begin{subequations}
\label{basis}
\begin{equation}
\vec{w}_{b{\mib R}}^{\,\prime}({\mib r})=
{\rm e}^{iI_{{\mib r}{\mib R}}}\vec{w}_{b{\mib R}}({\mib r}) \, ,
\label{basis0}
\end{equation}
where $I_{{\mib r}{\mib R}}$ is
defined by
\begin{equation}
I_{{\mib r}{\mib R}}\equiv \frac{e}{\hbar c}
\int_{{\mib R}}^{{\mib r}}{\mib A}({\mib r}')\cdot {\rm d}{\mib r}' \, ,
\label{phase}
\end{equation}
\end{subequations}
with ${\rm d}{\mib r}'$ taken
along the straight line path from ${\mib R}$ to ${\mib r}$.
We assume that the functions $\{\vec{w}_{b{\mib R}}^{\,\prime}\}$ 
form a complete set, although they are not orthonormal in a finite magnetic field.
This latter inconvenience can be removed by considering the linear 
combination:
\begin{equation}
\vec{\varphi}_{b{\mib R}}({\mib r})=\sum_{b'}\sum_{{\mib R}'}
{\cal S}_{b'{\mib R}',b{\mib R}}\,\vec{w}_{b'{\mib R}'}^{\,\prime}({\mib r}) \, ,
\label{varphi}
\end{equation}
and orthonormalize them as 
$\langle
\vec{\varphi}_{b{\mib R}}|\vec{\varphi}_{b'{\mib R}'}\rangle\!=\!\delta_{bb'}
\delta_{{\mib R}{\mib R}'}$; this will be performed shortly below.

We now transform eq.\ (\ref{Hamiltonian}) into a matrix
form by using eq.\ (\ref{varphi}).
Let us introduce the matrices:
\begin{subequations}
\begin{equation}
\underline{{\cal S}}=({\cal S}_{b{\mib R},b'{\mib R}'}) \, ,
\label{calS}
\end{equation}
\begin{equation}
\underline{{\cal O}}=(\langle \vec{w}_{b{\mib R}}^{\,\prime}|\vec{w}_{b'{\mib R}'}^{\,\prime}\rangle) \, ,
\label{calN'}
\end{equation}
\begin{equation}
\underline{\cal H}'=(\langle \vec{w}_{b{\mib R}}^{\,\prime}|\hat{\cal H}|
\vec{w}_{b'{\mib R}'}^{\,\prime}\rangle) \, .
\label{calH'}
\end{equation}
\end{subequations}
Note that $\underline{{\cal O}}$ is positive-definite Hermitian
and reduces to a unit matrix as ${\mib B}\!\rightarrow\!{\mib 0}$.
Now, the orthonormality $\langle
\vec{\varphi}_{b{\mib R}}|\vec{\varphi}_{b'{\mib R}'}\rangle\!=\!\delta_{bb'}
\delta_{{\mib R}{\mib R}'}$ reads
\begin{equation}
\underline{{\cal S}}^{\dagger}\underline{{\cal O}}\,\underline{{\cal S}}
= \underline{1} \, ,
\label{calSO-0}
\end{equation}
where $\underline{1}$ is the unit matrix:
$\underline{1}\!=\!(\delta_{bb'}\delta_{{\mib R}{\mib R}'})$.
Equation (\ref{calSO-0}) is solved easily by
choosing $\underline{{\cal S}}$ as Hermitian
$\underline{{\cal S}}^{\dagger}\!=\!\underline{{\cal S}}$
as
\begin{equation}
\underline{{\cal S}}=
\underline{{\cal O}}^{-1/2}\, .
\label{calSO}
\end{equation}
We next express the eigenstate of eq.\ (\ref{Hamiltonian}) 
as a linear combination of eq.\ (\ref{varphi}) as
$\vec{\psi}({\mib r})\!=\! \sum_{b{\mib R}}c_{b}({\mib R})
\vec{\varphi}_{b{\mib R}}({\mib r})$.
Then the eigenvalue problem of eq.\ 
(\ref{Hamiltonian}) is transformed into
\begin{equation}
\sum_{b'{\mib R}'}{\cal H}_{b{\mib R},b'{\mib R}'}\, c_{b'}({\mib R}')
={\cal E} \hspace{0.1mm} c_{b}({\mib R}) \, ,
\label{Schroedinger}
\end{equation}
with
\begin{equation}
\underline{\cal H}\equiv \underline{{\cal S}}\;
\underline{\cal H}'\,\underline{{\cal S}} \, .
\label{calH}
\end{equation}

A couple of comments are in order on eqs.\ (\ref{varphi}) 
and (\ref{calSO}) in connection with the treatment by Roth.\cite{Roth62}
First, our basis functions (\ref{varphi}) with the phase integral (\ref{phase})
have an advantage over Roth's basis functions that 
an extension to a non-uniform magnetic field can be performed easily. 
Whereas Roth had to assume that 
the field is uniform from the beginning, there is no limitation at this stage
on the form of ${\mib A}({\mib r})$ in eq.\ (\ref{phase}).
Second, eq.\ (\ref{calSO}) is different from Roth's choice.
Indeed, Roth expressed $\underline{\cal S}$ as a sum of Hermitian and anti-Hermitian
matrices and fixed the anti-Hermitian part by imposing
that $\underline{\cal H}$ be semi-diagonal up to the order of $B^{2}$.
However, eq.\ (\ref{calSO}) has the advantage that the formulation becomes 
more transparent. 
It helps to avoid unnecessary complications at the one-particle level and make
the extension to interacting systems easier.
Once the orthonormality is endowed as eq.\ (\ref{calSO}), 
the semi-diagonalization may be performed by a similarity transformation.
It should be noted, however, that the physical results are irrelevant of whether
one carries out such a similarity transformation at an intermediate
step.
Indeed, our procedure without further similarity transformations
leads exactly the same expression 
for the non-interacting magnetic susceptibility 
as that of Roth.

\subsection{Overlap integral}
\label{subsec:OI}

We first concentrate on the overlap
$\langle \vec{w}_{b{\mib R}}^{\,\prime}|\vec{w}_{b'{\mib R}'}^{\,\prime}\rangle$,
where the phase factor in eq.\ (\ref{basis}) yields
a line integral from ${\mib R}'$ to ${\mib R}$ via ${\mib r}$.
We transform it as follows:
\begin{eqnarray}
&&\hspace{-5mm} I_{{\mib R}{\mib r}}+I_{{\mib r}{\mib R}'}
= I_{{\mib R}{\mib R}'}+\frac{e}{\hbar c}
\int_{C}{\mib A}({\mib r}')\cdot {\rm d}{\mib r}'
\nonumber \\
&&\hspace{-5mm}=  I_{{\mib R}{\mib R}'} +\frac{e}{\hbar c}(
{\mib r}_{{\mib R}}\!\times\!{\mib r}_{{\mib R}'})\cdot\int_{0}^{1}\!\! {\rm d}u
\int_{0}^{1-u}\!\! {\rm d}v \,{\mib B}
\nonumber \\
&&\hspace{-5mm}=  I_{{\mib R}{\mib R}'} +\frac{e}{2\hbar c}
{\mib B} \cdot ({\mib r}_{{\mib R}}\! \times \! {\mib r}_{{\mib R}'}) \, .
\label{phase2}
\end{eqnarray}
Here $C$ denotes closed path 
${\mib R}'\!\rightarrow\!{\mib r}\!\rightarrow\!{\mib R}\!\rightarrow\!
{\mib R}'$ along the triangle, 
${\mib r}_{{\mib R}}$ is defined by
\begin{equation}
{\mib r}_{{\mib R}}\!\equiv\! {\mib r}-{\mib R} \, ,
\label{I_n'n-def}
\end{equation}
and the second line follows via Stokes' theorem
by writing the vector area element as ${\rm d}{\mib S}\!=\!
({\mib R}'\!-\!{\mib R}){\rm d}u \!\times \!({\mib r}\!-\!{\mib R}){\rm d}v$.
The transformation (\ref{phase2}) was used previously
in deriving a transport equation
for superconductors with Hall terms.\cite{Kita01}
Although we have assumed a uniform magnetic field here, 
an extension to a non-uniform field can be performed easily
as eqs.\ (31) and (33) of Ref.\ \citen{Kita01}.

Following Roth,\cite{Roth62} let us introduce the antisymmetric tensors:
\begin{equation}
h_{\alpha\beta}=\epsilon_{\alpha\beta\gamma}\frac{e}{2\hbar c}B_{\gamma}\, ,
\hspace{5mm}
\hat{\sigma}^{\alpha\beta}=\epsilon_{\alpha\beta\gamma}
\frac{g\hbar^{2}}{2m}\hat{s}^{\gamma} \, ,
\label{sigma-def}
\end{equation}
where $\epsilon_{\alpha\beta\gamma}$ $(\alpha, \beta, \gamma \!=\! x,y,z)$
is the third-rank completely antisymmetric tensor,
and summations over the repeated index $\gamma$ are implied.
The quantity $h_{\alpha\beta}$ is essentially inverse 
of the magnetic length squared.
Also useful is the following identity
for an arbitrary function $f({\mib r}_{{\mib R}},{\mib r}_{{\mib R}'})$:
\begin{eqnarray}
&&\hspace{-9mm}
f({\mib r}_{{\mib R}},{\mib r}_{{\mib R}'})\vec{w}_{b{\mib R}}^{\dagger}({\mib r})
\vec{w}_{b'{\mib R}'}({\mib r})
\nonumber \\
&&\hspace{-9mm}
=\frac{1}{N_{\rm c}}
\sum_{{\mib k}{\mib k}'}f({\mib r}_{{\mib R}},{\mib r}_{{\mib R}'})
\vec{u}_{b{\mib k}}^{\dagger}({\mib r})
\vec{u}_{b'{\mib k}'}({\mib r})\,
{\rm e}^{-i{\mib k}\cdot{\mib r}_{{\mib R}}+i{\mib k}'\cdot{\mib r}_{{\mib R}'}}
\nonumber \\
&&\hspace{-9mm}
=\frac{1}{N_{\rm c}}
\sum_{{\mib k}{\mib k}'}\vec{u}_{b{\mib k}}^{\dagger}({\mib r})
\vec{u}_{b'{\mib k}'}({\mib r})
f(i{\mib \nabla}_{\!{\mib k}},-i{\mib \nabla}_{\!{\mib k}'})\,
{\rm e}^{-i{\mib k}\cdot{\mib r}_{{\mib R}}+i{\mib k}'\cdot{\mib r}_{{\mib R}'}}
\nonumber \\
&&\hspace{-9mm}
=\frac{1}{N_{\rm c}}
\sum_{{\mib k}{\mib k}'}\left[
f(-i{\mib \nabla}_{\!{\mib k}},i{\mib \nabla}_{\!{\mib k}'})
\vec{u}_{b{\mib k}}^{\dagger}({\mib r})
\vec{u}_{b'{\mib k}'}({\mib r})\right]
\nonumber \\
&&\hspace{5mm}
\times{\rm e}^{-i{\mib k}\cdot{\mib r}_{{\mib R}}+i{\mib k}'\cdot{\mib r}_{{\mib R}'}}
\, ,
\label{k-diff}
\end{eqnarray}
where we substituted eqs.\ (\ref{Bloch}) and (\ref{Wannier})
and performed partial integrations over ${\mib k}$ and ${\mib k}'$.

Using eqs.\ (\ref{phase2})-(\ref{k-diff})
and the periodicity 
$\vec{u}_{b{\mib k}}({\mib r}+{\mib R})\!=\!
\vec{u}_{b{\mib k}}({\mib r})$,
we obtain an expression for eq.\ (\ref{calN'}) as
\begin{eqnarray}
&&\hspace{-9mm}
{\cal O}_{b{\mib R},b'{\mib R}'}=
{\rm e}^{iI_{{\mib R}{\mib R}'}}\!\!\int\! \exp(ih_{\alpha\beta} \,
r_{{\mib R}}^{\alpha} r_{{\mib R}'}^{\beta})\,
\vec{w}_{b{\mib R}}^{\dagger}({\mib r})
\vec{w}_{b'{\mib R}'}({\mib r})\,
{\rm d}{\mib r}
\nonumber \\
&&\hspace{5mm}=
\frac{\displaystyle{\rm e}^{iI_{{\mib R}{\mib R}'}}}{N_{\rm c}}\sum_{{\mib k}}
{\rm e}^{i{\mib k}\cdot({\mib R}-{\mib R}')}
{\cal O}_{bb'}({\mib k},{\mib B}) \, ,
\label{Norm}
\end{eqnarray}
where ${\cal O}_{bb'}({\mib k},{\mib B})$ is given by
\begin{subequations}
\label{Norm-kt}
\begin{equation}
{\cal O}_{bb'}({\mib k},{\mib B})\equiv\left.
\!\int
{\rm e}^{ih_{\alpha\beta} 
\nabla_{{\mib k}}^{\alpha} \nabla_{{\mib k}'}^{\beta}}
\,\vec{u}_{b{\mib k}}^{\dagger}({\mib r})
\vec{u}_{b'{\mib k}'}({\mib r})\,\right|_{{\mib k}'={\mib k}}
{\rm d}{\mib r} \, .
\label{Norm-k}
\end{equation}
The expression (\ref{Norm}) has a general form for any matrix element between
eq.\ (\ref{basis}), i.e., the Peierls phase factor\cite{Peierls33} 
${\rm e}^{iI_{{\mib R}{\mib R}'}}$ times
a summation over ${\mib k}$ for the product of
${\rm e}^{i{\mib k}\cdot({\mib R}-{\mib R}')}$ and some function of ${\mib k}$.

It is useful for practical purposes 
to expand eq.\ (\ref{Norm-k}) in terms of ${\mib B}$ as
\begin{equation}
{\cal O}_{bb'}({\mib k},{\mib B})=\delta_{bb'}+
\sum_{j=1}^{\infty}{\cal O}_{bb'}^{(j)}({\mib k},{\mib B}) \, ,
\label{Norm-kE}
\end{equation}
\end{subequations}
where ${\cal O}_{bb'}^{(j)}$ is the 
quantity of the order $B^{j}$.
The terms for $j\!=\! 1,2$ are given explicitly by
\begin{subequations}
\label{Norm0-2}
\begin{equation}
{\cal O}_{bb'}^{(1)}=
ih_{\alpha\beta}
\langle \nabla_{\! {\mib k}}^{\alpha}
\vec{u}_{b{\mib k}}|
\nabla_{\! {\mib k}}^{\beta}\vec{u}_{b'{\mib k}}\rangle
= ih_{\alpha\beta} 
\sum_{b''}x_{bb''}^{\alpha}({\mib k})x_{b''b'}^{\beta}({\mib k})
\, ,
\label{Norm(1)}
\end{equation}
\begin{equation}
{\cal O}_{bb'}^{(2)}=
-\frac{1}{2}h_{\alpha\beta} h_{\alpha'\beta'} 
\langle\nabla_{\! {\mib k}}^{\alpha'}\nabla_{\! {\mib k}}^{\alpha}
\vec{u}_{b{\mib k}}^{\dagger}|
\nabla_{\! {\mib k}}^{\beta'}\nabla_{\! {\mib k}}^{\beta}\vec{u}_{b'{\mib k}}
\rangle  
\, ,
\label{Norm(2)}
\end{equation}
\end{subequations}
with
\begin{eqnarray}
&&\hspace{-9mm}
x_{bb'}^{\alpha}({\mib k})\equiv i\langle
\vec{u}_{b{\mib k}}|
\nabla_{\! {\mib k}}^{\alpha}\vec{u}_{b'{\mib k}}\,
\rangle =-i\langle 
\nabla_{\! {\mib k}}^{\alpha}\vec{u}_{b{\mib k}}|
\vec{u}_{b'{\mib k}}\rangle
\nonumber \\
&&\hspace{2.5mm}
=\sum_{\mib R}{\rm e}^{i{\mib k}\cdot{\mib R}}
\langle \vec{w}_{b{\mib 0}}|\,r_{\mib R}^{\alpha}\,|
\vec{w}_{b'{\mib R}}\rangle \, .
\label{x-def}
\end{eqnarray}
The second expressions of eqs.\ (\ref{Norm(1)}) and (\ref{x-def}) 
have been derived by noting
$\vec{u}_{b{\mib k}}({\mib r}+{\mib R})\!=\!
\vec{u}_{b{\mib k}}({\mib r})$,
reducing the integral into a unit cell, and
using the completeness and orthonormality
of $\{\vec{u}_{b{\mib k}}\}$ over the unit cell.\cite{Roth62}
The last expression of eq.\ (\ref{x-def}) is given entirely with respect to the
Wannier functions of eq.\ (\ref{Wannier}).
Hence it may be suitable for the evaluation of
the matrix element $x_{bb'}^{\alpha}({\mib k})$ between the core states.
Equations (\ref{Norm-k}) and (\ref{Norm0-2}) are exactly
those obtained by Roth.\cite{Roth62}

A couple of comments are in order before closing 
the section. First, the expansion
of eq.\ (\ref{Norm-kt}) is convergent if we choose 
$\vec{\psi}_{b{\mib k}}({\mib r})$ as the quasi-Bloch functions
of des Cloizeaux\cite{desCloizeaux64a,desCloizeaux64b,Nenciu83,Nenciu91}
which are analytic in ${\mib k}$.
This property can be satisfied for a simple band by the eigenfunctions
of eq.\ (\ref{Hamiltonian}), but not for
a complex band with band crossings.
This statement holds for every expansion we shall encounter below.
Second, matrix elements like eqs.\ (\ref{Norm0-2}) and
(\ref{x-def}) cannot be determined
uniquely due to the gauge degree of freedom:
$$\vec{u}_{b{\mib k}}({\mib r})\!\rightarrow\!
{\rm e}^{i\phi_{b{\mib k}}}\vec{u}_{b{\mib k}}({\mib r})\, .$$
However, obsevable quantities such as the magnetic susceptibility
do not depend on the choice of the phase $\phi_{b{\mib k}}$.
See also the discussions of Roth\cite{Roth62} on this point, and
those by Resta\cite{Resta94} for the electronic polarization.

\subsection{Expression of $\underline{\cal S}$}
\label{subsec:MatS}

We next derive an expression of $\underline{{\cal S}}$
in powers of ${\mib B}$ starting from eq.\ (\ref{calSO-0})
and $\underline{\cal S}^{\dagger}\!=\!\underline{\cal S}$.
Equation (\ref{Norm}) suggests that we may expand
the matrix element as
\begin{equation}
{\cal S}_{b{\mib R},b'{\mib R}'}=
\frac{\displaystyle{\rm e}^{iI_{{\mib R}{\mib R}'}}}{N_{\rm c}}\sum_{{\mib k}}
{\rm e}^{i{\mib k}\cdot({\mib R}-{\mib R}')}
{\cal S}_{bb'}({\mib k},{\mib B}) \, ,
\label{calSRk}
\end{equation}
with $\underline{\cal S}^{\dagger}({\mib k},{\mib B})\!=\!\underline{\cal S}({\mib k},{\mib B})$.
Let us substitute eqs.\ (\ref{Norm}) and (\ref{calSRk}) into eq.\
(\ref{calSO-0}).
We then encounter a summation
over ${\mib R}''$ with both
the Peierls phase factor ${\rm e}^{iI_{{\mib R}{\mib R}''}+iI_{{\mib R}''{\mib R}'}}$
and the Bloch phase factor 
${\rm e}^{i{\mib k}\cdot({\mib R}-{\mib R}'')
+i{\mib k}'\cdot({\mib R}''-{\mib R}')}$.
This summation can also be transformed 
with the procedures of eqs.\ (\ref{phase2}) and (\ref{k-diff}).
For example, we obtain
\begin{eqnarray}
&&\hspace{-9mm}
\sum_{b''{\mib R}''}{\cal S}_{b{\mib R},b''{\mib R}''}
{\cal O}_{b''{\mib R}'',b'{\mib R}'}
\nonumber \\
&&\hspace{-9mm}
=\frac{{\rm e}^{iI_{{\mib R}{\mib R}'}}}{N_{\rm c}}\sum_{b''{\mib k}}
{\rm e}^{i{\mib k}\cdot({\mib R}-{\mib R}')}{\cal S}_{bb''}({\mib k})\!\otimes\!
{\cal O}_{b''b'}({\mib k}) \, ,
\label{MR}
\end{eqnarray}
where the operator $\otimes$ is defined by
\begin{equation}
\underline{\cal S}({\mib k})\!\otimes\!
\underline{\cal O}({\mib k})\equiv \left.
{\rm e}^{ih_{\alpha\beta} 
\nabla_{{\mib k}'}^{\alpha} \nabla_{{\mib k}}^{\beta}}\,
\underline{\cal S}({\mib k}')\,\underline{\cal O}({\mib k})
\right|_{{\mib k}'={\mib k}}\, .
\label{otimes}
\end{equation}
Equation (\ref{otimes}) is exactly the multiplication theorem
of Roth.\cite{Roth62}

It is worth summarizing the properties of the operator $\otimes$.
First, we realize from eq.\ (\ref{MR}) that 
it is associative as
\begin{subequations}
\label{otimes-p}
\begin{equation}
{\cal A}({\mib k})\!\otimes\![{\cal B}({\mib k})\!\otimes\!{\cal C}({\mib k})]
\!=\![{\cal A}({\mib k})\!\otimes\!{\cal B}({\mib k})]\!\otimes\!{\cal C}({\mib k}) \, .
\label{otimes-p1}
\end{equation}
Second, it satisfies
\begin{equation}
\sum_{\mib k}{\cal A}({\mib k})\!\otimes\!{\cal B}({\mib k})=
\sum_{\mib k}{\cal A}({\mib k})\,{\cal B}({\mib k}) \, ,
\label{otimes-p2}
\end{equation}
\end{subequations}
as can be shown with partial integrations over ${\mib k}$
and the antisymmetry $h_{\alpha\beta}\!=\!-h_{\beta\alpha}$.

By using eqs.\ (\ref{Norm}), (\ref{calSRk}) and (\ref{MR}), 
eq.\ (\ref{calSO-0}) with $\underline{\cal S}^{\dagger}\!=\!\underline{\cal S}$ 
is transformed into
\begin{subequations}
\label{calS-kt}
\begin{equation}
\underline{\cal S}({\mib k},{\mib B})\otimes
\underline{\cal O}({\mib k},{\mib B})\otimes\underline{\cal S}({\mib k},{\mib B})
=\underline{1}
\, ,
\label{calS-k}
\end{equation}
with $\underline{1}\!\equiv\!(\delta_{bb'})$.
Let us expand $\underline{{\cal S}}$ in powers of ${\mib B}$ as
\begin{equation}
\underline{\cal S}({\mib k},{\mib B})=\underline{1}+
\sum_{j=1}^{\infty}\underline{\cal S}^{(j)}({\mib k},{\mib B}) \, .
\label{calS-kE}
\end{equation}
\end{subequations}
We then find from eqs.\ (\ref{Norm-kE}), (\ref{calS-k}), and
(\ref{calS-kE}) that $\underline{\cal S}^{(j)}$
for $j\!=\! 1,2$ are given by
\begin{subequations}
\label{calS0-2}
\begin{equation}
\underline{\cal S}^{(1)}=-\frac{1}{2}\underline{\cal O}^{(1)}
\, ,
\label{calS(1)}
\end{equation}
\begin{equation}
\underline{\cal S}^{(2)}=-\frac{1}{2}\underline{\cal O}^{(2)}
+\frac{3}{8}\,\underline{\cal O}^{(1)}\underline{\cal O}^{(1)}
\, .
\label{calS(2)}
\end{equation}
\end{subequations}

\subsection{Matrix element of Hamiltonian}
\label{subsec:MatH}

We next consider
${\cal H}'_{b{\mib R},b'{\mib R}'}\!\equiv\!\langle \vec{w}_{b{\mib R}}^{\,\prime}|
\hat{\cal H}|\vec{w}_{b'{\mib R}'}^{\,\prime}\rangle$
with $\hat{\cal H}$ and $\vec{w}_{b{\mib R}}^{\,\prime}$ given by
eqs.\ (\ref{Hamiltonian}) and (\ref{basis}), respectively.
The following identity is useful for this purpose:\cite{Luttinger51}
\begin{eqnarray}
&&\hspace{-9mm}
{\mib \nabla}I_{{\mib r}{\mib R}}
=\frac{e}{\hbar c}{\mib A}({\mib r})+
\frac{e}{\hbar c}{\mib r}_{{\mib R}}\!\times\!
\int_{0}^{1}
{\mib B}\,u \,{\rm d}u
\nonumber \\
&&\hspace{1mm}
= \frac{e}{\hbar c}{\mib A}({\mib r})-\frac{e}{2\hbar c}{\mib B}
\!\times\!{\mib r}_{{\mib R}} \, .
\label{dI}
\end{eqnarray}
It is worth noting that this identity can also be extended easily
to a non-uniform magnetic field.\cite{Luttinger51}
Using eqs.\ (\ref{phase2}), (\ref{k-diff}) and (\ref{dI})
as well as
$\vec{u}_{b{\mib k}}({\mib r}\!+\!{\mib R})
\!=\!\vec{u}_{b{\mib k}}({\mib r})$,
the above matrix element is transformed into
\begin{equation}
{\cal H}'_{b{\mib R},b'{\mib R}'}
=\frac{\displaystyle{\rm e}^{iI_{{\mib R}{\mib R}'}}}{N_{\rm c}}\sum_{{\mib k}}
{\rm e}^{i{\mib k}\cdot({\mib R}-{\mib R}')}
{\cal H}_{bb'}^{\prime}({\mib k},{\mib B}) \, ,
\label{Hamil-M2}
\end{equation}
where ${\cal H}_{bb'}^{\prime}$ is defined by
\begin{subequations}
\label{Hamil-kt}
\begin{eqnarray}
&&\hspace{-9mm}
{\cal H}_{bb'}^{\prime}({\mib k},{\mib B})
\nonumber \\
&&\hspace{-9mm}
\equiv\!
\int {\rm e}^{ih_{\alpha\beta} 
\nabla_{{\mib k}'}^{\alpha} \nabla_{{\mib k}}^{\beta}} \,
\vec{u}_{b{\mib k}'}^{\dagger}({\mib r})
\biggl[ \hat{\cal H}^{(0)}_{{\mib k}}\!
-h_{\alpha'\beta'}
(i\nabla_{\!{\mib k}}^{\alpha'}\hat{\pi}^{\beta'}_{{\mib k}}
+\hat{\sigma}^{\alpha'\beta'})
\nonumber \\
&&\hspace{-5mm}
-\frac{\hbar^{2}}{2m}h_{\alpha'\beta'}h_{\alpha'\gamma'}
\nabla_{\!{\mib k}}^{\beta'}\nabla_{\!{\mib k}}^{\gamma'}\biggr]
\vec{u}_{b'{\mib k}}({\mib r})\,\biggr|_{{\mib k}'={\mib k}} {\rm d}{\mib r} 
\, ,
\label{Hamil-k}
\end{eqnarray}
with $\hat{\cal H}^{(0)}_{{\mib k}}\!\equiv\!\hat{\cal H}
(\hat{\mib p}\!\rightarrow\!\hat{\mib p}\!+\!\hbar{\mib k},{\mib A}\!=\!{\mib 0})$,
$\hat{\mib \pi}_{{\mib k}}\!\equiv\!\partial \hat{\cal H}^{(0)}_{{\mib k}}/\partial {\mib k}
\!= \!\frac{\hbar}{m}\!\left[\hat{\mib p}\!+\!\hbar{\mib k}
\!-\! g\frac{e\hbar}{2mc^{2}}\hat{\mib s}\!\times\! {\mib \nabla}{\cal V}({\mib r})\right]$,
and $\hat{\sigma}^{\alpha\beta}$ given in eq.\ (\ref{sigma-def}).

We now expand eq.\ (\ref{Hamil-k}) formally in terms of ${\mib B}$ as
\begin{equation}
{\cal H}_{bb'}^{\prime}({\mib k},{\mib B})
={\cal H}_{bb'}^{(0)}({\mib k})+
\sum_{j=1}^{\infty}{\cal H}_{bb'}^{\prime(j)}({\mib k},{\mib B}) \, ,
\label{Hamil-kH}
\end{equation}
\end{subequations}
with 
\begin{subequations}
\begin{equation}
{\cal H}^{(0)}_{bb'}({\mib k})\equiv 
\langle \vec{u}_{b{\mib k}}|
\hat{\cal H}^{(0)}_{\mib k}|\vec{u}_{b'{\mib k}}\rangle \, .
\label{barH(0)}
\end{equation}
To write down ${\cal H}_{bb'}^{\prime(j)}({\mib k},{\mib B})$ 
in eq.\ (\ref{Hamil-kH}) explicitly,
we introduce the following matrices:
\begin{equation}
\tilde{\mib v}_{bb'}({\mib k})\equiv \frac{1}{\hbar}\,
\frac{\partial {\cal H}^{(0)}_{bb'}({\mib k})}{
\partial {\mib k}} \, ,
\label{v-def}
\end{equation}
\begin{equation}
{\mib \pi}_{bb'}({\mib k})\equiv \biggl< \!\vec{u}_{b{\mib k}}\left|
\frac{\hbar}{m}\!\left[\hat{\mib p}+\hbar{\mib k}
- g\frac{e\hbar}{2mc^{2}}\hat{\mib s}\!\times\! {\mib \nabla}{\cal V}({\mib r})\right]
\right|\vec{u}_{b'{\mib k}}\!\biggr> \, ,
\label{pi}
\end{equation}
\begin{equation}
\sigma_{bb'}^{\alpha\beta}({\mib k})\equiv \langle \vec{u}_{b{\mib k}}|
\hat{\sigma}^{\alpha\beta}|\vec{u}_{b'{\mib k}}\rangle \, .
\label{sigma}
\end{equation}
\end{subequations}
Note $\hbar \tilde{\underline{v}}^{\alpha}\!=\! \underline{\pi}^{\alpha}+
i(\underline{x}^{\alpha}\underline{\cal H}^{(0)}-
\underline{\cal H}^{(0)}\underline{x}^{\alpha})$,
as can be shown from
$\hat{\mib \pi}_{{\mib k}}\!=\!\partial \hat{\cal H}^{(0)}_{{\mib k}}/\partial {\mib k}$.
Next, we express $\hat{\cal H}^{(0)}_{{\mib k}}|
\vec{u}_{b'{\mib k}}\rangle$ and 
$ih_{\alpha'\beta'}\nabla_{\!{\mib k}}^{\alpha'}\hat{\pi}^{\beta'}_{{\mib k}}|
\vec{u}_{b'{\mib k}}\rangle$ in eq.\ (\ref{Hamil-k}) as
\begin{eqnarray*}
\hat{\cal H}^{(0)}_{{\mib k}}|
\vec{u}_{b'{\mib k}}\rangle \!=\!\sum_{b''}|
\vec{u}_{b''{\mib k}}\,\rangle {\cal H}^{(0)}_{b''b'}({\mib k}) \, ,
\end{eqnarray*}
$$
ih_{\alpha'\beta'}\nabla_{\!{\mib k}}^{\alpha'}\hat{\pi}^{\beta'}_{{\mib k}}|
\vec{u}_{b'{\mib k}}\rangle\! =h_{\alpha'\beta'}\!
\sum_{b''}\! |
\vec{u}_{b''{\mib k}}\rangle\!\left[\underline{\pi}^{\beta'}\!({\mib k})\,
\underline{x}^{\alpha'}\!({\mib k})\right]_{b''b'}  .
$$
We then expand eq.\ (\ref{Hamil-k}) in powers of ${\mib B}$
and perform straightforward calculations order by order
to obtain $\underline{\cal H}^{\prime(j)}$ in eq.\ (\ref{Hamil-kH}).

The first-order term can be written down easily.
Averaging the resulting expression
with its Hermitian conjugate, we obtain
\begin{subequations}
\label{Hamil0-2}
\begin{equation}
\underline{\cal H}^{\prime(1)}\!
=\{\underline{\cal O}^{(1)}\!,\underline{\cal H}^{(0)}\}
-h_{\alpha\beta}\bigl( \{\underline{x}^{\alpha},
\hbar \tilde{\underline{v}}^{\beta}\!+\!\underline{\pi}^{\beta}\}
\!+\!\underline{\sigma}^{\alpha\beta}\bigr) \, ,
\label{Hamil(1)}
\end{equation}
with $\underline{\cal O}^{(1)}$ given by eq.\ (\ref{Norm(1)})
and
$\{\underline{\cal A},\underline{\cal B}\}\!\equiv\!
\frac{1}{2}(\underline{\cal A}\,\underline{\cal B}
\!+\! \underline{\cal B}\,\underline{\cal A})$.

To obtain $\underline{\cal H}^{\prime(2)}$ compactly,
on the other hand, we carry out partial integrations using
the antisymmetry $h_{\alpha\beta}\!=\!-h_{\beta\alpha}$.
For example, one of the relevant terms are transformed as
\begin{eqnarray*}
&&\hspace{-9mm}
-\frac{1}{2}h_{\alpha\beta}h_{\alpha'\beta'}\langle 
\nabla_{\!{\mib k}}^{\alpha'} \nabla_{\!{\mib k}}^{\alpha} \vec{u}_{b{\mib k}}|
\nabla_{\!{\mib k}}^{\beta'} \nabla_{\!{\mib k}}^{\beta} 
\hat{\cal H}^{(0)}_{{\mib k}}\vec{u}_{b'{\mib k}}\rangle
\nonumber \\
&&\hspace{-9mm}
=\sum_{b''}
\biggl\{\!{\cal O}_{bb''}^{(2)}{\cal H}^{(0)}_{b''b'}
-\biggl[
\nabla_{\!{\mib k}}^{\beta}\!\left(\!
\langle 
\nabla_{\!{\mib k}}^{\alpha'} \nabla_{\!{\mib k}}^{\alpha} \vec{u}_{b{\mib k}}|
\vec{u}_{b''{\mib k}}\rangle \hbar \tilde{v}_{b''b'}^{\beta'} \!\right)
\nonumber \\
&&\hspace{-5mm}
+ \frac{1}{2}\langle 
\nabla_{\!{\mib k}}^{\alpha'} \nabla_{\!{\mib k}}^{\alpha} \vec{u}_{b{\mib k}}|
\vec{u}_{b''{\mib k}}\rangle \nabla_{\!{\mib k}}^{\beta'}\nabla_{\!{\mib k}}^{\beta}
{\cal H}^{(0)}_{b''b'}({\mib k})\biggr] h_{\alpha\beta}h_{\alpha'\beta'}\!
\biggr\}\, ,
\end{eqnarray*}
with
\begin{eqnarray*}
&&\hspace{-9mm}
\langle 
\nabla_{\!{\mib k}}^{\alpha'} \nabla_{\!{\mib k}}^{\alpha} \vec{u}_{b{\mib k}}|
\vec{u}_{b'{\mib k}}\rangle = i\nabla_{\!{\mib k}}^{\alpha'}x_{bb'}^{\alpha}
-\sum_{b''}x_{bb''}^{\alpha}x_{b''b'}^{\alpha'}
\nonumber \\
&&\hspace{17.5mm}
= i\nabla_{\!{\mib k}}^{\alpha}x_{bb'}^{\alpha'}
-\sum_{b''}x_{bb''}^{\alpha'}x_{b''b'}^{\alpha}
\, ,
\end{eqnarray*}
and $h_{\alpha\beta}\nabla_{\!{\mib k}}^{\beta}[
(\nabla_{\!{\mib k}}^{\alpha}x_{bb''}^{\alpha'})
\hbar \tilde{v}_{b''b'}^{\beta'}]\!=\!
h_{\alpha\beta}(\nabla_{\!{\mib k}}^{\alpha}x_{bb''}^{\alpha'})
\nabla_{\!{\mib k}}^{\beta}\nabla_{\!{\mib k}}^{\beta'}{\cal H}^{(0)}_{b''b'}$.
Similar calculations are performed for the other terms
to make $\underline{\cal H}^{\prime(2)}$
compact and symmetric.
We finally perform the averaging with the Hermitian conjugate.
We thereby obtain
\begin{eqnarray}
&&\hspace{-9mm}\underline{\cal H}^{\prime(2)}
= \{\underline{\cal O}^{(2)},\underline{\cal H}^{(0)}\}
+h_{\alpha\beta}h_{\alpha'\beta'}
\biggl\{\!\nabla_{\!{\mib k}}^{\beta}\biggl[\bigl\{\!
\{\underline{x}^{\alpha}\!,\underline{x}^{\alpha'}\},
\hbar\tilde{\underline{v}}^{\beta'}\bigr\}
\nonumber \\
&&\hspace{2mm}+\frac{1}{2}
(
\underline{x}^{\alpha}\underline{\pi}^{\beta'}\underline{x}^{\alpha'}\!+\!
\underline{x}^{\alpha'}\underline{\pi}^{\beta'}\underline{x}^{\alpha}
)
+\{\underline{x}^{\alpha},\underline{\sigma}^{\alpha'\beta'}\}\biggr]
\nonumber \\
&&\hspace{2mm}+\frac{1}{2}
\bigl\{\{\underline{x}^{\alpha},\underline{x}^{\alpha'}\},
{\textstyle\frac{\hbar^{2}}{m}}\delta_{\beta\beta'}\underline{1}
-\nabla_{\!{\mib k}}^{\beta}
\nabla_{\!{\mib k}}^{\beta'}\,\underline{\cal H}^{(0)}\,
\bigr\}
\nonumber \\
&&\hspace{2mm}+\frac{i}{4}\nabla_{\!{\mib k}}^{\beta}\nabla_{\!{\mib k}}^{\beta'}
[\,\underline{x}^{\alpha'},\hbar\tilde{\underline{v}}^{\alpha}]
\biggr\}
 \, ,
\label{Hamil(2)}
\end{eqnarray}
\end{subequations}
with
$\{\underline{A},\underline{B}\}\!\equiv\!\frac{1}{2}(\underline{A}\,\underline{B}
\!+\! \underline{B}\,\underline{A})$ and
$[\,\underline{A},\underline{B}\,]\!\equiv\!\underline{A}\,\underline{B}
\!-\! \underline{B}\,\underline{A}$.
Equation (\ref{Hamil(2)}) is in agreement with eq.\ (57) of Roth.\cite{Roth62}

\subsection{Effective Hamiltonian}
\label{subsec:EH}

We are ready to derive an effective one-particle 
Hamiltonian in a finite magnetic field.
Let us substitute eqs.\ (\ref{calSRk}) and (\ref{Hamil-M2}) into
eq.\ (\ref{calH}).
Using eq.\ (\ref{MR}) twice, we obtain
\begin{equation}
{\cal H}_{b{\mib R},b'{\mib R}'}=
\frac{\displaystyle{\rm e}^{iI_{{\mib R}{\mib R}'}}}{N_{\rm c}}\sum_{{\mib k}}
{\rm e}^{i{\mib k}\cdot({\mib R}-{\mib R}')}
{\cal H}_{bb'}({\mib k},{\mib B}) \, ,
\label{calH-1a}
\end{equation}
with
\begin{subequations}
\label{calH-2t}
\begin{eqnarray}
&&\hspace{-10mm}
\underline{\cal H}({\mib k},{\mib B})=
\underline{\cal S}({\mib k},{\mib B})\!\otimes\!
\underline{\cal H}'({\mib k},{\mib B})\!\otimes\!
\underline{\cal S}({\mib k},{\mib B})\, .
\label{calH-2a}
\end{eqnarray}
Let us expand eq.\ (\ref{calH-2a}) in powers of ${\mib B}$ as
\begin{equation}
\underline{\cal H}({\mib k},{\mib B})=
\underline{\cal H}^{(0)}({\mib k})+
\sum_{j=1}^{\infty}
\underline{\cal H}^{(j)}({\mib k},{\mib B}) \, ,
\label{BarCalH-E}
\end{equation}
\end{subequations}
with $\underline{\cal H}^{(0)}({\mib k})$ given by eq.\ (\ref{barH(0)}).
The expressions of $\underline{\cal H}^{(j)}$
can be obtained order by order by
substituting eqs.\ (\ref{calS-kE}), (\ref{Hamil-kH}) and (\ref{BarCalH-E})
into eq.\ (\ref{calH-2a}) and using eqs.\ (\ref{calS0-2}) and (\ref{Hamil0-2}).
The first-order term is obtained from
$\underline{\cal H}^{(1)}\!=\!
\underline{\cal H}^{\prime(1)}
\!+\!2\{\underline{S}^{(1)},\underline{\cal H}^{(0)}\}$
as
\begin{subequations}
\label{barH}
\begin{equation}
\underline{\cal H}^{(1)}({\mib k})=
-h_{\alpha\beta}
\bigl(\{\underline{x}^{\alpha},\hbar\tilde{\underline{v}}^{\beta}
\!+\!\underline{\pi}^{\beta}\}
\!+\!\underline{\sigma}^{\alpha\beta}\bigr) \, ,
\label{barH(1)}
\end{equation}
with
$\{\underline{A},\underline{B}\}\!\equiv\!\frac{1}{2}(\underline{A}\,\underline{B}
\!+\! \underline{B}\,\underline{A})$.
The second-order contribution is calculated from
$\underline{\cal H}^{(2)}\!=\!
\underline{\cal H}^{\prime(2)}
\!+\!2\{\underline{S}^{(2)},\underline{\cal H}^{(0)}\}
\!+\!2\{\underline{S}^{(1)},\underline{\cal H}^{\prime(1)}\}
+\underline{S}^{(1)}\underline{\cal H}^{(0)}\underline{S}^{(1)}
-ih_{\alpha\beta}[\nabla_{\!{\mib k}}^{\beta}\underline{S}^{(1)},
\nabla_{\!{\mib k}}^{\alpha}\underline{\cal H}^{(0)}]$
as
\begin{eqnarray}
&&\hspace{-9mm} \underline{\cal H}^{(2)}({\mib k})=
h_{\alpha\beta}h_{\alpha'\beta'}
\biggl\{\!\nabla_{\!{\mib k}}^{\beta}\biggl[\bigl\{\!
\{\underline{x}^{\alpha}\!,\underline{x}^{\alpha'}\},
\hbar\tilde{\underline{v}}^{\beta'}\bigr\}
\nonumber \\
&&\hspace{7mm}
+ \frac{1}{2}\,
(
\underline{x}^{\alpha}\underline{\pi}^{\beta'}\underline{x}^{\alpha'}\!+\!
\underline{x}^{\alpha'}\underline{\pi}^{\beta'}\underline{x}^{\alpha}
)+\{\underline{x}^{\alpha},\underline{\sigma}^{\alpha'\beta'}\}
\biggr]
\nonumber \\
&&\hspace{7mm}
+\frac{1}{2}\,
\bigl\{\{\underline{x}^{\alpha},\underline{x}^{\alpha'}\},
{\textstyle\frac{\hbar^{2}}{m}}\delta_{\beta\beta'}\underline{1}
-\!\nabla_{\!{\mib k}}^{\beta}
\nabla_{\!{\mib k}}^{\beta'}\underline{\cal H}^{(0)}
\bigr\}
\nonumber \\
&&\hspace{7mm}
+\,\frac{i}{4}\nabla_{\!{\mib k}}^{\beta}\nabla_{\!{\mib k}}^{\beta'}
[\underline{x}^{\alpha'},\hbar\tilde{\underline{v}}^{\alpha}]\biggr\}
\nonumber \\
&&\hspace{7mm}
-\bigl\{\underline{\cal O}^{(1)},
\underline{\cal H}^{(1)}\bigr\}
+\frac{1}{8}\bigl[\,\underline{\cal O}^{(1)},
\bigl[\,\underline{\cal O}^{(1)},
\underline{\cal H}^{(0)}\,\bigr]\bigr] 
\nonumber \\
&&\hspace{7mm}
+\frac{i}{2}h_{\alpha\beta}\nabla_{\!{\mib k}}^{\beta}
\bigl[\,\underline{\cal O}^{(1)},\hbar\tilde{\underline{v}}^{\alpha}\,\bigr]
\, ,
\label{barH(2)}
\end{eqnarray}
\end{subequations}
with
$\{\underline{A},\underline{B}\}\!\equiv\!\frac{1}{2}(\underline{A}\,\underline{B}
\!+\! \underline{B}\,\underline{A})$ and
$[\,\underline{A},\underline{B}\,]\!\equiv\!\underline{A}\,\underline{B}
\!-\! \underline{B}\,\underline{A}$.

\subsection{Schr\"odinger equation and Green's function}
\label{subsec:Schrodinger}

We finally write down an effective Schr\"odinger equation and
the corresponding Green's function for
non-interacting Bloch electrons in a magnetic field. 
To this end, it is useful to
introduce the transfer integral in terms of eq.\ (\ref{calH-2t}) as
\begin{equation}
\underline{t}({\mib R},{\mib B})\equiv \frac{1}{N_{\rm c}}\sum_{{\mib k}}
{\rm e}^{i{\mib k}\cdot{\mib R}}\,
\underline{\cal H}({\mib k},{\mib B}) \, .
\label{transfer}
\end{equation}
Substituting eq.\ (\ref{calH-1a}) and using
eq.\ (\ref{transfer}), 
the Schr\"odinger equation (\ref{Schroedinger}) is transformed into
\begin{equation}
\sum_{b'{\mib R}'}{\rm e}^{iI_{{\mib R}{\mib R}'}} \,
t_{bb'}({\mib R}-{\mib R}',{\mib B})\,c_{b'}({\mib R}')
={\cal E}\hspace{0.1mm} c_{b}({\mib R}) \, .
\label{Schroedinger2a}
\end{equation}
Thus, the Hamiltonian is given as a product of the transfer integral
and the Peierls phase factor. Note that the transfer integral
also depends on the magnetic field here.
We can provide an alternative expression to
eq.\ (\ref{Schroedinger2a}).
Let us change the summation in the above equation
from ${\mib R}'$ to ${\mib R}''\!\equiv\!{\mib R}\!-\!{\mib R}'$,
express $c_{b'}({\mib R}\!-\!{\mib R}'')\!=\!
{\rm e}^{-{\mib R}''\cdot{\mib \nabla}_{\!{\mib R}}}c_{b'}({\mib R})$,
and use the following identity
proved in Appendix II of Ref.\ \citen{Luttinger51}:
\begin{eqnarray*}
\exp(iI_{{\mib R},{\mib R}-{\mib R}''})
\exp\left(-{\mib R}''\!\cdot\!{\mib \nabla}_{\! {\mib R}}\right)
=\exp\left(-{\mib R}''\!\cdot\!{\mib \partial}_{\mib R}\right) \, ,
\end{eqnarray*}
with ${\mib \partial}_{\mib R}$ defined by
\begin{equation}
{\mib \partial}_{\mib R}\equiv \frac{\partial}{\partial {\mib R}}
-i\frac{e}{\hbar c}{\mib A}({\mib R}) \, .
\label{partial}
\end{equation}
Equation (\ref{Schroedinger2a}) is thereby transformed into
\begin{equation}
\sum_{b'}{\cal H}_{bb'}(-i{\mib \partial}_{\mib R},{\mib B})\,
c_{b'}({\mib R}) ={\cal E}\hspace{0.1mm} c_{b}({\mib R}) \, ,
\label{Schroedinger2b}
\end{equation}
where ${\cal H}_{bb'}(-i{\mib \partial}_{\mib R},{\mib B})$ is an operator
symmetrized with respect to ${\mib\partial}_{\mib R}$ as
\begin{equation}
\underline{\cal H}(-i{\mib \partial}_{\mib R},{\mib B})
\equiv \sum_{{\mib R}''}\underline{t}({\mib R}'',{\mib B})\,
{\rm e}^{i{\mib R}''\cdot\,(-i{\mib\partial}_{\mib R})} \, .
\label{calH-2b}
\end{equation}
Equation (\ref{Schroedinger2b}) extends the well-known rule
${\cal E}({\mib k})\!\rightarrow\! {\cal E}(-i{\mib \partial}_{\mib R})$
in a finite magnetic field\cite{Luttinger51, Onsager52}
to include the 
change of the energy-band structures in ${\mib B}$.

Next, Dyson's equation corresponding to eq.\ (\ref{Schroedinger2b})
is given by
\begin{equation}
\sum_{b''}\!\left[i\varepsilon_{n}\delta_{b'b''}\!-\!
{\cal H}_{b'b''}(-i{\mib \partial}_{{\mib R}'},{\mib B})\right]
g_{b''{\mib R}',b{\mib R}}(\varepsilon_{n}) 
=\delta_{b'b}\delta_{{\mib R}'{\mib R}} \, ,
\label{Dyson0}
\end{equation}
where $g_{b''{\mib R}',b{\mib R}}(\varepsilon_{n})$ is the non-interacting
Matsubara Green's function with
$\varepsilon_{n}\!\equiv\!(2n\!+\! 1)\pi T$  $(n\!=\!0,\pm 1,\cdots)$.
Suggested by eq.\ (\ref{Schroedinger2a}),
we write the Green's function as
\begin{subequations}
\label{g}
\begin{equation}
g_{b'{\mib R}',b{\mib R}}(\varepsilon_{n}) 
={\rm e}^{iI_{{\mib R}'{\mib R}}}\,
\tilde{g}_{b'{\mib R}',b{\mib R}}(\varepsilon_{n}) 
\, ,
\label{g1}
\end{equation}
and substitute it into eq.\ (\ref{Dyson0}). 
Using eq.\ (\ref{dI}), 
we then find an equation for $\underline{\tilde{g}}$ which 
is also given as eq.\ (\ref{Dyson0}) with a replacement:
\begin{eqnarray*}
{\mib \partial}_{{\mib R}'}\,\, \longrightarrow \,\,
\tilde{\mib \partial}_{{\mib R}'}\equiv\frac{\partial}{\partial {\mib R}'} 
-i\frac{e}{2\hbar c}{\mib B}
\!\times\! ({\mib R}'\!-\!{\mib R}) \, .
\end{eqnarray*}
It hence follows that $\tilde{g}_{b'{\mib R}',b{\mib R}}(\varepsilon_{n})$ 
depends only on ${\mib R}'\!-\!{\mib R}$ and
can be expanded as
\begin{equation}
\tilde{g}_{b'{\mib R}',b{\mib R}}(\varepsilon_{n}) 
 =\frac{1}{N_{\rm c}}
\sum_{{\mib k}}{\rm e}^{i{\mib k}\cdot({\mib R}'-{\mib R})}
g_{b'b}(\varepsilon_{n},{\mib k}) \, .
\label{g2}
\end{equation}
\end{subequations}
Let us substitute eq.\ (\ref{g2}) into the equation for $\underline{\tilde{g}}$,
express $({\mib R}'\!-\!{\mib R}){\rm e}^{i{\mib k}\cdot({\mib R}'-{\mib R})}
\!=\! -i{\mib\nabla}_{{\mib k}}{\rm e}^{i{\mib k}\cdot({\mib R}'-{\mib R})}$,
and perform partial integrations over ${\mib k}$.
We thereby obtain an equation for $\underline{g}(\varepsilon_{n},{\mib k})$ as
\begin{equation}
\sum_{{\mib k}'}\delta_{{\mib k}{\mib k}'}\!\left[i\varepsilon_{n}\underline{1}-
\underline{\cal H}({\mib \kappa},{\mib B})\right]
\underline{g}(\varepsilon_{n},{\mib k}') 
=\underline{1} \, ,
\label{Dyson0k}
\end{equation}
where the operator ${\mib \kappa}$ is defined by
\begin{equation}
{\mib \kappa}\equiv {\mib k}-i\frac{e}{2\hbar c}{\mib B}
\!\times\! {\mib \nabla}_{\!{\mib k}'} \, .
\label{kappa}
\end{equation}
Note that eq.\ (\ref{Dyson0k}) may be written alternatively 
by using the operator of eq.\ (\ref{otimes}) as 
$\left[i\varepsilon_{n}\underline{1}-
\underline{\cal H}({\mib k},{\mib B})\right]\otimes
\underline{g}(\varepsilon_{n},{\mib k}) 
=\underline{1}$.
This latter result may have been obtained directly by writing Dyson's equation
(\ref{Dyson0}) in terms of the Hamiltonian in eq.\ (\ref{Schroedinger2a}),
substituting eqs.\ (\ref{transfer}) and (\ref{g}) into it, and 
using eq.\ (\ref{MR}).

\section{Interaction Effects}

We now include the two-body interaction:
\begin{equation}
{\cal U}({\mib r}\!-\!{\mib r}')=\frac{1}{V}\sum_{\mib q}
{\cal U}_{\mib q}\,{\rm e}^{i{\mib q}\cdot({\mib r}-{\mib r}')} 
\label{U_q}
\end{equation}
into our consideration, where $V$ is the volume of the system.

The total Hamiltonian $\hat{\cal H}_{\rm tot}$
is given in second quantization by 
using the basis function (\ref{varphi}) and the notation $\nu\!\equiv\! b{\mib R}$
as
\begin{equation}
\hat{\cal H}_{\rm tot}=\sum_{\nu\nu'}{\cal H}_{\nu\nu'}c_{\nu}^{\dagger}c_{\nu'} 
+\frac{1}{2}\sum_{\nu_{1}\nu_{2}\nu_{1}'\nu_{2}'}
{\cal U}_{\nu_{1}\nu_{2};\nu_{1}'\nu_{2}'}
c_{\nu_{1}}^{\dagger}c_{\nu_{2}}^{\dagger}c_{\nu_{2}'}c_{\nu_{1}'} \, .
\label{Htot}
\end{equation}
Here $c_{\nu}$ is the fermion operator and ${\cal H}_{\nu\nu'}$ is given by
eq.\ (\ref{calH-1a}). The quantity
${\cal U}_{\nu_{1}\nu_{2};\nu_{1}'\nu_{2}'}$ is defined by
\begin{equation}
{\cal U}_{\nu_{1}\nu_{2};\nu_{1}'\nu_{2}'}\equiv
\sum_{\nu_{3}\nu_{4}\nu_{3}'\nu_{4}'}{\cal S}_{\nu_{1}\nu_{3}}
{\cal S}_{\nu_{2}\nu_{4}}
{\cal U}_{\nu_{3}\nu_{4};\nu_{3}'\nu_{4}'}'
{\cal S}_{\nu_{3}'\nu_{1}'}{\cal S}_{\nu_{4}'\nu_{2}'} \, ,
\label{U}
\end{equation}
where ${\cal S}_{\nu\nu'}$ is given by eq.\ (\ref{calSRk}) and
${\cal U}_{\nu_{3}\nu_{4};\nu_{3}'\nu_{4}'}'$ denotes
\begin{eqnarray}
&&\hspace{-15mm}
{\cal U}_{\nu_{3}\nu_{4};\nu_{3}'\nu_{4}'}'
\nonumber \\
&&\hspace{-15mm}
\equiv\! \int\!\! {\rm d}{\mib r}\!\!\int\!\! {\rm d}{\mib r}'
\vec{w}_{\nu_{3}}^{\prime\dagger}({\mib r})\vec{w}_{\nu_{3}'}^{\prime}({\mib r})
{\cal U}({\mib r}\!-\!{\mib r}')
\vec{w}_{\nu_{4}}^{\prime\dagger}({\mib r}')\vec{w}_{\nu_{4}'}^{\prime}({\mib r}')
\, ,
\label{U'}
\end{eqnarray}
with $\vec{w}_{\nu}'({\mib r})$ defined by eq.\ (\ref{basis}).

\subsection{Bare vertex}
\label{subsec:bare-vertex}

As shown in Appendix A,
eq.\ (\ref{U}) can be written alternatively
with respect to the Bloch states $\{\psi_{b{\mib k}}({\mib r})\}$ in
zero field as
\begin{eqnarray}
&&\hspace{-9mm}
{\cal U}_{\nu_{1}\nu_{2};\nu_{1}'\nu_{2}'}
=
\frac{{\rm e}^{iI_{{\mib R}_{1}{\mib R}_{1}'}+iI_{{\mib R}_{2}{\mib R}_{2}'}}}
{N_{\rm c}^{2}}\!\sum_{{\mib k}{\mib k}'}\!
{\rm e}^{i{\mib k}\cdot({\mib R}_{1}-{\mib R}_{1}')
+i{\mib k}'\cdot({\mib R}_{2}-{\mib R}_{2}')}
\nonumber \\
&&\hspace{10mm}
\times 
\frac{1}{V}\!
\sum_{{\mib q}}{\rm e}^{i{\mib q}\cdot({\mib R}_{1}-{\mib R}_{2})}
\,{\cal U}_{{\mib q}}\,
\Lambda_{b_{1}b_{1}'}({\mib k},{\mib q},{\mib B})
\nonumber \\
&&\hspace{10mm}
\times 
\Lambda_{b_{2}b_{2}'}({\mib k}',-{\mib q},{\mib B})\, .
\label{UExp}
\end{eqnarray}
Here $I_{{\mib R}{\mib R}'}$ and
${\cal U}_{{\mib q}}$ are given in eqs.\ (\ref{phase}) and (\ref{U_q}),
respectively.
The quantity
$\underline{\Lambda}({\mib k},{\mib q},{\mib B})$ may be called a vertex
function which is
defined in terms of the operator $\otimes$ in eq.\ (\ref{otimes}), 
$\underline{{\cal S}}$ in eq.\ (\ref{calS-kt}) and 
$\underline{\Lambda}'$ of eq.\ (\ref{Lambda'}) by
\begin{subequations}
\label{Lambda}
\begin{equation}
\underline{\Lambda}({\mib k},{\mib q},{\mib B})\equiv
\underline{\cal S}(\overline{{\mib k}\!+\!{\mib q}},{\mib B})\!\otimes\!
\underline{\Lambda}^{\!\prime}({\mib k},{\mib q},{\mib B})
\!\otimes\! \underline{\cal S}({\mib k},{\mib B}) \, ,
\label{Gamma-D}
\end{equation}
with $\overline{{\mib k}\!+\!{\mib q}}$ the
wave vector in the first Brillouin zone
corresponding to ${\mib k}\!+\!{\mib q}$ in the extended zone scheme.
Thus, ${\mib k}$ in $\underline{\Lambda}({\mib k},{\mib q},{\mib B})$ belongs to
the incoming electron, ${\mib q}$ 
is an additional wave vector from the interaction, 
and $\overline{{\mib k}\!+\!{\mib q}}$ specifies 
the outgoing electron.
In zero field without 
the Peierls phase factors,
the interaction (\ref{UExp}) in 
${\mib k}$ space may be expressed diagrammatically as
Fig.\ 1. We shall see below that this diagram in zero field
suffices for the diagrammatic calculations of the properties 
in a finite magnetic field.

\begin{figure}[tb]
\begin{center}
  \includegraphics[width=5cm]{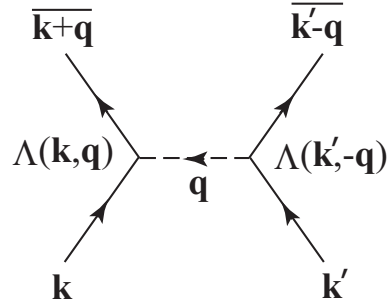}
\end{center}
  \caption{A diagrammatic representation of the interaction (\ref{UExp})
  in the absence of the Peierls phase factors.
  The vertex $\Lambda({\mib k},{\mib q})$ is given by eq.\ (\ref{Lambda}),
 and $\overline{{\mib k}\!+\!{\mib q}}$ denotes the
wave vector in the first Brillouin zone
corresponding to ${\mib k}+{\mib q}$ in the extended zone scheme.}
  \label{fig:1}
\end{figure}

It is also possible to expand eq.\ (\ref{Gamma-D}) in powers of ${\mib B}$ as
\begin{equation}
\underline{\Lambda}({\mib k},{\mib q},{\mib B})
=\underline{\Lambda}^{\!(0)}({\mib k},{\mib q})+\sum_{j=1}^{\infty}
\underline{\Lambda}^{\!(j)}({\mib k},{\mib q},{\mib B})\, .
\label{Gamma-E}
\end{equation}
\end{subequations}
The expressions of $\underline{\Lambda}^{\!(j)}$ may be obtained
with the same procedure as that of deriving eq.\ (\ref{barH}).
The quantity $\underline{\Lambda}^{\!(0)}$ is given by
\begin{subequations}
\label{Lambda0-2}
\begin{equation}
\Lambda_{bb'}^{\!(0)}({\mib k},{\mib q})=\langle 
\vec{u}_{b\overline{{\mib k}+{\mib q}}}|
{\rm e}^{i{\mib K}_{{\mib k}+{\mib q}}\cdot{\mib r}}|\vec{u}_{b'{\mib k}}\rangle \, ,
\label{Lambda0}
\end{equation}
with ${\mib K}_{{\mib k}+{\mib q}}\!\equiv\! 
{\mib k}\!+\!{\mib q}\!-\!\overline{{\mib k}\!+\!{\mib q}}$
 a reciprocal lattice vector.
The first-order term is obtained from
$\underline{\Lambda}^{\!(1)}({\mib k},{\mib q})\!=\!
\underline{\Lambda}^{\!\prime(1)}({\mib k},{\mib q})
+\underline{S}^{(1)}(\overline{{\mib k}\!+\!{\mib q}})
\underline{\Lambda}^{(0)}({\mib k},{\mib q})
+
\underline{\Lambda}^{\!(0)}({\mib k},{\mib q})
\underline{S}^{(1)}({\mib k})$ together with eqs.\
(\ref{calS0-2}) and (\ref{Lambda'1}) as
\begin{eqnarray}
&&\hspace{-14mm}
\underline{\Lambda}^{\!(1)}({\mib k},{\mib q})
=
-\frac{1}{2}h_{\alpha\beta}\,
\underline{x}^{\alpha}(\overline{{\mib k}\!+\!{\mib q}})\,
\nabla_{\!{\mib k}}^{\beta}\underline{\Lambda}^{\!(0)}({\mib k},{\mib q})
\nonumber \\
&&\hspace{5mm}
-\frac{1}{2}h_{\alpha\beta}\!
\left[\nabla_{\!{\mib k}}^{\beta}
\underline{\Lambda}^{\!(0)}({\mib k},{\mib q})
\right]\underline{x}^{\alpha}({\mib k}) \, .
\label{Lambda1}
\end{eqnarray}
Note that the argument of $\underline{x}^{\alpha}$ is 
${\mib k}$ ($\overline{{\mib k}\!+\!{\mib q}}$)
when it appears to the right (left) of 
$\underline{\Lambda}^{\!(0)}({\mib k},{\mib q})$;
keeping this rule in mind and dropping the argument of
$\underline{x}^{\alpha}$, eq.\ (\ref{Lambda1}) may be written
exactly as the $\hbar \underline{\tilde{v}}^{\beta}$ contribution
of eq.\ (\ref{barH(1)}).
This rule also applies to higher-order terms.

The second-order term is calculated similarly. We obtain 
an expression analogous to the $\hbar \underline{\tilde{\mib v}}$
contributions of eq.\ (\ref{barH(2)}) as
\begin{eqnarray}
&&\hspace{-9mm}
\underline{\Lambda}^{\!(2)}({\mib k},{\mib q})
\nonumber \\
&&\hspace{-9mm}
=\frac{1}{2}h_{\alpha\beta}h_{\alpha'\beta'}\nabla_{\!{\mib k}}^{\beta}\biggl[
\bigl\{\underline{x}^{\alpha}(\overline{{\mib k}\!+\!{\mib q}}),
\underline{x}^{\alpha'}(\overline{{\mib k}\!+\!{\mib q}})\bigr\}
\nabla_{\!{\mib k}}^{\beta'}\!\underline{\Lambda}^{\!(0)}\!({\mib k},{\mib q})
\nonumber \\
&&\hspace{17mm}
+\bigl[
\nabla_{\!{\mib k}}^{\beta'}\!\underline{\Lambda}^{\!(0)}\!({\mib k},{\mib q})\bigr]
\bigl\{\underline{x}^{\alpha}({\mib k}),
\underline{x}^{\alpha'}({\mib k})\bigr\}\biggr]
\nonumber \\
&&\hspace{-9mm}
-\frac{1}{4}h_{\alpha\beta}h_{\alpha'\beta'}\biggl[
\bigl\{\underline{x}^{\alpha}(\overline{{\mib k}\!+\!{\mib q}}),
\underline{x}^{\alpha'}(\overline{{\mib k}\!+\!{\mib q}})\bigr\}
\nabla_{\!{\mib k}}^{\beta}\nabla_{\!{\mib k}}^{\beta'}
\!\underline{\Lambda}^{\!(0)}\!({\mib k},{\mib q})
\nonumber \\
&&\hspace{11mm}
+\bigl[\nabla_{\!{\mib k}}^{\beta}\nabla_{\!{\mib k}}^{\beta'}
\!\underline{\Lambda}^{\!(0)}\!({\mib k},{\mib q})
\bigr]
\bigl\{\underline{x}^{\alpha}({\mib k}),
\underline{x}^{\alpha'}({\mib k})\bigr\}\biggr]
\nonumber \\
&&\hspace{-9mm}
+\frac{i}{4}h_{\alpha\beta}h_{\alpha'\beta'}
\nabla_{\!{\mib k}}^{\beta}\nabla_{\!{\mib k}}^{\beta'}
\biggl[
\underline{x}^{\alpha'}(\overline{{\mib k}\!+\!{\mib q}})\nabla_{\!{\mib k}}^{\alpha}
\!\underline{\Lambda}^{\!(0)}\!({\mib k},{\mib q})
\nonumber \\
&&\hspace{20mm}
-\nabla_{\!{\mib k}}^{\alpha}
\!\underline{\Lambda}^{\!(0)}\!({\mib k},{\mib q})\,
\underline{x}^{\alpha'}({\mib k})\biggr] 
\nonumber \\
&&\hspace{-9mm}
-\frac{1}{2}\left[\underline{\cal O}^{(1)}(\overline{{\mib k}\!+\!{\mib q}})
\underline{\Lambda}^{\!(1)}({\mib k},{\mib q})+\underline{\Lambda}^{\!(1)}({\mib k},{\mib q})
\underline{\cal O}^{(1)}({\mib k})\right]
\nonumber \\
&&\hspace{-9mm}
+\frac{1}{8}\left[\underline{\cal O}^{(1)}(\overline{{\mib k}\!+\!{\mib q}})
\underline{\cal O}^{(1)}(\overline{{\mib k}\!+\!{\mib q}})
\underline{\Lambda}^{\!(0)}({\mib k},{\mib q})\right.
\nonumber \\
&&\hspace{-2mm}
-2\underline{\cal O}^{(1)}(\overline{{\mib k}\!+\!{\mib q}})
\underline{\Lambda}^{\!(0)}({\mib k},{\mib q})
\underline{\cal O}^{(1)}({\mib k})
\nonumber \\
&&\hspace{-3mm}
 \left. 
+\underline{\Lambda}^{\!(0)}({\mib k},{\mib q})
\underline{\cal O}^{(1)}({\mib k})\underline{\cal O}^{(1)}({\mib k})\right]
\nonumber \\
&&\hspace{-9mm}
+\frac{i}{2}h_{\alpha\beta}\nabla_{\!{\mib k}}^{\beta}
\left\{\underline{\cal O}^{(1)}(\overline{{\mib k}\!+\!{\mib q}})
\nabla_{\!{\mib k}}^{\alpha}\underline{\Lambda}^{\!(0)}({\mib k},{\mib q})
\right.
\nonumber \\
&&\hspace{8mm}
 \left.
-\left[\nabla_{\!{\mib k}}^{\alpha}\underline{\Lambda}^{\!(0)}({\mib k},{\mib q})\right]
\underline{\cal O}^{(1)}({\mib k})\right\}\, .
\label{Lambda2}
\end{eqnarray}
\end{subequations}

\subsection{Perturbation expansion}

With these preliminaries, we proceed to the calculation of 
the thermodynamic potential $\Omega\!
=\!-T \ln{\rm Tr}\,{\rm e}^{-\hat{\cal H}_{\rm tot}/T}$, 
the self-energy $\Sigma_{\nu\nu'}$ and
Green's function $G_{\nu'\nu}$
in the framework of the conserving approximation.\cite{BK61,Baym62}

Let us define the Matsubara Green's function by
\begin{equation}
G_{\nu'\nu}(\varepsilon_{n})=-\int_{0}^{1/T}\!\!\!\!
\langle\, T_{\tau}\,c_{\nu'}(\tau)\,c_{\nu}^{\dagger} \, \rangle\, 
{\rm e}^{i\varepsilon_{n}\tau} \, {\rm d}\tau \, ,
\label{Green}
\end{equation}
which as ${\cal U}\!\rightarrow\! 0$ reduces to $g_{\nu'\nu}(\varepsilon_{n})$ 
of eq.\ (\ref{Dyson0}).
As shown by Luttinger and Ward,\cite{LW60}
$\Omega$
can be written as a functional of $\underline{G}$ as
\begin{equation}
\Omega =  
-T \sum_{n} {\rm Tr}
\bigl\{
\ln \bigl[ \underline{\cal H}+\underline{\Sigma}
-i\varepsilon_{n}\underline{1} \bigr]+
\underline{\Sigma}\,
\underline{G}  \bigr\}{\rm e}^{i\varepsilon_{n}0_{+}}
+ \Phi[\underline{G}]  \, .
\label{Omega}
\end{equation}
Here $0_{+}$ is an infinitesimal positive constant and 
$\Phi$ denotes contributions of all the skeleton diagrams
in the bare perturbation expansion for $\Omega$
with $\underline{G}$ used as the propagator.
The self-energy is obtained from $\Phi$ by
\begin{subequations}
\label{SG}
\begin{equation}
\Sigma_{\nu\nu'}(\varepsilon_{n})
=\frac{1}{T}\,\frac{\delta\Phi}{\delta G_{\nu'\nu}(\varepsilon_{n})} \, .
\label{Sigma}
\end{equation}
With this relation, $\Omega$ is stationary with respect
to a variation in $\underline{G}$ satisfying Dyson's equation:
\begin{equation}
\left[i\varepsilon_{n}\underline{1}-\underline{\cal H}
-\underline{\Sigma}(\varepsilon_{n})\right]\underline{G}(\varepsilon_{n})
=\underline{1} \, .
\label{Dyson}
\end{equation}
\end{subequations}
The conserving approximation denotes approximating $\Phi$ by
some selected diagrams and determining $\underline{G}$
and $\underline{\Sigma}$ self-consistently by eq.\ (\ref{SG}).\cite{BK61,Baym62}
Its advantages may be summarized as follows:
(i) both the equilibrium and dynamical 
properties can be described within the same approximation
with various conservation laws automatically satisfied;
(ii) the Laudau Fermi-liquid corrections,\cite{Landau56,BP91} 
or vertex corrections in a different terminology,
are automatically included.
For example, the lowest-order conserving approximation is nothing but
the Hartree-Fock approximation where $\Phi^{\rm HF}$
and $\Sigma_{\nu\nu'}^{\rm HF}$ are given by
\begin{subequations}
\begin{eqnarray}
&&\hspace{-9mm}
\Phi^{\rm HF}=\frac{T^{2}}{2}
\sum_{n_{1},n_{2}}\sum_{\nu_{1}\nu_{2}\nu_{1}'\nu_{2}'}
{\cal U}_{\nu_{1}\nu_{2};\nu_{1}'\nu_{2}'}[G_{\nu_{1}'\nu_{1}}(\varepsilon_{n_{1}})
G_{\nu_{2}'\nu_{2}}(\varepsilon_{n_{2}})
\nonumber \\
&&\hspace{1.5mm}
-G_{\nu_{1}'\nu_{2}}(\varepsilon_{n_{1}})
G_{\nu_{2}'\nu_{1}}(\varepsilon_{n_{2}})]
{\rm e}^{i\varepsilon_{n_{1}}\!0_{+}+i\varepsilon_{n_{2}}\!0_{+}} \, ,
\label{PhiHF}
\end{eqnarray}
\begin{equation}
\Sigma_{\nu\nu'}^{\rm HF}=T\sum_{n_{1}}\sum_{\nu_{1}\nu_{1}'}
({\cal U}_{\nu\nu_{1};\nu'\nu_{1}'}\!
-{\cal U}_{\nu\nu_{1};\nu_{1}'\nu'})G_{\nu_{1}'\nu_{1}}(\varepsilon_{n_{1}})
{\rm e}^{i\varepsilon_{n_{1}}\! 0_{+}} \, ,
\label{SigmaHF}
\end{equation}
\end{subequations}
respectively.

It follows from eqs.\ (\ref{calH-1a}) and (\ref{UExp})
that
$\Sigma_{\nu\nu'}(\varepsilon_{n})$ 
and $G_{\nu'\nu}(\varepsilon_{n})$ can be expanded as
\begin{subequations}
\label{SGbar}
\begin{equation}
\Sigma_{b{\mib R},b'{\mib R}'}(\varepsilon_{n})=
\frac{{\rm e}^{iI_{{\mib R}{\mib R}'}}}{N_{\rm c}}\sum_{{\mib k}}
{\rm e}^{i{\mib k}\cdot({\mib R}-{\mib R}')}
\Sigma_{bb'}(\varepsilon_{n},{\mib k}) \, ,
\label{Sbar}
\end{equation}
\begin{equation}
G_{b'{\mib R}',b{\mib R}}(\varepsilon_{n})=
\frac{{\rm e}^{iI_{{\mib R}'{\mib R}}}}{N_{\rm c}}\sum_{{\mib k}}
{\rm e}^{i{\mib k}\cdot({\mib R}'-{\mib R})}G_{b'b}(\varepsilon_{n},{\mib k}) \, .
\label{Gbar}
\end{equation}
\end{subequations}
This is proved by induction as follows:
First, the non-interacting Green's function 
$g_{\nu'\nu}(\varepsilon_{n})$ is already given in 
the form of eq.\ (\ref{Gbar}) as eq.\ (\ref{g}).
We next assume the expression (\ref{Gbar}) and substitute it 
into eq.\ (\ref{Sigma}) to calculate $\Sigma_{\nu\nu'}(\varepsilon_{n})$ 
order by order by using eq.\ (\ref{MR}).
We then find that the self-energy can also be written as
eq.\ (\ref{Sbar}), i.e., the same expression as the non-interacting
Hamiltonian (\ref{calH-1a}).
It hence follows that $G_{\nu'\nu}(\varepsilon_{n})$ 
may be written as eq.\ (\ref{Gbar}).
This completes the proof.

\begin{figure}[tb]
\begin{center}
  \includegraphics[width=0.95\linewidth]{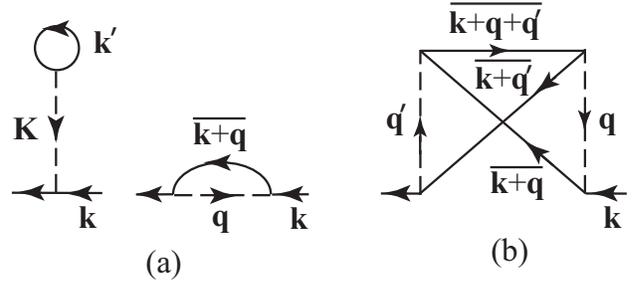}
  \end{center}
  \caption{Some typical diagrams for the self-energy:
  (a) the Hartree and Fock terms;
  (b) the second-order exchange contribution.}
  \label{fig:2}
\end{figure}

To see how to calculate $\underline{\Sigma}(\varepsilon_{n},{\mib k})$
practically, we present expressions of 
$\underline{\Sigma}^{\rm HF}\hspace{-0.5mm}({\mib k})$ obtained
from eq.\ (\ref{SigmaHF}), see also Fig.\ 2(a), and 
$\underline{\Sigma}^{\rm 2e}\hspace{-0.3mm}(\varepsilon_{n},{\mib k})$
which corresponds to the diagram of Fig.\ 2(b):
\begin{subequations}
\label{Sigma-HF2e}
\begin{eqnarray}
&&\hspace{-9mm}
\underline{\Sigma}^{\rm HF}\hspace{-0.5mm}({\mib k})
\nonumber \\
&&\hspace{-9mm}
=
\frac{T}{V}\sum_{n}\biggl[\sum_{{\mib K}}
{\cal U}_{{\mib K}} \underline{\Lambda}({\mib k},{\mib K})\!
\sum_{{\mib k}'} \!{\rm Tr}
\underline{\Lambda}({\mib k}',-{\mib K})\underline{G}(\varepsilon_{n},{\mib k}')
\nonumber \\
&&\hspace{-5mm}
-\sum_{{\mib q}}{\cal U}_{{\mib q}}\,
\underline{\Lambda}(\overline{{\mib k}\!+\!{\mib q}},-{\mib q})
\!\otimes\!
\underline{G}(\varepsilon_{n},\overline{{\mib k}\!+\!{\mib q}})
\!\otimes\!
\underline{\Lambda}({\mib k},{\mib q})\biggr]
{\rm e}^{i\varepsilon_{n}\! 0_{+}} \, ,
\nonumber \\
\label{Sigma-HF}
\end{eqnarray}
\begin{eqnarray}
&&\hspace{-10mm}
\underline{\Sigma}^{\rm 2e}\hspace{-0.3mm}(\varepsilon_{n},{\mib k})
=
\frac{T^{2}}{V^{2}}\!
\sum_{n_{1}n_{2}}
\sum_{{\mib q}{\mib q}'}
{\cal U}_{{\mib q}}\,{\cal U}_{{\mib q}'}\bigl[\,
\underline{\Lambda}(\overline{{\mib k}\!+\!{\mib q}'},-{\mib q}')
\nonumber\\
&&\hspace{10mm}
\otimes\,
\underline{G}(\varepsilon_{n+n_{2}-n_{1}},\overline{{\mib k}\!+\!{\mib q}'})
\otimes\underline{\Lambda}(\overline{{\mib k}\!+\!{\mib q}\!+\!{\mib q}'},-{\mib q})
\nonumber\\
&&\hspace{10mm}
\otimes\, \underline{G}(\varepsilon_{n_{2}},
\overline{{\mib k}\!+\!{\mib q}\!+\!{\mib q}'})
\otimes \underline{\Lambda}(\overline{{\mib k}\!+\!{\mib q}},{\mib q}')
\nonumber\\
&&\hspace{10mm}
\otimes\, \underline{G}(\varepsilon_{n_{1}},\overline{{\mib k}\!+\!{\mib q}})
\otimes \underline{\Lambda}({\mib k},{\mib q})\, \bigr] \, ,
\label{Sigma-2e}
\end{eqnarray}
\end{subequations}
where $\otimes$ operates on the ${\mib k}'$ or 
${\mib k}$ dependences of the adjacent functions. 
Note that
removing the operator $\otimes$ in the above expressions yields the self-energies
at ${\mib B}\!=\!{\mib 0}$.
We hence have a simple rule to calculate the ${\mib k}$-space self-energy for
$B\!>\! 0$ using only the zero-field Feynman diagrams
as follows: (i)
Label each of the connected electron lines with
the same wave vector, like ${\mib k}$ or ${\mib k}'$ in the above examples, and 
insert the operator $\otimes$ of eq.\ (\ref{otimes}) between every
adjacent $\underline{\Lambda}$ and $\underline{G}$.
(ii)
We can remove any single operator $\otimes$ within every closed electron loop,
as in the first term of eq.\ (\ref{Sigma-HF}),
because of the absence of the Peierls phase factor in the final calculation 
of using eq.\ (\ref{MR});
this may also be proved directly in ${\mib k}$ space 
by using eq.\ (\ref{otimes-p2}).
The rules (i) and (ii) apply also to the calculations
of $\Omega$ and $\underline{G}(\varepsilon_{n},{\mib k})$.

We now provide ${\mib k}$-space expressions of eqs. (\ref{Omega}) and (\ref{SG}).
Noting the rule (ii) above, we realize that $\Phi$ in 
eq.\ (\ref{Omega}) is given alternatively as a functional 
of $\underline{G}(\varepsilon_{n},{\mib k})$ as
\begin{equation}
\Phi = 
T\sum_{n}\sum_{\ell=1}^{\infty}\frac{1}{2\ell}\sum_{{\mib k}}{\rm Tr}\,
\underline{\Sigma}^{\ell}\!(\varepsilon_{n},{\mib k},{\mib B})\,
\underline{G}(\varepsilon_{n},{\mib k},{\mib B})
{\rm e}^{i\varepsilon_{n}0_{+}} \, ,
\label{Phi-2}
\end{equation}
with $\underline{\Sigma}^{\ell}$ the contribution of the $\ell$th-order
skeleton diagrams to $\underline{\Sigma}$.
We hence have
\begin{subequations}
\label{SG-2}
\begin{equation}
\Sigma_{bb'}(\varepsilon_{n},{\mib k},{\mib B},\{\underline{G}\}) 
= \frac{1}{T}\,
\frac{\delta\Phi}{\delta G_{b'b}(\varepsilon_{n},{\mib k},{\mib B})} \, .
\label{Sigma-k}
\end{equation}
Next, Dyson's equation of eq.\ (\ref{Dyson}) is transformed into
$\left[i\varepsilon_{n}\underline{1}\!-\!\underline{\cal H}({\mib k},{\mib B})
\!-\!\underline{\Sigma}(\varepsilon_{n},{\mib k},{\mib B})\right]
\otimes\underline{G}(\varepsilon_{n},{\mib k},{\mib B})
\!=\!\underline{1}$, as shown by
using eqs.\ (\ref{calH-1a}), (\ref{SGbar}), and (\ref{MR}).
It may be written alternatively as
\begin{equation}
\sum_{{\mib k}'} \delta_{{\mib k}{\mib k}'}\!
\left[i\varepsilon_{n}\underline{1}\!-\!
\underline{\cal H}({\mib\kappa},{\mib B})
\!-\!\underline{\Sigma}(\varepsilon_{n},{\mib\kappa},{\mib B})\right]
\underline{G}(\varepsilon_{n},{\mib k}',{\mib B})
=\underline{1} \, ,
\label{Dyson-2}
\end{equation}
\end{subequations}
where ${\mib \kappa}$ is defined by eq.\ (\ref{kappa}).
Hence the functional (\ref{Omega}) is given in ${\mib k}$ space by
\begin{eqnarray}
&&\hspace{-9mm}
\Omega 
=-T \sum_{n} {\rm Tr}\biggl[{\rm Tr}_{\mib k}
\ln \!\left\{\left[\,
 \underline{\cal H}({\mib \kappa},{\mib B})
\!+\!\underline{\Sigma}(\varepsilon_{n},{\mib \kappa},{\mib B})
\!-\!i\varepsilon_{n}\underline{1}  \,\right] \right\}
\nonumber \\
&&\hspace{-1.5mm}
+\sum_{\mib k}\underline{\Sigma}(\varepsilon_{n},{\mib k},{\mib B})\,
\underline{G}(\varepsilon_{n},{\mib k},{\mib B})  \biggr]\,
{\rm e}^{i\varepsilon_{n}0_{+}}
+ \Phi[\underline{G}]  \, ,
\label{Omega-2}
\end{eqnarray}
with Tr$_{\mib k}$ denoting the trace in ${\mib k}$ space.
It satisfies 
\begin{equation}
\frac{\delta \Omega}{\delta G_{b'b}(\varepsilon_{n},{\mib k},{\mib B})}=0\, ,
\label{dPhidG}
\end{equation}
as shown by using eq.\ (\ref{SG-2}).

Finally, eqs.\ (\ref{Phi-2}) and
(\ref{Sigma-k}) may be expanded with respect to the explicit ${\mib B}$
dependences originating from the vertex $\Lambda$ and the operator $\otimes$ as
\begin{equation}
\Phi({\mib B},\{ \underline{G} \})=
\Phi^{(0)}(\{\underline{G} \})+\sum_{j=1}^{\infty}
\Phi^{(j)}({\mib B},\{\underline{G} \}) \, ,
\label{Phi-Exp}
\end{equation}
\begin{eqnarray}
&&\hspace{-9mm}
\underline{\Sigma}(\varepsilon_{n},{\mib k},{\mib B},\{ \underline{G} \})=
\underline{\Sigma}^{(0)}(\varepsilon_{n},{\mib k},\{\underline{G} \})
\nonumber \\
&&\hspace{20mm}
 +\sum_{j=1}^{\infty}
\underline{\Sigma}^{(j)}(\varepsilon_{n},{\mib k},{\mib B},\{\underline{G} \}) \, ,
\label{Sigma-Exp}
\end{eqnarray}
where $\Phi^{(j)}$ and $\underline{\Sigma}^{(j)}$ are quantities of the order $B^j$.
Note that $\Phi^{(0)}$ and $\underline{\Sigma}^{(0)}$ 
are different from
those at ${\mib B}\!=\!{\mib 0}$ due to the implicit ${\mib B}$
dependences through $\underline{G}$.
It follows from eq.\ (\ref{Phi-2}) that $\Phi^{(j)}$ is connected with
$\underline{\Sigma}^{(j)}$ as
\begin{equation}
\Phi^{(j)}
=T\sum_{n\ell}\frac{1}{2\ell}\sum_{{\mib k}}{\rm Tr}
\underline{\Sigma}^{\ell(j)}\!(\varepsilon_{n},{\mib k},{\mib B})
\underline{G}(\varepsilon_{n},{\mib k},{\mib B})\,
{\rm e}^{i\varepsilon_{n}0_{+}} \, .
\label{Phi-2(j)}
\end{equation}

\subsection{Hartree-Fock approximation}
\label{subsec:pert-HF}

We here concentrate on the Hartree-Fock approximation.
Since there is
no $\varepsilon_{n}$ dependence in the self-energy 
$\underline{\Sigma}^{{\rm HF}}({\mib k})$,
eqs.\ (\ref{Phi-2})-(\ref{dPhidG}) can be simplified further
by performing every summation over $n$ 
with the Fermi distribution function $f(\varepsilon)\!\equiv\!
1/({\rm e}^{\varepsilon/T}\!+\! 1)$.\cite{LW60}
The potentials $\Omega$ and $\Phi$ are thereby transformed into the
functionals of the occupation number defined by
\begin{equation}
\underline{n}({\mib k},{\mib B})\equiv T\sum_{n}
\underline{G}(\varepsilon_{n},{\mib k},{\mib B})\,{\rm e}^{i\varepsilon_{n} 0_{+}} \, .
\label{n_k}
\end{equation}
Indeed, eq.\ (\ref{Omega-2}) is now approximated
(dropping the ${\mib B}$ dependences for simplicity) by
\begin{eqnarray}
&&\hspace{-9mm}
\Omega^{{\rm HF}}
= -T \,{\rm Tr}\,{\rm Tr}_{{\mib k}}\ln 
\bigl\{\underline{1}+{\rm e}^{-[\underline{\cal H}({\mib \kappa})
+\underline{\Sigma}^{{\rm HF}}({\mib \kappa})]/T} \bigr\}
\nonumber \\
&&\hspace{2.5mm}
 - \sum_{{\mib k}}{\rm Tr} \,
\underline{n}({\mib k})\,
\underline{\Sigma}^{{\rm HF}}\!({\mib k}) 
+ \Phi^{{\rm HF}}\, ,
\label{Omega-HF}
\end{eqnarray}
with
\begin{eqnarray}
&&\hspace{-9mm}
\Phi^{{\rm HF}}
\nonumber \\
&&\hspace{-9mm}
=\!
\frac{1}{2V}\!\left\{ \sum_{{\mib k}{\mib k}'}\!\sum_{{\mib K}}
{\cal U}_{{\mib K}}
{\rm Tr}[\,\underline{n}({\mib k})
\underline{\Lambda}({\mib k},{\mib K})]\,{\rm Tr}[
\,\underline{n}({\mib k}')
\underline{\Lambda}({\mib k}',-{\mib K})]
\right.
\nonumber \\
&&\hspace{-5mm}
-\!\sum_{{\mib k}{\mib q}}{\cal U}_{{\mib q}}{\rm Tr}
\underline{\Lambda}(\overline{{\mib k}\!+\!{\mib q}},-{\mib q})
\otimes
\underline{n}(\overline{{\mib k}\!+\!{\mib q}})
\otimes 
\underline{\Lambda}({\mib k},{\mib q})
\otimes \underline{n}({\mib k}) \biggr\} .
\nonumber \\
\label{Phi-HF}
\end{eqnarray}
The self-energy $\underline{\Sigma}^{{\rm HF}}\!({\mib k})$ of eq.\
(\ref{Sigma-HF}) is obtained from $\Phi^{{\rm HF}}$ by
\begin{subequations}
\label{Sn-HF}
\begin{equation}
\Sigma_{bb'}^{{\rm HF}}\!({\mib k})=\frac{\delta \Phi^{{\rm HF}}}{\delta 
n_{b'b}({\mib k})} \, .
\label{S-HF}
\end{equation}
It hence follows that $\Omega^{{\rm HF}}$ is stationary with respect to
a variation in $\underline{n}({\mib k})$ satisfying
\begin{equation}
\underline{n}({\mib k})=\frac{1}{\displaystyle 
{\rm e}^{[\underline{\cal H}({\mib k})
+\underline{\Sigma}^{{\rm HF}}({\mib k})]/T}
+\underline{1}} \, .
\label{n-HF}
\end{equation}
\end{subequations}
Equations (\ref{S-HF}) and (\ref{n-HF}) constitute a closed set of self-consistent
equations.

Finally, $\Phi^{{\rm HF}}$ and $\underline{\Sigma}^{{\rm HF}}$ can be expanded
with respect to the explicit ${\mib B}$ dependences as eqs.\ (\ref{Phi-Exp}) and
(\ref{Sigma-Exp}), respectively. 
The expressions of $\Phi^{{\rm HF}(j)}$ and
$\underline{\Sigma}^{{\rm HF}(j)}$
are given in Appendix B to clarify some of their fundamental properties.

\subsection{Density functional theory}
\label{subsec:LDF}

Extensive theoretical studies have been carried out for more than three decades
to describe atoms and solids quantitatively based on the
DFT.\cite{JG89,Kohn99}
It is thereby established now as one of the most efficient and reliable
methods for the quantitative understanding of solids.
Hence it is well worth applying the present method to the density functional theory.

We consider the cases where the exchange-correlation energy $E_{\rm xc}$ is 
given as a functional of the spin density:
\begin{equation}
n_{\pm}({\mib r})=T\sum_{n}{\rm Tr} \,\frac{1}{2}
(\underline{1}\pm\underline{\tau}_{3})\,
\underline{G}({\mib r},{\mib r},\varepsilon_{n})\, {\rm e}^{i\varepsilon_{n}0_{+}} \, ,
\label{n_sigma}
\end{equation}
where $\underline{1}$, $\underline{\tau}_{3}$, and $\underline{G}$ are 
$2\!\times\! 2$ matrices corresponding to the spin degrees of freedom 
with $\underline{\tau}_{3}$ denoting the third
Pauli matrix.
Let us expand $\underline{G}({\mib r},{\mib r},\varepsilon_{n})$ in eq.\ (\ref{n_sigma})
with respect to the basis functions of eq.\ (\ref{varphi}) and 
transform the resulting expression 
with the procedures
of eqs.\ (\ref{phase2}), (\ref{k-diff}) and (\ref{MR}).
We thereby obtain 
\begin{equation}
n_{\pm}({\mib r})=\frac{1}{N_{\rm c}}\sum_{{\mib k}}{\rm Tr} \,\underline{n}({\mib k})\,
\underline{\rho}^{\pm}({\mib k},{\mib r})\, ,
\label{n_sigma2}
\end{equation}
where $\underline{n}({\mib k})$ is given by eq.\ (\ref{n_k}) and
$\underline{\rho}^{\pm}({\mib k},{\mib r})$ is defined by
\begin{equation}
\underline{\rho}^{\pm}({\mib k},{\mib r})\equiv \underline{\cal S}({\mib k})
\otimes \underline{\rho}^{\prime\, \pm}({\mib k},{\mib r}) \otimes
\underline{\cal S}({\mib k}) \, ,
\label{rho}
\end{equation}
with
\begin{equation}
{\rho}_{bb'}^{\prime\, \pm}({\mib k},{\mib r})\equiv \,
{\rm e}^{ih_{\alpha\beta} 
\nabla_{{\mib k}}^{\alpha} \nabla_{{\mib k}'}^{\beta}}\!\left.
\vec{u}_{b{\mib k}}^{\dagger}({\mib r})
\frac{N_{\rm c}}{2}
(\underline{1}\pm\underline{\tau}_{3})
\vec{u}_{b'{\mib k}'}({\mib r})\right|_{{\mib k}'={\mib k}} .
\label{rho'}
\end{equation}
The factor $N_{\rm c}$ is introduced in eq.\ (\ref{rho'}) to make 
$\rho_{bb'}^{\prime\, \pm}({\mib k},{\mib r})$ finite in the thermodynamic limit.
Note that $\underline{\rho}^{\pm}({\mib k},{\mib r})$ in eq.\ (\ref{n_sigma2})
is just a matrix element with no information on the occupation number.
It hence follows that
the DFT as a functional
of $n_{\pm}({\mib r})$ can be written alternatively as a functional of
$\underline{n}({\mib k})$.

The thermodynamic potential of the DFT is given in a form
similar to eq.\ (\ref{Omega-HF}) of the Hartree-Fock theory as
\begin{eqnarray}
&&\hspace{-9mm}
\Omega =
 -T \,{\rm Tr}\,{\rm Tr}_{{\mib k}}\ln 
\bigl\{\underline{1}+{\rm e}^{-[\underline{\cal H}({\mib \kappa})
+\underline{\Sigma}^{{\rm DFT}}({\mib \kappa})]/T} \bigr\}
\nonumber \\
&&\hspace{-1.5mm}
- \sum_{{\mib k}}{\rm Tr} \,
\underline{n}({\mib k})\,
\underline{\Sigma}^{{\rm DFT}}\!({\mib k}) 
+ \Phi^{\rm DFT}\, .
\label{Omega-DFT}
\end{eqnarray}
The quantity $\Phi^{\rm DFT}$ is defined by
\begin{eqnarray}
\Phi^{\rm DFT}=\Phi^{\rm H}[n({\mib r})]
+E_{\rm xc}[n_{\sigma}({\mib r})]\, ,
\label{Phi-DFT}
\end{eqnarray}
where $\Phi^{\rm H}[n({\mib r})]$ and $E_{\rm xc}[n_{\sigma}({\mib r})]$ are
the classic Coulomb energy and the exchange-correlation energy, respectively,
with $n({\mib r})\!\equiv\!n_{+}({\mib r})+n_{-}({\mib r})$ and $\sigma\!=\!\pm$.
The former is given explicitly by
\begin{equation}
\Phi^{\rm H}[n({\mib r})]=
\frac{e^{2}}{2} \! \int \! \frac{n({\mib r})n({\mib r}')}{|{\mib r}\!-\!{\mib r}'|}
{\rm d}{\mib r}{\rm d}{\mib r}' \, ,
\label{Coulomb}
\end{equation}
which is equivalent to the Hartree term
in eq.\ (\ref{Phi-HF}) with $U_{\mib K}\!=\! 4\pi e^{2}/K^{2}$;
the ${\mib K}\!=\!{\mib 0}$ term should be removed due 
to the charge neutrality of the system.

The self-energy $\underline{\Sigma}^{{\rm DFT}}({\mib k})$ is
obtained from $\Phi^{\rm DFT}$ by
\begin{subequations}
\label{KS}
\begin{equation}
\Sigma_{bb'}^{{\rm DFT}}({\mib k})=\frac{\delta\Phi^{\rm DFT}}
{\delta n_{b'b}({\mib k})}
\, .
\label{Sigma-DFT}
\end{equation}
The condition
$\delta\Omega^{\rm DFT}/\delta n_{b'b}({\mib k})\!=\! 0$
for the thermodynamic equilibrium then yields
\begin{equation}
\underline{n}({\mib k})=
\frac{1}{{\rm e}^{[\underline{\cal H}({\mib k})
+\underline{\Sigma}^{{\rm DFT}}({\mib k})]/T}+\underline{1}} \, ,
\label{KS2}
\end{equation}
\end{subequations}
which is the Kohn-Sham equation\cite{KS65} in disguise.
Equations (\ref{Sigma-DFT}) and (\ref{KS2}) should be solved self-consistently
for a given $E_{\rm xc}$.

The equivalence between eqs.\ (\ref{Omega-2}) and (\ref{Omega-DFT}) 
for the exact functional $E_{\rm xc}$ can be established
by an argument of using the coupling-constant integral.
Suppose we change the interaction of eq.\ (\ref{U_q}) as ${\cal U}\!\rightarrow\!\lambda{\cal U}$
and express the corresponding thermodynamic potential as $\Omega^{\lambda}$.
The two expressions (\ref{Omega-2}) and (\ref{Omega-DFT}) satisfy the same 
differential 
equation\cite{LW60,JG89}
$\partial \Omega^{\lambda}/\partial \lambda\!=\!\langle \hat{\cal H}^{\lambda}_{\rm int}
\rangle /\lambda$
and the same initial condition $\Omega^{\lambda=0}\!=\!\Omega_{0}$, where
$\langle \hat{\cal H}^{\lambda}_{\rm int}\rangle$ is
the thermodynamic average of the interaction in eq.\ (\ref{Htot}) 
and $\Omega_{0}$ denotes
the non-interacting thermodynamic potential.
It hence follows that eqs.\ (\ref{Omega-2}) and (\ref{Omega-DFT}) are equivalent.

It is also possible to expand $\Phi^{{\rm DFT}}$ and $\Sigma^{{\rm DFT}}$
with respect to the explicit ${\mib B}$ dependences as 
eqs.\ (\ref{Phi-Exp}) and (\ref{Sigma-Exp}), respectively. 
The expressions of $\Phi^{{\rm DFT}(j)}$
and $\underline{\Sigma}^{{\rm DFT}(j)}$
are given in Appendix C.

\section{Susceptibility}

We here study the susceptibility $\chi_{\alpha'\alpha}$ of
${\mib B}\!\rightarrow\! {\mib 0}$ based on eq.\ (\ref{Omega-2}).
We first establish a general procedure to calculate $\chi_{\alpha'\alpha}$ 
for a given functional $\Phi$.
We then specialize to the Hartree-Fock approximation and
the density-functional theory to derive explicit expressions 
of $\chi_{\alpha'\alpha}$ within those approximations.
Unlike the treatments by Buot\cite{Buot76} and Misra {\em et al}.\ \cite{MMM82}
where they jumped directly to the expression
$\chi_{\alpha'\alpha}\!\equiv\!-
\partial^{2}\Omega/\partial B_{\alpha'}\partial B_{\alpha}$,
our consideration will proceed in two stages by first 
calculating the magnetization $M_{\alpha}
\!=\!-\partial\Omega/\partial B_{\alpha}$
and then performing another differentiation with respect to
$B_{\alpha'}$.
This approach has an advantage that vertex corrections can be
incorporated explicitly in the formula.

\subsection{Exact expression}

To derive a formally exact expression of $\chi_{\alpha'\alpha}$,
let us start with classifying various ${\mib B}$ dependences of 
the thermodynamic potential $\Omega$ in Eq.\ (\ref{Omega-2})
into three groups. 
The first category is that through ${\mib\kappa}$ in eq.\ (\ref{kappa}).
The second category is the explicit ${\mib B}$ dependences in 
$\underline{\cal H}({\mib k},{\mib B})$ 
and $\underline{\Sigma}(\varepsilon_{n},{\mib k},{\mib B},
\{\underline{G}\})$.
Those in $\underline{\Sigma}$ originate 
from the ${\mib B}$ dependence in the vertex $\Lambda$ of eq.\ (\ref{Lambda}) and
the operator $\otimes$ of eq.\ (\ref{otimes});
see eq.\ (\ref{Sigma-HF2e}), for example.
The third category is the implicit ${\mib B}$ dependence through $\underline{G}$,
which is
responsible for the Fermi-liquid corrections.\cite{Landau56,BP91}
When calculating ${\mib M}\!\equiv\!-\partial\Omega/\partial {\mib B}$, however,
we need not consider
this third category because of eq.\ (\ref{dPhidG}).

The dependence through ${\mib \kappa}$ needs a special treatment due to
the property (\ref{otimes-p2}).
As shown in Appendix D, it does not mix with the other categories 
for ${\mib B}\!\rightarrow\! {\mib 0}$,
yielding the Landau-Peierls diamagnetic susceptibility\cite{Landau30,Peierls33} as
\begin{eqnarray}
&&\hspace{-9mm}
\chi_{\alpha'\alpha}^{\rm LP}=
\frac{\epsilon_{\alpha'\beta'\gamma'}\epsilon_{\alpha\beta\gamma}}{6}\!
\left(\frac{e}{2\hbar c}\right)^{\! 2}
T\sum_{n} {\rm e}^{i\varepsilon_{n}0_{+}}  \! \sum_{{\mib k}}{\rm Tr}\biggl\{
\underline{G}^{2}(\varepsilon_{n},{\mib k})
\nonumber \\
&&\hspace{2mm}
\times
[\nabla_{\!{\mib k}}^{\beta'}\nabla_{\!{\mib k}}^{\beta} 
\underline{G}^{-1}(\varepsilon_{n},{\mib k})]
[\nabla_{\!{\mib k}}^{\gamma'}\nabla_{\!{\mib k}}^{\gamma} 
\underline{G}^{-1}(\varepsilon_{n},{\mib k})] \biggr\} \, .
\label{chi-LP}
\end{eqnarray}
This expression agrees exactly with eq.\ (4.15) of Buot.\cite{Buot76}

We next consider the second category. Differentiating eq.\ (\ref{Omega-2})
with respect to the explicit ${\mib B}$ dependences in 
$\underline{\cal H}({\mib k},{\mib B})$,
$\underline{\Sigma}(\varepsilon_{n},{\mib k},{\mib B},\{\underline{G} \})$, and 
$\Phi({\mib B},\{\underline{G}\})$, we obtain
the relevant magnetization as
\begin{eqnarray}
&&\hspace{-9mm}
M_{\alpha}'=-T
\sum_{n}{\rm e}^{i\varepsilon_{n}0_{+}}\sum_{{\mib k}}{\rm Tr}\,
\underline{G}(\varepsilon_{n},{\mib k},{\mib B})
\frac{\partial \underline{\cal H}({\mib k},{\mib B})}
{\partial B_{\alpha}}
\nonumber \\
&&\hspace{1mm}
-\,\frac{\partial \Phi({\mib B},\{\underline{G}\})}{\partial B_{\alpha}} \, .
\label{M'}
\end{eqnarray}
We further differentiate eq.\ (\ref{M'}) with respect to 
$B_{\alpha'}$ and put ${\mib B}\!=\! {\mib 0}$,
where we also need to consider the implicit ${\mib B}$
dependence through $\underline{G}$.
Using eq.\ (\ref{Sigma-k}), we obtain
\begin{eqnarray}
&& \hspace{-10mm}
\chi_{\alpha'\alpha}'=-T
\sum_{n}{\rm e}^{i\varepsilon_{n}0_{+}}\sum_{{\mib k}}{\rm Tr}\,
\underline{G}\,
\frac{\partial^{2} \underline{\cal H}^{(2)}}
{\partial B_{\alpha'}\partial B_{\alpha}}
-\,\frac{\partial \Phi^{(2)}}
{\partial B_{\alpha'}\partial B_{\alpha}} 
\nonumber \\
&& \hspace{-2mm}
-T \sum_{n}{\rm e}^{i\varepsilon_{n}0_{+}}\sum_{{\mib k}}{\rm Tr}\!
\left(\!\frac{\partial \underline{\cal H}^{(1)}}
{\partial B_{\alpha}}\!+\!
\frac{\partial \underline{\Sigma}^{(1)}}
{\partial B_{\alpha}}\!\right) \!
\frac{d \underline{G}}{d B_{\alpha'}}\, ,
\label{chi'}
\end{eqnarray}
where $\underline{\cal H}^{(j)}$, $\Phi^{(j)}$, and 
$\underline{\Sigma}^{(j)}$ $(j\!=\! 1,2)$ are quantities 
of the order $B^{j}$ with respect to the explicit ${\mib B}$
dependences as given by eqs.\ (\ref{barH}), (\ref{Phi-Exp}),
and (\ref{Sigma-Exp}), respectively.
To find an expression for 
${d \underline{G}}/{d B_{\alpha'}}$ in eq.\ (\ref{chi'}),
we differentiate $\underline{G}^{-1}\underline{G}
\!=\! \underline{1}$
with respect to $B_{\alpha'}$.
We thereby obtain a leading-order equation 
for ${d \underline{G}}/{d B_{\alpha'}}$ given in terms of
${d \underline{G}}^{-1}/{d B_{\alpha'}}$ as
\begin{equation}
\frac{d \underline{G}}{d B_{\alpha'}}=\underline{G}
\left(\!\frac{\partial \underline{\cal H}^{(1)}}
{\partial B_{\alpha'}}+
\frac{\partial \underline{\Sigma}^{(1)}}
{\partial B_{\alpha'}}
+
\frac{d \underline{\Sigma}^{(0)}}{d B_{\alpha'}}
\!\right)\underline{G} \, ,
\label{dGdB}
\end{equation}
with $\underline{G}\!=\!\underline{G}({\mib B}\!=\!{\mib 0})$ on the right-hand side.
The last term in the bracket of eq.\ (\ref{dGdB})
originates from the implicit ${\mib B}$ dependence though $\underline{G}$,
giving rise to the Fermi-liquid (or vertex) corrections.\cite{Landau56,BP91}
To find an equation for ${d \underline{\Sigma}^{(0)}}/d B_{\alpha'}$,
we introduce a vertex function as
\begin{equation}
\Gamma_{b_{1}b_{1}',b_{2}b_{2}'}^{(0)}(1,2)
\equiv
\frac{N_{\rm c}}{T^{2}}\,\frac{\delta^{2}\Phi^{(0)}}{\delta G_{b_{1}'b_{1}}(1)
\delta G_{b_{2}'b_{2}}(2)} \, ,
\label{Gamma}
\end{equation}
with $1\!\equiv\! (\varepsilon_{n_{1}},{\mib k}_{1})$. 
Noting eq.\ (\ref{Sigma-k}), we then obtain an
expression for $d \underline{\Sigma}^{(0)}\!/{d B_{\alpha'}}$
in terms of $d \underline{G}/{d B_{\alpha'}}$ as
\begin{equation}
\frac{d \Sigma_{b_{1}b_{1}'}^{(0)}(1)}{d B_{\alpha'}}
=\frac{T}{N_{\rm c}}\sum_{b_{2}b_{2}'}\sum_{n_{2}{\mib k}_{2}}
\Gamma^{(0)}_{b_{1}b_{1}',b_{2}b_{2}'}(1,
2)
\frac{d G_{b_{2}'b_{2}}(2)} 
{d B_{\alpha'}}\, .
\label{dSdB}
\end{equation}
Equations (\ref{dGdB}) and (\ref{dSdB}) form a closed set of equations
to determine ${d \underline{G}}/{d B_{\alpha'}}$.
Introducing the matrix $\underline{\cal I}$ by
\begin{equation}
{\cal I}_{b_{1}'b_{1}1,b_{2}'b_{2}2}\!\equiv\!
\frac{T}{N_{\rm c}}\sum_{b_{3}b_{3}'}
G_{b_{1}'b_{3}}(1)G_{b_{3}'b_{1}}(1)
\Gamma^{(0)}_{b_{3}b_{3}';b_{2}b_{2}'}(1,2) \, ,
\label{calM}
\end{equation}
the coupled equations are solved formally as
\begin{eqnarray}
&&\hspace{-12mm}
\frac{d G_{b_{1}'b_{1}}(1)}{d B_{\alpha'}}
=
\sum_{b_{2}b_{2}'}\sum_{n_{2}{\mib k}_{2}}
\left[(\underline{1}-\underline{\cal I})^{-1}\right]_{b_{1}'b_{1}1,b_{2}'b_{2}2}
\nonumber \\
&&\hspace{-1mm}
\times\!
\left[\underline{G}(2)\!
\left(\!\frac{\partial \underline{\cal H}^{(1)}\!({\mib k}_{2})}
{\partial B_{\alpha'}}\!+\!
\frac{\partial \underline{\Sigma}^{(1)}\!(2)}
{\partial B_{\alpha'}}\!\right)
\!\underline{G}(2)\right]_{b_{2}'b_{2}} \! ,
\label{dGdB2}
\end{eqnarray}
with $\underline{1}\!\equiv\! (\delta_{b_{1}'b_{2}'}\delta_{b_{1}b_{2}}
\delta_{{\mib k}_{1}{\mib k}_{2}}\delta_{n_{1}n_{2}})$.
Substituting eq.\ (\ref{dGdB2}) into eq.\ (\ref{chi'}), one can check that the symmetry
$\chi'_{\alpha'\alpha}\!=\!\chi'_{\alpha\alpha'}$ is satisfied as required.

Equation (\ref{chi'}) with eqs.\ (\ref{Gamma}), (\ref{calM}), and (\ref{dGdB2})
enables us to calculate the relevant susceptibility $\chi'_{\alpha'\alpha}$
once the functional $\Phi$ is given explicitly.
The total susceptibility is then obtained by
\begin{equation}
\chi_{\alpha'\alpha}=\chi_{\alpha'\alpha}^{\rm LP}+\chi_{\alpha'\alpha}' \, ,
\label{chi}
\end{equation}
with $\chi_{\alpha'\alpha}^{\rm LP}$ given by eq.\ (\ref{chi-LP}).
The factor $(\underline{1}-\underline{\cal I})^{-1}$ in eq.\ (\ref{dGdB2}) 
originates from vertex corrections for both the orbital and spin parts,
which have been derived naturally in our treatment.
It includes the Stoner enhancement factor as the intra-band contribution 
of the spin part.

\subsection{Hartree-Fock approximation}

We now specialize to the Hartree-Fock approximation.
Due to the absence of $\varepsilon_{n}$ dependence in the self-energy 
$\underline{\Sigma}^{{\rm HF}}({\mib k})$,
every summation over $n$ can be performed\cite{LW60}
by using the Fermi distribution function:
\begin{equation}
f(\varepsilon)\equiv
\frac{1}{{\rm e}^{\varepsilon/T}+ 1} \, .
\label{f}
\end{equation}
Thus, it is possible to simplify the expression of $\chi_{\alpha'\alpha}$
further.

Let us adopt the representation of
diagonalizing the Hartree-Fock energy in zero field as
\begin{equation}
{\cal H}_{bb'}({\mib k},{\mib B}\!=\!{\mib 0})
+ \Sigma_{bb'}^{{\rm HF}}({\mib k},{\mib B}\!=\!{\mib 0})
= \delta_{bb'}\xi_{b{\mib k}} \, .
\label{xi}
\end{equation}
Then the Landau-Peierls susceptibility 
of eq.\ (\ref{chi-LP}) is transformed into
\begin{equation}
\chi_{\alpha'\alpha}^{{\rm LP-HF}}\!=\! \frac{1}{6} \!
\left(\! \frac{e\hbar}{2c}\!\right)^{\! 2} \sum_{b{\mib k}} 
\frac{\partial f(\xi_{b{\mib k}})}{\partial \xi_{b{\mib k}}} \,
\frac{\epsilon_{\alpha'\beta'\gamma'}\epsilon_{\alpha\beta\gamma}}
{m^{*}_{\beta'\beta}(b{\mib k})\,m^{*}_{\gamma'\gamma}(b{\mib k})}
\, ,
\label{chi-LPHF}
\end{equation}
with $m^{*}_{\beta'\beta}(b{\mib k})\!\equiv\! \hbar^{2}/
(\nabla_{\!{\mib k}}^{\beta'}\nabla_{\!{\mib k}}^{\beta}\xi_{b{\mib k}})$.
Thus, 
only the states near the Fermi energy are relevant to
the Landau-Peierls susceptibility with the effective mass in place of 
the bare electron mass.
However, this simple result no longer holds in those
approximations where the self-energy has 
$\varepsilon_{n}$ dependence.
Equation (\ref{chi-LPHF}) may be regarded as an extension of the 
Philippas-McClure result for the electron gas\cite{PM72}
to Bloch electrons.

Next, eq.\ (\ref{chi'}) can be written with respect to
$\underline{n}({\mib k})$ of eq.\ (\ref{n_k}) as
\begin{eqnarray}
&&\hspace{-15.5mm}
\chi_{\alpha'\alpha}^{\prime{\rm HF}}=-\sum_{{\mib k}}{\rm Tr}\,
\underline{n}({\mib k})\,
\frac{\partial^{2} \underline{\cal H}^{(2)}\hspace{-0.3mm}({\mib k})}
{\partial B_{\alpha'}\partial B_{\alpha}}
-\,\frac{\partial \Phi^{{\rm HF}(2)}}
{\partial B_{\alpha'}\partial B_{\alpha}} 
\nonumber \\
&&\hspace{-4mm}
-\sum_{{\mib k}}{\rm Tr}\!
\left[\frac{\partial \underline{\cal H}^{(1)}\hspace{-0.3mm}({\mib k})}
{\partial B_{\alpha}}\!+\!
\frac{\partial \underline{\Sigma}^{{\rm HF}(1)}\hspace{-0.3mm}({\mib k})}
{\partial B_{\alpha}}\right] \!
\frac{d \underline{n}({\mib k})}{d B_{\alpha'}}\, .
\label{chi'-HF}
\end{eqnarray}
The quantity
${d \underline{n}}/{d B_{\alpha'}}$ is obtained from eq.\ (\ref{dGdB}) 
by adopting the representation (\ref{xi}) and 
carrying out the summation over $n$.
We thereby arrive at the expression:
\begin{eqnarray}
&&\hspace{-9mm}
\frac{d n_{b'b}({\mib k})}{d B_{\alpha'}}=
\frac{f(\xi_{b'{\mib k}})\!-\!f(\xi_{b{\mib k}})}
{\xi_{b'{\mib k}}\!-\!\xi_{b{\mib k}}}\!
\left[\frac{\partial {\cal H}_{b'b}^{(1)}({\mib k})}
{\partial B_{\alpha'}}+
\frac{\partial {\Sigma}_{b'b}^{{\rm HF}(1)}({\mib k})}
{\partial B_{\alpha'}}\right.
\nonumber \\
&&\hspace{25mm}
\left. +
\frac{d \Sigma^{{\rm HF}(0)}_{b'b}({\mib k})}{d B_{\alpha'}}
\right] \, .
\label{dGdB-HF}
\end{eqnarray}
To find an expression of ${d \Sigma^{{\rm HF}(0)}_{b'b}({\mib k})}/{d B_{\alpha'}}$,
we introduce a vertex function in terms of eq.\ (\ref{Phi-HF}) as
\begin{eqnarray}
&&\hspace{-9mm}
\Gamma^{{\rm HF}(0)}_{b_{1}b_{1}',b_{2}b_{2}'}({\mib k}_{1},{\mib k}_{2})
\equiv
N_{\rm c} \frac{\delta^{2}\Phi^{{\rm HF}(0)}}{\delta n_{b_{1}'b_{1}}({\mib k}_{1})
\delta n_{b_{2}'b_{2}}({\mib k}_{2})}
\nonumber \\
&&\hspace{-9mm}
= \frac{N_{\rm c}}{V}
\sum_{\mib K}\!\left[\,{\cal U}_{\mib K}\Lambda_{b_{1}b_{1}'}^{\!(0)}({\mib k}_{1},{\mib K})
\Lambda_{b_{2}b_{2}'}^{\!(0)}({\mib k}_{2},-{\mib K})
-{\cal U}_{{\mib k}_{2}-{\mib k}_{1}+{\mib K}}\right.
\nonumber \\
&&\hspace{-5mm}
\left.
\times\Lambda_{b_{1}b_{2}'}^{\!(0)}({\mib k}_{2},{\mib k}_{1}\!-\!{\mib k}_{2}\!-\!{\mib K})
\Lambda_{b_{2}b_{1}'}^{\!(0)}({\mib k}_{1},
{\mib k}_{2}\!-\!{\mib k}_{1}\!+\!{\mib K})\right]  .
\label{Gamma-HF}
\end{eqnarray}
Noting eq.\ (\ref{S-HF}), we then obtain
\begin{equation}
\frac{d \Sigma^{{\rm HF}(0)}_{b'b}({\mib k})}{d B_{\alpha'}}
=\frac{1}{N_{\rm c}}\sum_{b_{1}b_{1}'{\mib k}_{1}}
\Gamma_{b'b,b_{1}b_{1}'}^{{\rm HF}(0)}({\mib k},{\mib k}_{1})
\frac{d n_{b_{1}'b_{1}}({\mib k}_{1})}
{d B_{\alpha'}} \, .
\label{dS-HFdB}
\end{equation}
Equations (\ref{dGdB-HF}) and (\ref{dS-HFdB}) form a closed set of equations
to determine ${d {n}_{b'b}({\mib k})}/{d B_{\alpha'}}$.
Defining the matrix:
\begin{equation}
{\cal I}_{b_{1}'b_{1}{\mib k}_{1},b_{2}'b_{2}{\mib k}_{2}}^{\rm HF}\!\equiv
\frac{1}{N_{\rm c}} \frac{f(\xi_{b_{1}'{\mib k}_{1}})\!-\!
f(\xi_{b_{1}{\mib k}_{1}})}{\xi_{b_{1}'{\mib k}_{1}}\!-\!\xi_{b_{1}{\mib k}_{1}}}
\Gamma^{{\rm HF}(0)}_{b_{1}'b_{1},b_{2}b_{2}'}\!({\mib k}_{1},{\mib k}_{2}) \, ,
\label{calM-HF}
\end{equation}
the coupled equations are solved formally as
\begin{eqnarray}
&&\hspace{-15mm}
\frac{d n_{b'b}({\mib k})}{d B_{\alpha'}}
=
\sum_{b_{1}'b_{1}{\mib k}_{1}}
\left[(\underline{1}-\underline{\cal I}^{\rm HF})^{-1}\right]_{b'b{\mib k},
b_{1}'b_{1}{\mib k}_{1}}
\nonumber \\
&&\hspace{-13mm}
\times
\frac{f(\xi_{b_{1}'{\mib k}_{1}})\!-\!f(\xi_{b_{1}{\mib k}_{1}})}
{\xi_{b_{1}'{\mib k}_{1}}\!-\!\xi_{b_{1}{\mib k}_{1}}}\!
\left[\frac{\partial {\cal H}_{b_{1}'b_{1}}^{(1)}({\mib k}_{1})}
{\partial B_{\alpha'}}\! +\!
\frac{\partial {\Sigma}_{b_{1}'b_{1}}^{{\rm HF}(1)}({\mib k}_{1})}
{\partial B_{\alpha'}}\right] \! ,
\label{dndB2}
\end{eqnarray}
with $\underline{1}\!\equiv\! (\delta_{b'b_{1}'}\delta_{bb_{1}}
\delta_{{\mib k}{\mib k}_{1}})$.

Equations (\ref{chi'-HF}) and (\ref{dndB2})
enables us to calculate $\chi^{\prime{\rm HF}}_{\alpha'\alpha}$.
The total susceptibility is then obtained by
\begin{equation}
\chi^{{\rm HF}}_{\alpha'\alpha}=\chi^{{\rm LP-HF}}_{\alpha'\alpha}+
\chi^{\prime{\rm HF}}_{\alpha'\alpha} \, ,
\label{chi-HF}
\end{equation}
with $\chi^{\rm LP-HF}_{\alpha'\alpha}$ given by eq.\ (\ref{chi-LPHF}).

\subsection{Density functional theory}
\label{subsec:chi-DFT}

From a practical viewpoint, it is perhaps
most useful at present to derive
an expression of the magnetic susceptibility
within the density functional theory.
We hence consider it in most detail.
Since the self-energy does not depend on the Matsubara frequency, however,
we can exactly follow the procedure 
of the previous Hartree-Fock treatment.

\begin{table}[b]
\caption{\label{tab:table1}Quantities necessary to
calculate the magnetic susceptibility $\chi^{\rm DFT}_{\alpha'\alpha}$
of the density functional theory given by
eq.\ (\ref{chi-DFT}). Those with underlines are matrices 
with respect to the band-spin index $b$.}

\vspace{1.5mm}
\begin{tabular}{cccccc}
\vspace{0.5mm}
$\vec{u}_{b{\mib k}}({\mib r})$ & $\vec{w}_{b{\mib R}}({\mib r})$ & $\xi_{b{\mib k}}$ 
& $\underline{v}^{\alpha}({\mib k})$ & $m^{*}_{\beta\beta'}(b{\mib k})$
& $\epsilon_{\alpha\beta\gamma}$ 

\\
\hline
\vspace{0.5mm}
(\ref{Bloch}) & (\ref{Wannier}) & (\ref{xi-DFT}) & (\ref{v-2}) & (\ref{m*-DFT})
& (\ref{sigma-def})
\end{tabular}   
\begin{tabular}{cccccc}
\vspace{0.5mm}
$\underline{x}^{\alpha}({\mib k})$ 
& $\underline{\pi}^{\alpha}({\mib k})$
& $\underline{\sigma}^{\alpha}({\mib k})$ 
& $\underline{\cal O}^{(\alpha)}$ 
& $\underline{\cal H}^{{\rm DFT}(\alpha)}$
& $\underline{\Gamma}^{{\rm DFT}(0)}$
\\
\hline
(\ref{x-def}) 
& (\ref{pi}) & (\ref{sigma}) & (\ref{Oalpha}) & (\ref{H-DFT-alpha}) 
& (\ref{Gamma-DFT})

\end{tabular}
\begin{tabular}{cccccc}
\vspace{0.5mm}
$\underline{\cal I}^{{\rm DFT}}$
& $\underline{\Lambda}^{(0)}$
& $\underline{\rho}^{\pm(0)}$
& $\tilde{\Lambda}^{(\alpha)}_{\mib K}$
& $\tilde{n}^{(\alpha)}_{\sigma}({\mib r})$
& $F^{(2)}_{\sigma\sigma'}({\mib r})$
\\
\hline
(\ref{calM-DFT}) & (\ref{Lambda0}) & (\ref{rho(0)})
& (\ref{tilde-Lambda}) & (\ref{tilde-n}) & (\ref{F(2)}) 

\end{tabular}
\end{table}

Let us adopt the representation of 
diagonalizing the Kohn-Sham single-particle energy
in zero field as
\begin{equation}
{\cal H}_{bb'}^{(0)}({\mib k})
+ \Sigma_{bb'}^{{\rm H}(0)}({\mib k})
+ \Sigma_{bb'}^{{\rm xc}(0)}({\mib k})
= \delta_{bb'}\xi_{b{\mib k}} \, ,
\label{xi-DFT}
\end{equation}
where ${\cal H}_{bb'}^{(0)}$, $\Sigma_{bb'}^{{\rm H}(0)}$, and $
\Sigma_{bb'}^{{\rm xc}(0)}$ are given by eqs.\ (\ref{barH(0)}),
(\ref{Sigma-H(0)}), and (\ref{Sigma-xc(0)}), respectively,
with ${n}_{bb'}({\mib k})\!=\! \delta_{bb'}f(\xi_{b{\mib k}})$.
The vertex function corresponding to eq.\ (\ref{Gamma-HF})
is defined in terms of eq.\ (\ref{Phi-DFT}) by
\begin{eqnarray}
&&\hspace{-14mm}
\Gamma^{{\rm DFT}(0)}_{b_{1}b_{1}',b_{2}b_{2}'}({\mib k}_{1},{\mib k}_{2})
\equiv
N_{\rm c} \frac{\delta^{2}\Phi^{{\rm DFT}(0)}}{\delta n_{b_{1}'b_{1}}({\mib k}_{1})
\delta n_{b_{2}'b_{2}}({\mib k}_{2})}
\nonumber \\
&&\hspace{-14mm}=\frac{N_{\rm c}}{V}
\sum_{{\mib K}\neq{\mib 0}}
\Lambda_{b_{1}b_{1}'}^{\!(0)}({\mib k}_{1},{\mib K})
\frac{4\pi e^{2}}{K^{2}}
\Lambda_{b_{2}b_{2}'}^{\!(0)}({\mib k}_{2},-{\mib K})
\nonumber \\
&&\hspace{-11mm}
+\frac{1}{N_{\rm c}}\sum_{\sigma,\sigma'=\pm}\int \!
\rho_{b_{1}b_{1}'}^{\sigma(0)}({\mib k}_{1},{\mib r})F^{(2)}_{\sigma\sigma'}({\mib r})
\rho^{\sigma'(0)}_{b_{2}b_{2}'}({\mib k}_{2},{\mib r}) \, {\rm d}{\mib r} \, ,
\label{Gamma-DFT}
\end{eqnarray}
with $\rho_{bb'}^{\sigma(0)}$ and $F^{(2)}_{\sigma\sigma'}$ given by eqs.\ 
(\ref{rho(0)}) and (\ref{F(2)}), respectively.
We also introduce a matrix:
\begin{equation}
{\cal I}_{b_{1}'b_{1}{\mib k}_{1},b_{2}'b_{2}{\mib k}_{2}}^{\rm DFT}\!\equiv
\frac{1}{N_{\rm c}} \frac{f(\xi_{b_{1}'{\mib k}_{1}})\!-\!
f(\xi_{b_{1}{\mib k}_{1}})}{\xi_{b_{1}'{\mib k}_{1}}\!-\!\xi_{b_{1}{\mib k}_{1}}}
\Gamma^{{\rm DFT}(0)}_{b_{1}'b_{1},b_{2}b_{2}'}\!({\mib k}_{1},{\mib k}_{2}) \, ,
\label{calM-DFT}
\end{equation}
with $f$ the Fermi distribution function of eq.\ (\ref{f}).

It is also useful to define a couple of 
quantities in terms of eqs.\ (\ref{Norm(1)}), (\ref{barH(1)}), 
(\ref{Sigma-H(1)}), and (\ref{Sigma-xc(1)}) as
\begin{equation}
\underline{\cal O}^{(\alpha)}({\mib k})
\equiv \frac{\partial \underline{\cal O}^{(1)}({\mib k})}
{\partial B_{\alpha}}= i\epsilon_{\alpha\beta\gamma}\frac{e}{2\hbar c}
\underline{x}^{\beta}({\mib k})\underline{x}^{\gamma}({\mib k}) \, ,
\label{Oalpha}
\end{equation}
\begin{eqnarray}
&&\hspace{-5mm} 
\underline{\cal H}^{{\rm DFT}(\alpha)}({\mib k})
\equiv \frac{\partial }
{\partial B_{\alpha}}\!
\left[\underline{\cal H}^{(1)}({\mib k})
\!+\!\underline{\Sigma}^{{\rm H}(1)}({\mib k})
\!+\! \underline{\Sigma}^{{\rm xc}(1)}({\mib k})\right]
\nonumber \\
&&\hspace{-5mm}
=-\epsilon_{\alpha\beta\gamma}\frac{e}{2\hbar c}\bigl(
\{\underline{x}^{\beta},\hbar \underline{v}^{\gamma}+\underline{\pi}^{\gamma}\}
+\underline{\sigma}^{\beta\gamma} \bigr) 
\nonumber \\
&&\hspace{-1mm}
+\frac{N_{\rm c}}{V}
\sum_{{\mib K}\neq{\mib 0}}
\underline{\Lambda}^{\!(0)}({\mib k},{\mib K})\,\frac{4\pi e^{2}}{K^{2}}\,
\tilde{\Lambda}^{(\alpha)}_{-{\mib K}}
\nonumber \\
&&\hspace{-1mm}
+\frac{1}{N_{\rm c}}\sum_{\sigma,\sigma'=\pm}\int\!
\underline{\rho}^{\sigma(0)}({\mib k},{\mib r})\, F^{(2)}_{\sigma\sigma'}({\mib r})
\, \tilde{n}_{\sigma'}^{(\alpha)}({\mib r}) \, {\rm d}{\mib r}\, .
\label{H-DFT-alpha}
\end{eqnarray}
Here $\underline{v}^{\alpha}$
is the renormalized velocity:
\begin{equation}
v^{\alpha}_{bb'}
\equiv  \delta_{bb'} \frac{1}{\hbar} \,
\frac{\partial \xi_{b{\mib k}}}{\partial k_{\alpha}} \, ,
\label{v}
\end{equation}
and $\underline{\Lambda}^{\!(0)}({\mib k},{\mib K})$, 
$\underline{\rho}^{\pm(0)}({\mib k},{\mib r})$,
and $F^{(2)}_{\sigma\sigma'}({\mib r})$
are given by eqs.\ (\ref{Lambda0}), (\ref{rho(0)}) and (\ref{F(2)}),
respectively. The quantities $\tilde{\Lambda}_{\mib K}^{(\alpha)}$ and
$\tilde{n}_{\pm}^{(\alpha)}({\mib r})$ are defined by
\begin{subequations}
\begin{eqnarray}
&&\hspace{-10mm}
\tilde{\Lambda}_{\mib K}^{(\alpha)}
\equiv
\frac{1}{N_{\rm c}}\!\sum_{b{\mib k}}
f(\xi_{b{\mib k}})\!\left[
{\Lambda}^{\prime(\alpha)}_{bb}({\mib k},{\mib K})\right.
\nonumber \\
&&\hspace{1mm}\left.
-\{\underline{{\cal O}}^{(\alpha)}({\mib k}),
\underline{\Lambda}^{\! (0)}({\mib k},{\mib K})\}_{bb}\right] \, ,
\label{tilde-Lambda}
\end{eqnarray}
\begin{eqnarray}
&&\hspace{-10mm}
\tilde{n}_{\pm}^{(\alpha)}({\mib r})\! \equiv \! \frac{1}{N_{\rm c}}\!
\sum_{b{\mib k}}
\hspace{-0.5mm}
f(\xi_{b{\mib k}})\!
\left[
{\rho}^{\prime\pm(\alpha)}_{bb}({\mib k},{\mib r})
\right.
\nonumber \\
&&\hspace{4mm}\left.-
\{\underline{{\cal O}}^{(\alpha)}({\mib k}),
\underline{\rho}^{\pm (0)}({\mib k},{\mib r})\}_{bb}\right]
,
\label{tilde-n}
\end{eqnarray}
\end{subequations}
with
\begin{subequations}
\begin{equation}
\underline{\Lambda}^{\prime(\alpha)}({\mib k},{\mib K})
\equiv i\epsilon_{\alpha\beta\gamma}
\frac{e}{2\hbar c}\underline{x}^{\beta}({\mib k})
\underline{\Lambda}^{\! (0)}({\mib k},{\mib K})
\underline{x}^{\gamma}({\mib k}) \, ,
\label{Lambda-alpha}
\end{equation}
\begin{equation}
\underline{\rho}^{\prime\pm(\alpha)}({\mib k},{\mib r})
\equiv i\epsilon_{\alpha\beta\gamma}
\frac{e}{2\hbar c}\underline{x}^{\beta}({\mib k})
\underline{\rho}^{\pm(0)}({\mib k},{\mib r})
\underline{x}^{\gamma}({\mib k}) \, .
\label{rho-alpha}
\end{equation}
\end{subequations}
Note $\underline{\Lambda}^{\prime(\alpha)}
\!=\!\partial \underline{\Lambda}^{\prime (1)}/\partial
B_{\alpha}$ and
$\underline{\rho}^{\prime\pm(\alpha)}
\!=\!\partial \underline{\rho}^{\prime \,\pm(1)}/\partial
B_{\alpha}$ with 
$\underline{\Lambda}^{\prime (1)}$ and
$\underline{\rho}^{\prime \,\pm(1)}$ given by eqs.\ (\ref{Lambda'1})
and (\ref{rho'(1)}), respectively.

With these preliminaries, 
the susceptibility $\chi_{\alpha'\alpha}^{\rm DFT}$
can be calculated easily by eq.\ (\ref{chi-HF}) with
the replacement of the superscript HF$\rightarrow$DFT.
It may be written as a sum of three contributions as
\begin{equation}
\chi_{\alpha'\alpha}^{{\rm DFT}}=\chi_{\alpha'\alpha}^{{\rm LP-DFT}}
+\chi_{\alpha'\alpha}^{{\rm PvV-DFT}}+\chi_{\alpha'\alpha}^{{\rm (2)-DFT}} \, .
\label{chi-DFT}
\end{equation}

The first term denotes the Landau-Peierls diamagnetism corresponding 
to eq.\ (\ref{chi-LPHF}), given explicitly by
\begin{subequations}
\label{chi-DFT0-2}
\begin{equation}
\chi_{\alpha'\alpha}^{{\rm LP-DFT}}\!=\! \frac{1}{6} \!
\left(\! \frac{e\hbar}{2c}\!\right)^{\!\! 2} \sum_{b{\mib k}} 
\frac{\partial f(\xi_{b{\mib k}})}{\partial \xi_{b{\mib k}}} \,
\frac{\epsilon_{\alpha'\beta'\gamma'}\epsilon_{\alpha\beta\gamma}}
{m^{*}_{\beta'\beta}(b{\mib k})\,m^{*}_{\gamma'\gamma}(b{\mib k})}
\, ,
\label{chi-LPDFT}
\end{equation}
with
$m^{*}_{\beta'\beta}(b{\mib k})\!\equiv\! \hbar^{2}/
(\nabla_{\!{\mib k}}^{\beta'}\nabla_{\!{\mib k}}^{\beta}\xi_{b{\mib k}})$.

The second term in eq.\ (\ref{chi-DFT}) comes from the last term
in eq.\ (\ref{chi'-HF}), i.e., the second-order perturbation
with respect to $\underline{\cal H}^{(1)}\!+\!\underline{\Sigma}^{{\rm DFT}(1)}$.
Its intra- and inter-band contributions yield
the Pauli and van Vleck paramagnetism,
respectively.
It is calculated as
\begin{eqnarray}
&&\hspace{-5mm}
\chi_{\alpha'\alpha}^{{\rm PvV-DFT}}=-
\sum_{b_{1}b_{1}'{\mib k}_{1}}\sum_{b_{2}b_{2}'{\mib k}_{2}}
{\cal H}^{{\rm DFT}(\alpha')}_{b_{1}b_{1}'}({\mib k}_{1})
\nonumber \\
&&\hspace{0mm}
\times[(\underline{1}-\underline{\cal I}^{\rm DFT})^{-1}]_{b_{1}'b_{1}{\mib k}_{1},
b_{2}'b_{2}{\mib k}_{2}}
\frac{f(\xi_{b_{2}'{\mib k}_{2}})\!-\!f(\xi_{b_{2}{\mib k}_{2}})}
{\xi_{b_{2}'{\mib k}_{2}}\!-\!\xi_{b_{2}{\mib k}_{2}}}
\nonumber \\
&&\hspace{0mm}
\times{\cal H}^{{\rm DFT}(\alpha)}_{b_{2}'b_{2}}({\mib k}_{2}) \, ,
\label{chi-DFT-PvV}
\end{eqnarray}
where $\underline{\cal I}^{\rm DFT}$ and $\underline{\cal H}^{{\rm DFT}(\alpha)}$ are
given by eqs.\ (\ref{calM-DFT}) and (\ref{H-DFT-alpha}), respectively,
and $[{f(\xi_{b_{2}'{\mib k}_{2}})\!-\!f(\xi_{b_{2}{\mib k}_{2}})}]/
({\xi_{b_{2}'{\mib k}_{2}}\!-\!\xi_{b_{2}{\mib k}_{2}}})$ should be replaced by 
$\partial f(\xi_{b_{2}{\mib k}_{2}})/\partial \xi_{b_{2}{\mib k}_{2}}$
for $\xi_{b_{2}'{\mib k}_{2}}\!=\!\xi_{b_{2}{\mib k}_{2}}$.
Thus, the factor $(\underline{1}-\underline{\cal I}^{\rm DFT})^{-1}$ from
vertex corrections is explicitly included in our formula.

The third term in eq.\ (\ref{chi-DFT}) is due to
the first two terms in eq.\ (\ref{chi'-HF}).
It is calculated from eq.\ (\ref{barH(2)}), (\ref{Phi-H(2)}), and (\ref{Exc(2)})
as
\begin{eqnarray}
&&\hspace{-8.5mm}
\chi_{\alpha'\alpha}^{{\rm (2)-DFT}}\! =
\epsilon_{\alpha'\beta'\gamma'}
\epsilon_{\alpha\beta\gamma}
\frac{e^{2}}{2c^{2}}\! \sum_{b{\mib k}}\biggl\{\!
\frac{\partial f(\xi_{b{\mib k}})}{\partial \xi_{b{\mib k}}}
\biggl[
\{\underline{x}^{\beta'}\!,\underline{x}^{\beta}\}_{bb}\,
v_{b}^{\gamma'}\! v_{b}^{\gamma}
\nonumber \\
&&\hspace{-1mm}
+ \frac{1}{4\hbar}
(\underline{x}^{\beta}\underline{\pi}^{\gamma'}\underline{x}^{\beta'}\!+\!
\underline{x}^{\beta'}\underline{\pi}^{\gamma'}\underline{x}^{\beta}
)_{bb} v_{b}^{\gamma}
\!+\!\frac{1}{2\hbar}\{\underline{x}^{\beta},\underline{\sigma}^{\beta'\gamma'}\}_{bb}
v_{b}^{\gamma}
\nonumber \\
&&\hspace{-1mm}
+ \frac{1}{4\hbar}
(\underline{x}^{\beta'}\underline{\pi}^{\gamma}\underline{x}^{\beta}\!+\!
\underline{x}^{\beta}\underline{\pi}^{\gamma}\underline{x}^{\beta'}
)_{bb} v_{b}^{\gamma'}
\!+\!\frac{1}{2\hbar}\{\underline{x}^{\beta'},\underline{\sigma}^{\beta\gamma}\}_{bb}
v_{b}^{\gamma'}
\biggr]
\nonumber \\
&&\hspace{-1mm}
-\frac{1}{2m}f(\xi_{b{\mib k}})
\{\underline{x}^{\beta'},\underline{x}^{\beta}\}_{bb}
\biggl[ \delta_{\gamma'\gamma}-\frac{m}
{m^{*}_{\gamma'\gamma}(b{\mib k})}\biggr] \biggr\}
\nonumber \\
&&\hspace{-1mm}
+\sum_{b{\mib k}}f(\xi_{b{\mib k}})
\left[
\{\underline{\cal O}^{(\alpha)}({\mib k}),
\underline{\cal H}^{(\alpha')}\!({\mib k})\}_{bb}\right.
\nonumber \\
&&\hspace{16.5mm}\left.
+\{\underline{\cal O}^{(\alpha')}({\mib k}),
\underline{\cal H}^{(\alpha)}\!({\mib k})\}_{bb}\right]
\nonumber \\
&&\hspace{-1mm}
+\frac{1}{4}\sum_{bb'{\mib k}}f(\xi_{b{\mib k}})
(\xi_{b'{\mib k}}\!-\!\xi_{b{\mib k}})
({\cal O}^{(\alpha')}_{bb'}{\cal O}_{b'b}^{(\alpha)}+
{\cal O}^{(\alpha)}_{bb'}{\cal O}_{b'b}^{(\alpha')})
\nonumber \\
&&\hspace{-1mm}
-\frac{N_{\rm c}^{2}}{V}\sum_{{\mib K}\neq{\mib 0}}
\tilde{\Lambda}_{\mib K}^{(\alpha')}\, \frac{4\pi e^{2}}{K^{2}} \, 
\tilde{\Lambda}_{-{\mib K}}^{(\alpha)}
\nonumber \\
&&\hspace{-1mm}
-2\!\!\! \sum_{\sigma,\sigma'=\pm}\int\!
\tilde{n}_{\sigma}^{(\alpha)}({\mib r})F^{(2)}_{\sigma\sigma'}({\mib r})
\tilde{n}_{\sigma'}^{(\alpha')}({\mib r}) \, {\rm d}{\mib r}\, .
\label{chi-DFT2}
\end{eqnarray}
\end{subequations}

Equation (\ref{chi-DFT}) with eq.\ (\ref{chi-DFT0-2}) is one of the main results
of the paper. It extends the result of Roth\cite{Roth62} 
to take both the Coulomb and
exchange-correlation effects into account 
with appropriate vertex corrections.
Indeed, one can check that $\chi^{\rm DFT}$ reduces 
in the non-interacting limit 
to the expression obtained by Roth.\cite{comment}
The formula enables us to perform
non-empirical calculations of
the magnetic susceptibility in solids 
based on the density-functional theory.
Table I summarizes the quantities necessary for the calculation of
$\chi^{\rm DFT}_{\alpha'\alpha}$.

For practical purposes, it may be worth transforming eq.\ (\ref{v}) and 
$m^{*}_{\beta'\beta}(b{\mib k})\!\equiv\! \hbar^{2}/
(\nabla_{\!{\mib k}}^{\beta'}\nabla_{\!{\mib k}}^{\beta}\xi_{b{\mib k}})$
in eq.\ (\ref{chi-LPDFT})
into expressions without differentiations with respect to ${\mib k}$.
To this end, let us define a transfer integral in terms of eq.\ (\ref{xi-DFT}) as
\begin{equation}
\bar{t}_{b{\mib R}}\equiv \frac{1}{N_{\rm c}}\sum_{\mib k}
{\rm e}^{i{\mib k}\cdot{\mib R}}\,\xi_{b{\mib k}} \, .
\label{bar-t}
\end{equation}
Then $v_{b}^{\alpha}({\mib k})$ can be calculated 
by either of the two expressions:
\begin{subequations}
\label{vm}
\begin{equation}
v_{b}^{\alpha}({\mib k})=\frac{1}{\hbar}\frac{\partial \xi_{b{\mib k}}}
{\partial k_{\alpha}}=
\frac{-i}{\hbar}\sum_{\mib R}
{\rm e}^{-i{\mib k}\cdot{\mib R}}\,\bar{t}_{b{\mib R}}\,R_{\alpha} \, .
\label{v-2}
\end{equation}
Also $1/m^{*}_{\beta'\beta}(b{\mib k})$ is obtained by
\begin{equation}
\frac{1}{m^{*}_{\beta'\beta}(b{\mib k})}=\frac{1}{\hbar^{2}}\frac{\partial^{2} \xi_{b{\mib k}}}
{\partial k_{\beta}\partial k_{\beta'}}=-\frac{1}{\hbar^{2}}\sum_{\mib R}
{\rm e}^{-i{\mib k}\cdot{\mib R}}\,\bar{t}_{b{\mib R}}\,R_{\beta}R_{\beta'}\, .
\label{m*-DFT}
\end{equation}
\end{subequations}
Together with eq.\ (\ref{x-def}), one may perform 
a calculation of $\chi^{\rm DFT}_{\alpha'\alpha}$ without recourse to
numerical differentiations in ${\mib k}$ space.

Three comments are in order before closing the section.
First, our susceptibility is defined
in terms of the mean flux density ${\mib B}$
by $M_{\alpha'}\!=\!\chi_{\alpha'\alpha}B_{\alpha}$,
which is different from the conventional definition
$M_{\alpha'}\!=\!\chi_{\alpha'\alpha}^{H} H_{\alpha}$
of using the external field ${\mib H}$.
However, $\chi_{\alpha'\alpha}^{H}$ is found easily
from the thermodynamic relation ${\mib H}\!=\!{\mib B}\!-\!
4\pi{\mib M}/V$.
Indeed, 
$\chi_{\alpha'\alpha}^{H}$ in the crystallographic coordinates is given by
\begin{equation}
\chi_{\alpha'\alpha}^{H}=\delta_{\alpha'\alpha}
\frac{\chi_{\alpha\alpha}}{1-4\pi\chi_{\alpha\alpha}/V} \, .
\end{equation}
However, the difference between $\chi_{\alpha\alpha}$ and
$\chi_{\alpha\alpha}^{H}$ is negligible for most of the non-magnetic materials.
It is also worth noting that choosing ${\mib B}$ as an independent variable 
is more favorable when extending the theory to superconductors.\cite{Kita98,Kita98-2}
Secondly, an expression of the spin susceptibility 
within the density functional theory was already derived
by Vosko and Perdew\cite{VP75} incorporating vertex corrections,
which has been a basis of the number of theoretical works on the
spin susceptibility of metals.\cite{MV76,Gunnarsson76,Janak77}
Compared with it, however,
the present formula has a couple of advantages that
(i) the orbital contribution can be calculated on an equal footing with 
the spin part and
(ii) vertex corrections are incorporated explicitly in the formula
without any further approximations.
Third, any calculations of the orbital susceptibility based on a model Hamiltonian,
such as the Hubbard model, completely fail to incorporate the orbital contributions to
$\chi^{\rm PvV}_{\alpha'\alpha}$ and $\chi^{(2)}_{\alpha'\alpha}$.
This is because the key quantity
$\underline{x}^{\alpha}({\mib k})$, 
which originates from the change of the basis functions by the magnetic field,
is necessarily set equal to zero in those calculations. 
Thus, we need to include the field effect on the basis functions
for any practical calculations of the orbital magnetism.

\section{De Haas-Van Alphen Oscillation}

We finally study many-body effects on the dHvA oscillation in metals,
limiting our consideration to clean systems without impurity scatterings.
A definite advantage here over the previous 
studies\cite{Luttinger61,Gorkov62,ES70,WS96} is
that the structure of the thermodynamic
potential is known explicitly as eq.\ (\ref{Omega-2}).

Following the original work by Luttinger and Ward,\cite{LW60}
let us regard eq.\ (\ref{Omega-2}) as a functional of
$\underline{\Sigma}(\varepsilon_{n},{\mib k})$ instead of 
$\underline{G}(\varepsilon_{n},{\mib k})$.
It is also stationary with respect to a variation in 
$\underline{\Sigma}$ satisfying Dyson's equation (\ref{Dyson-2}).
We then write $\underline{\Sigma}$ as a sum of the monotonic and
oscillatory parts as
\begin{equation}
\underline{\Sigma}=\underline{\Sigma}^{0}+\underline{\Sigma}^{\rm osc}\, .
\end{equation}
As shown by 
Luttinger within the Hartree-Fock approximation\cite{Luttinger61} 
and discussed more generally by Bychkov and Go'kov,\cite{Gorkov62}
the oscillatory part $\underline{\Sigma}^{\rm osc}$ for the spherical Fermi surface is smaller than
the monotonic part $\underline{\Sigma}^{0}$ by 
the order $(\hbar \omega_{\rm c}/\mu)^{3/2}$ ($\omega_{\rm c}$:
the cyclotron frequency).
We assume that the statement holds up to the infinite order,
as expected for three-dimentional Fermi surfaces.
Let us expand $\Omega(\underline{\Sigma})$ at $\underline{\Sigma}\!=\!
\underline{\Sigma}^{0}$ as
\begin{eqnarray}
&&\hspace{-9mm}
\Omega(\underline{\Sigma})=\Omega(\underline{\Sigma}^{0})
+\frac{1}{2}\!\!\left.\sum_{b_{1}b_{1}'b_{2}b_{2}'}\sum_{{\mib k}_{1}{\mib k}_{2}}\!
\frac{\delta^{2}\Omega}
{\delta \Sigma_{b_1 b_1'}({\mib k}_{1})\delta \Sigma_{b_2 b_2'}
({\mib k}_{2})}\right|_{0}
\nonumber \\
&&\hspace{3.5mm}
 \times
\Sigma_{b_1 b_1'}^{\rm osc}({\mib k}_{1})\Sigma_{b_2 b_2'}^{\rm osc}
({\mib k}_{2})+\cdots \, ,
\label{Omega-Exp}
\end{eqnarray}
where the term linear in $\underline{\Sigma}^{\rm osc}$ vanishes due
to Dyson's equation (\ref{Dyson-2}) for $\underline{\Sigma}\!=\!\underline{\Sigma}^{0}$
and $\underline{G}\!=\!\underline{G}^{0}$.
The second term on the right-hand side of eq.\ (\ref{Omega-Exp}) 
is of the order $(\hbar \omega_{\rm c}/\mu)^{3}$.
On the other hand, the first term also has an oscillatory
contribution originating from the operator ${\mib \kappa}$ in the logarithm
of eq.\ (\ref{Omega-2}); it is of the order 
$(\hbar \omega_{\rm c}/\mu)^{5/2}$ relative to the nomotonic part
of $\Omega$, as seen from the result of non-interacting systems.\cite{LK55}
 It hence follows that we may neglect the 
second contribution in eq.\ (\ref{Omega-Exp}).

The term relevant to the dHvA oscillation 
in $\Omega(\underline{\Sigma}^{0})$ is given by
\begin{eqnarray}
&&\hspace{-7mm}\Omega^{\rm osc}
=-T \! \sum_{n} {\rm Tr} \, {\rm Tr}_{\mib k} 
\ln \bigl[
 \underline{\cal H}({\mib \kappa},{\mib B})
\!+\!\underline{\Sigma}^{0}(\varepsilon_{n},{\mib \kappa},{\mib B})
\!-\!i\varepsilon_{n}\underline{1}  \bigr]
\nonumber \\
&&\hspace{6mm} \times {\rm e}^{i\varepsilon_{n}0_{+}} 
\nonumber \\
&&\hspace{-7mm}=- {\rm Tr}\, {\rm Tr}_{\mib k} 
\! \int_{-\infty}^{\infty}\! f(\varepsilon)\,
\bigl\{ \ln \bigl[
 \underline{\cal H}({\mib \kappa},{\mib B})
\!+\!\underline{\Sigma}^{0{\rm A}}(\varepsilon,{\mib \kappa},{\mib B})
\!-\!\varepsilon\underline{1}  \bigr]
\nonumber \\ 
&&\hspace{-3mm} -\ln \bigl[
 \underline{\cal H}({\mib \kappa},{\mib B})
\!+\!\underline{\Sigma}^{0{\rm A}*}(\varepsilon,{\mib \kappa},{\mib B})
\!-\!\varepsilon\underline{1}  \bigr]\bigr\}
\frac{{\rm d}\varepsilon}{2\pi i} \, ,
\label{Omega-osc2}
\end{eqnarray}
where $\underline{\Sigma}^{0{\rm A}}(\varepsilon,{\mib \kappa},{\mib B})\!\equiv\!
\underline{\Sigma}^{0}(\varepsilon_{n}\!\rightarrow\! 
-i\varepsilon\!- 0_{+},{\mib \kappa},{\mib B})$,
and we have used eq.\ (\ref{f}) to transform the summation over $n$ into an integration
on the real energy axis.
Since the oscillation is due to the states near the 
Fermi level, we only need to consider the region $\varepsilon\!\sim\! 0$
in the integral
where ${\rm Im} \underline{\Sigma}^{0{\rm A}}(\varepsilon)$ is
infinitesimal positive definite and may be approximated by 
${0}_{+}\underline{1}$.
The double trace in eq.\ (\ref{Omega-osc2}) is thereby simplified into
\begin{eqnarray}
&&\hspace{-9mm}
\Omega^{\rm osc}
=- \sum_{b} {\rm Tr}_{\mib k}
\! \int_{-\infty}^{\infty}\! f(\varepsilon)\,
\bigl\{ \ln \bigl[
h_{b}(\varepsilon,{\mib \kappa},{\mib B})\!-\!\mu
\!-\!\varepsilon+i0_{+} \bigr]
\nonumber \\ 
&&\hspace{2mm}
 -\ln \bigl[
h_{b}(\varepsilon,{\mib \kappa},{\mib B})\!-\!\mu
\!-\!\varepsilon -i0_{+} \bigr] \bigr\}
\frac{{\rm d}\varepsilon}{2\pi i} \, ,
\label{Omega-osc3}
\end{eqnarray}
where $h_{b}(\varepsilon,{\mib k},{\mib B})$ denotes
a characteristic value of the Hermitian matrix
$\underline{\cal H}({\mib k},{\mib B})
\!+\!{\rm Re}\underline{\Sigma}^{0{\rm A}}(\varepsilon,{\mib k},{\mib B})
\!+\!\mu\underline{1}$.

Next, we diagonalize $h_{b}(\varepsilon,{\mib \kappa},{\mib B})$
by solving the eigenvalue problem:
\begin{subequations}
\label{E_lambda}
\begin{equation}
h_{b}(\varepsilon,{\mib \kappa},{\mib B})\varphi_{\lambda}(\varepsilon,{\mib k},{\mib B})=
g_{\lambda}(\varepsilon,{\mib B})\varphi_{\lambda}(\varepsilon,{\mib k},{\mib B}) \, ,
\label{eigen-kappa}
\end{equation}
where $\varphi_{\lambda}$ is an eigenfunction and $g_{\lambda}$
is its eigenvalue. The subscript
$\lambda$ is composed of the band-spin index $b$, 
the Landau level $N$
together with an additional quantum number distinguishing its degeneracy,
and the wave vector $k_{z}$ parallel to the magnetic field 
${\mib B}\!\parallel\!{\mib z}$.
We then define the energy $E_{\lambda}\!=\!E_{\lambda}({\mib B})$ 
as the solution of the equation:
\begin{equation}
E_{\lambda}=g_{\lambda}(E_{\lambda}\!-\!\mu,{\mib B}) \, .
\label{xi-lambda}
\end{equation}
\end{subequations}
Note that the quantity $1\!-\!\left.\frac{\partial g_{\lambda}}
{\partial \varepsilon}\right|_{\varepsilon =0}$
for the state $E_{\lambda}\!=\! \mu$
is inverse of the discontinuity in the single-particle occupation
at the Fermi level so that it is positive.\cite{Luttinger60}
It hence follows that 
$\varepsilon-g_{\lambda}(\varepsilon)+\mu$ is a monotonically 
increasing function
for $\varepsilon\!\sim\! 0$ and $E_{\lambda}\!\sim\! \mu$.
Equation (\ref{Omega-osc3}) is thereby transformed into
\begin{eqnarray}
&&\hspace{-9mm}
\Omega^{\rm osc}
=- \sum_{\lambda} 
\! \int_{-\infty}^{\infty}\!f(\varepsilon)\,
\theta (\varepsilon+\mu-g_{\lambda}(\varepsilon))
\, {\rm d}\varepsilon \, ,
\nonumber \\
&&\hspace{-1.5mm}
=- T\sum_{\lambda} 
\ln \bigl[1+{\rm e}^{-(E_{\lambda}-\mu)/T}\bigr] \, ,
\label{Omega-osc4}
\end{eqnarray}
where $\theta$ is the step function.
Equation (\ref{Omega-osc4}) includes the main oscillatory part of the
thermodynamic potential.

Since $E_{\lambda}\!\sim\!\mu$ for the oscillatory part of eq.\ (\ref{Omega-osc4}), 
$E_{\lambda}$ may be calculated accurately by
a two-step semiclassical quantization scheme as follows.
We first determine the energy $E_{b{\mib k}}\!=\!E_{b{\mib k}}({\mib B})$
by solving
\begin{equation}
E_{b{\mib k}}=h_{b}(E_{b{\mib k}}\!-\!\mu,{\mib k},{\mib B}) \, .
\label{xi-ell}
\end{equation}
We then adopt
the Onsager-Lifshitz-Kosevich procedure:\cite{Onsager52,LK55}
\begin{equation}
S(E_{\lambda})=2\pi(N\!+\!\gamma) \frac{|e|B}{c} \, ,
\label{LK}
\end{equation}
with $N$ an integer,
$\gamma$ a constant of the order $1$, and
$S(E_{\lambda})$ denoting the area in ${\mib k}'$ space perpendicular to ${\mib B}$ 
specified by $E_{b{\mib k}'}\!\leq\! E_{\lambda}$ and $k'_{z}\!=\! k_{z}$.
Let us substitute those quantized energy levels into eq.\ (\ref{Omega-osc4}) and
follow the procedure of Lifshitz and Kosevich.\cite{LK55}
We thereby obtain a theory of the dHvA oscillation where
the Fermi surface by $E_{b{\mib k}}({\mib B})$ replaces
the non-interacting Fermi surface of Lifshitz and Kosevich.\cite{LK55}

There may be an alternative way to calculate $E_{\lambda}$ semiclassically.
We first determine the eigenvalue $g_{\lambda}(\varepsilon,{\mib B})$ 
in eq.\ (\ref{eigen-kappa}) as a function of $\varepsilon$
by 
\begin{equation}
S(\varepsilon,g_{\lambda})=2\pi(N\!+\!\gamma) \frac{|e|B}{c} \, ,
\label{LK2}
\end{equation}
where $S(\varepsilon,g_{\lambda})$ denotes the area in ${\mib k}'$ 
space perpendicular to 
${\mib B}\!\parallel\!{\mib z}$ 
specified by $h_{b}(\varepsilon,{\mib k}',{\mib B})\!\leq\! 
g_{\lambda}(\varepsilon,{\mib B})$
and $k'_{z}\!=\! k_{z}$.
We then obtain the quasiparticle energy $E_{\lambda}$
by solving eq.\ (\ref{xi-lambda}). Since the characteristic energy of 
${\mib \kappa}$ is $\hbar \omega_{\rm c}$, however,
the Fermi-surface structures determined by this latter procedure will not 
differ substantially from those of the first one.
It hence follows that we may adopt the first procedure
which is certainly more convenient.

Three comments are in order before closing the section.
First, the operator ${\mib \kappa}$ appears naturally in the argument of 
$\underline{\Sigma}^{0}$ as well as in $\underline{\cal H}$ of eq.\
(\ref{Omega-osc2}). This fact shows unambiguously the necessity of
considering the self-energy to make up the quantized energy levels.
Equation (\ref{Omega-osc2}) thus removes the confusion
on the many-body effects of the dHvA oscillation mentioned in Introduction;
it supports Luttinger's original argument, 
refuting the quantization procedure without the self-energy.\cite{ES70,WS96}
Second, the present theory can also treat changes
of the energy band structure with ${\mib B}$ by incorporating the 
explicit ${\mib B}$ dependences in
$\underline{\cal H}$ and $\underline{\Sigma}$.
Indeed, $\underline{\cal H}({\mib k},{\mib B})$ can be calculated by eq.\ (\ref{barH}), 
and $\underline{\Sigma}(\varepsilon_{n},{\mib k},{\mib B})$ may be obtained 
by the procedure of \S3.2 with dropping the oscillatory contribution of 
$\underline{G}(\varepsilon_{n},{\mib k},{\mib B})$.
This effect was neglected by Luttinger\cite{Luttinger61}
but can have a substantial importance,
particularly when approaching the magnetic breakdown.\cite{CF61}
It should be noted that this magnetic breakdown 
is beyond the description of the semiclassical
quantization procedure and we have to solve
eq.\ (\ref{E_lambda}) exactly
by taking the relevant multiple bands into account.
Third, we have used in eq.\ (\ref{Omega-osc3}) a characteristic value
$h_{b}(\varepsilon,{\mib k},{\mib B})$ 
and replaced ${\mib k}$ by the operator
${\mib \kappa}$.
This procedure is well defined for a simple band where
$h_{b}(\varepsilon,{\mib k},{\mib B})$ is analytic in ${\mib k}$.\cite{Nenciu91}
For a complex band with band crossings, however, 
$h_{b}(\varepsilon,{\mib k},{\mib B})$ may not be analytic in ${\mib k}$
so that the use of $h_{b}(\varepsilon,{\mib \kappa},{\mib B})$
may fail to describe some important effects such as band mixings
due to ${\mib \kappa}$. 
In this situation, one may be required to use a representation where
$\underline{\cal H}({\mib k},{\mib B})
\!+\!{\rm Re}\underline{\Sigma}^{0{\rm A}}(\varepsilon,{\mib k},{\mib B})
\!+\!\mu\underline{1}$ is analytic in ${\mib k}$
and directly solve the eigenvalue problem for 
$\underline{\cal H}({\mib \kappa},{\mib B})
\!+\!{\rm Re}\underline{\Sigma}^{0{\rm A}}(\varepsilon,{\mib \kappa},{\mib B})
\!+\!\mu\underline{1}$.

\section{Summary}

We have constructed a many-body perturbation theory for
Bloch electrons in a magnetic field on the basis of the energy band structure
in zero field. 
We have thereby clarified the structures of the thermodynamic
potential and the self-energy in a finite magnetic 
field, and provided a microscopic foundation
for the replacement procedure on the self-energy: 
$\Sigma^{(0)}(\varepsilon_{n},{\mib k})\!\rightarrow\!
\Sigma\!\left(\varepsilon_{n},-i{\mib \nabla}\!-\! \frac{e}{\hbar c}{\mib A}\right)$.
This perturbation theory is then applied to
obtain explicit expressions of the magnetic susceptibility $\chi$
at various approximation levels on the interaction.
The result for the density functional theory is given by
eq.\ (\ref{chi-DFT}) together with eq.\ (\ref{chi-DFT0-2}) and Table I.
It incorporates vertex corrections as well as interband transitions and 
core polarizations.
The expression enables us non-empirical calculations of $\chi$.
Thus, it will be useful to improve our quantitative understanding of 
the magnetic susceptibility in solids.
Finally, we have presented a many-body theory on the dHvA oscillation
in metals to show unambiguously that the Fermi surface structure with
interaction effects in zero field are indeed relevant to the phenomenon.

The present formulation may be extended easily to a non-uniform magnetic field.
Hence an application to superconductors will be fairly straightforward.
It is also desired to make up a many-body theory on the transport phenomena
of Bloch electrons in a magnetic field.

\section*{Acknowledgements}
This work is supported by
the 21st century COE program ``Topological Science and Technology,'' 
Hokkaido University.

\appendix

\section{Derivation of eq.\ (\ref{UExp})}
\label{app:bare-vertex}

We here transform eq.\ (\ref{U}) into eq.\ (\ref{UExp}). 
To start with,
let us consider eq.\ (\ref{U'}) and rewrite
$\vec{w}_{\nu_{3}}^{\prime\dagger}({\mib r})\vec{w}_{\nu_{3}'}^{\prime}({\mib r})$
in the integral with the procedure of eqs.\ 
(\ref{phase2}) and (\ref{k-diff}) as
\begin{eqnarray*}
&&\hspace{-5mm}
\vec{w}_{\nu_{3}}^{\prime\dagger}({\mib r})\vec{w}_{\nu_{3}'}^{\prime}({\mib r})
= \frac{{\rm e}^{iI_{{\mib R}_{3}{\mib R}_{3}'}}}{N_{\rm c}}
\sum_{{\mib k}_{3}{\mib k}_{3}'}{\rm e}^{-i{\mib k}_{3}\cdot({\mib r}-{\mib R}_3)
+i{\mib k}_{3}'\cdot({\mib r}-{\mib R}_3')}
\nonumber \\
&&\hspace{18mm}
\times {\rm e}^{ih_{\alpha\beta}{\mib \nabla}_{\!{\mib k}_{3}}^{\alpha}
{\mib \nabla}_{\!{\mib k}_{3}'}^{\beta}}\,
\vec{u}_{b_{3}{\mib k}_{3}}^{\dagger}({\mib r})
\vec{u}_{b_{3}'{\mib k}_{3}'}({\mib r}) \, .
\end{eqnarray*}
The quantity $\vec{w}_{\nu_{4}}^{\prime\dagger}({\mib r}')
\vec{w}_{\nu_{4}'}^{\prime}({\mib r}')$ may be expressed similarly.
Equation (\ref{U'}) is thereby transformed into
\begin{eqnarray}
&&\hspace{-8mm}
{\cal U}_{\nu_{3}\nu_{4};\nu_{3}'\nu_{4}'}'
\nonumber \\
&&\hspace{-8mm}
=
\frac{{\rm e}^{iI_{{\mib R}_{3}{\mib R}_{3}'}+iI_{{\mib R}_{4}{\mib R}_{4}'}}}
{N_{\rm c}^{2}V}\sum_{{\mib k}_{3}{\mib k}_{4}{\mib k}_{3}'{\mib k}_{4}'}
{\rm e}^{i{\mib k}_{3}\cdot{\mib R}_{3}-i{\mib k}_{3}'\cdot{\mib R}_{3}'
+i{\mib k}_{4}\cdot{\mib R}_{4}-i{\mib k}_{4}'\cdot{\mib R}_{4}'}
\nonumber \\
&& \hspace{-3mm} \times
\!\sum_{\mib q} {\cal U}_{{\mib q}}\,
\bar{\Lambda}_{b_{3}b_{3}'}'({\mib k}_{3},{\mib k}_{3}',{\mib q})
\bar{\Lambda}_{b_{4}b_{4}'}'({\mib k}_{4},{\mib k}_{4}',-{\mib q})\, ,
\label{U'Exp0}
\end{eqnarray}
where $\bar{\Lambda}_{bb'}'({\mib k},{\mib k}',{\mib q})$ is defined by
\begin{eqnarray*}
&&\hspace{-9mm}
\bar{\Lambda}_{bb'}'({\mib k},{\mib k}',{\mib q})
\equiv
\!\int
{\rm e}^{i(-{\mib k}+{\mib k}'+{\mib q})\cdot{\mib r}}\,
{\rm e}^{ih_{\alpha\beta}{\mib \nabla}_{\!{\mib k}}^{\alpha}
{\mib \nabla}_{\!{\mib k}'}^{\beta}}\,
\vec{u}_{b{\mib k}}^{\dagger}({\mib r})
\nonumber \\
&&\hspace{17mm}\times
\vec{u}_{b'{\mib k}'}({\mib r})\, {\rm d}{\mib r} \, .
\end{eqnarray*}
Note $\bar{\Lambda}_{bb'}'({\mib k},{\mib k}',{\mib q})
\!\propto \! \delta_{-{\mib k}+{\mib k}'+{\mib q},{\mib K}}$
with ${\mib K}$ a reciprocal lattice vector.
It hence follows that ${\mib k}$ 
in $\bar{\Lambda}_{bb'}'({\mib k},{\mib k}',{\mib q})$
can be written alternatively as ${\mib k}\!=\!\overline{{\mib k}'\!+\!{\mib q}}$ with
$\overline{{\mib k}'\!+\!{\mib q}}$ denoting the wave vector in the first Brillouin zone
corresponding to ${\mib k}'\!+\!{\mib q}$ in the extended zone scheme.
Using this notation and noting 
${\rm e}^{i(\overline{{\mib k}'+{\mib q}})\cdot{\mib R}}
\!=\!{\rm e}^{i({\mib k}'+{\mib q})\cdot{\mib R}}$, 
eq.\ (\ref{U'Exp0}) is simplified into
\begin{eqnarray}
&&\hspace{-9mm}
{\cal U}_{\nu_{3}\nu_{4};\nu_{3}'\nu_{4}'}'
=
\frac{{\rm e}^{iI_{{\mib R}_{3}{\mib R}_{3}'}+iI_{{\mib R}_{4}{\mib R}_{4}'}}}
{N_{\rm c}^{2}V}\!\sum_{{\mib k}{\mib k}'}\!
{\rm e}^{i{\mib k}\cdot({\mib R}_{3}-{\mib R}_{3}')
+i{\mib k}'\cdot({\mib R}_{4}-{\mib R}_{4}')}
\nonumber \\
&&\hspace{9mm}
\times \!
\sum_{{\mib q}}{\rm e}^{i{\mib q}\cdot({\mib R}_{3}-{\mib R}_{4})}
\,{\cal U}_{{\mib q}}\,
\Lambda_{b_{3}b_{3}'}'({\mib k},{\mib q},{\mib B})
\nonumber \\
&&\hspace{9mm}
\times 
\Lambda_{b_{4}b_{4}'}'({\mib k}',-{\mib q},{\mib B})\, ,
\label{U'Exp}
\end{eqnarray}
where $\Lambda_{bb'}'({\mib k},{\mib q},{\mib B})$ is defined by
\begin{subequations}
\label{Lambda'}
\begin{eqnarray}
&&\hspace{-12mm}
\Lambda_{bb'}'({\mib k},{\mib q},{\mib B})
\nonumber \\
&&\hspace{-12mm}
\equiv
\!\int\! 
{\rm e}^{i{\mib K}_{{\mib k}+{\mib q}}\cdot{\mib r}}\,
{\rm e}^{ih_{\alpha\beta}{\mib \nabla}_{\!{\mib k}'}^{\alpha}
{\mib \nabla}_{\!{\mib k}}^{\beta}}
\vec{u}_{b{\mib k}'}^{\dagger}({\mib r})
\vec{u}_{b'{\mib k}}({\mib r})\bigr|_{{\mib k}'=\overline{{\mib k}+{\mib q}}}\,\,
{\rm d}{\mib r} 
\nonumber \\
&&\hspace{-12mm}
=
\!\int\! 
{\rm e}^{ih_{\alpha\beta}{\mib \nabla}_{\!{\mib k}'}^{\alpha}
{\mib \nabla}_{\!{\mib k}}^{\beta}}
\vec{u}_{b{\mib k}'}^{\dagger}({\mib r})\,
{\rm e}^{i{\mib K}_{{\mib k}+{\mib q}}\cdot{\mib r}}\,
\vec{u}_{b'{\mib k}}({\mib r})\bigr|_{{\mib k}'=\overline{{\mib k}+{\mib q}}}\,\,
{\rm d}{\mib r} \, ,
\nonumber \\
\label{Lambda'-D}
\end{eqnarray}
with ${\mib K}_{{\mib k}+{\mib q}}\!\equiv\! 
{\mib k}\!+\!{\mib q}\!-\!\overline{{\mib k}\!+\!{\mib q}}$.
The third line follows from the fact that the discrete 
reciprocal vector
${\mib K}_{{\mib k}+{\mib q}}$ is independent of the infinitesimal changes in
${\mib k}$ or ${\mib k}\!+\!{\mib q}$.
The wave vector ${\mib k}$ in $\underline{\Lambda}'({\mib k},{\mib q},{\mib B})$ 
belongs to the incoming electron, ${\mib q}$ 
is an additional wave vector from the interaction, 
and $\overline{{\mib k}\!+\!{\mib q}}$ specifies 
the outgoing electron.

Equation (\ref{Lambda'-D}) can be expanded in powers of ${\mib B}$ as
\begin{equation}
\Lambda_{bb'}'({\mib k},{\mib q},{\mib B})\!
=\Lambda_{bb'}^{\!(0)}({\mib k},{\mib q})+\sum_{j=1}^{\infty}
\Lambda_{bb'}^{\!\prime(j)}({\mib k},{\mib q},{\mib B}) \, ,
\label{Lambda'-E}
\end{equation}
\end{subequations}
with $\Lambda_{bb'}^{\!(0)}$ given by eq.\ (\ref{Lambda0}).
The quantities $\Lambda_{bb'}^{\!\prime(j)}$ 
can be calculated with the same procedure as that for
obtaining eqs.\ (\ref{Hamil(1)}) and (\ref{Hamil(2)}).
To start with, let us rewrite
\begin{subequations}
\label{Lambda'0-2}
\begin{eqnarray*}
{\rm e}^{i{\mib K}_{{\mib k}+{\mib q}}\cdot{\mib r}}\,
|\vec{u}_{b'{\mib k}}\rangle= \sum_{b''}|\vec{u}_{b''\overline{{\mib k}+{\mib q}}}\rangle
\Lambda_{b''b'}^{\!(0)}({\mib k},{\mib q}) \, ,
\end{eqnarray*}
or equivalently,
\begin{eqnarray*}
\langle\vec{u}_{b\overline{{\mib k}+{\mib q}}}|\,{\rm e}^{i{\mib K}_{{\mib k}+{\mib q}}\cdot{\mib r}}
= \sum_{b''} 
\Lambda_{bb''}^{\!(0)}({\mib k},{\mib q})
\langle\vec{u}_{b''{\mib k}}| \, ,
\end{eqnarray*}
where use has been made of the completeness 
$1\!=\!\sum_{b''}|\vec{u}_{b''\overline{{\mib k}+{\mib q}}}\rangle
\langle\vec{u}_{b''\overline{{\mib k}+{\mib q}}}| =\sum_{b''}|\vec{u}_{b''{\mib k}}\rangle
\langle\vec{u}_{b''{\mib k}}| $ over the unit cell.
We then substitute the above identities into eq.\ (\ref{Lambda'-D}),
expand it in powers of ${\mib B}$ to obtain 
$\underline{\Lambda}^{\!(j)}({\mib k},{\mib q})$,
and average the resulting two different expressions.
We thereby obtain the first-order term as
\begin{eqnarray}
&&\hspace{-8.5mm}
\underline{\Lambda}^{\!\prime(1)}({\mib k},{\mib q})
=
\frac{1}{2}\biggl\{\underline{{\cal O}}^{(1)}(\overline{{\mib k}\!+\!{\mib q}})
\underline{\Lambda}^{\!(0)}({\mib k},{\mib q})+
\underline{\Lambda}^{\!(0)}({\mib k},{\mib q})
\underline{\cal O}^{(1)}({\mib k})
\nonumber \\
&&\hspace{17mm}
-h_{\alpha\beta}\,
\underline{x}^{\alpha}(\overline{{\mib k}\!+\!{\mib q}})\,
\nabla_{\!{\mib k}}^{\beta}\underline{\Lambda}^{\!(0)}({\mib k},{\mib q})
\nonumber \\
&&\hspace{17mm}
-h_{\alpha\beta}\!
\left[\nabla_{\!{\mib k}}^{\beta}
\underline{\Lambda}^{\!(0)}({\mib k},{\mib q})
\right]\underline{x}^{\alpha}({\mib k}) \biggr\}\, .
\label{Lambda'1}
\end{eqnarray}
Note that the arguments of $\underline{\cal O}^{(1)}$ and $\underline{x}^{\alpha}$
are 
${\mib k}$ ($\overline{{\mib k}\!+\!{\mib q}}$)
when they appear to the right (left) of 
$\underline{\Lambda}^{\!(0)}({\mib k},{\mib q})$.
Keeping this rule in mind and dropping the arguments of
$\underline{\cal O}^{(1)}$ and $\underline{x}^{\alpha}$, 
we can write eq.\ (\ref{Lambda'1})
exactly as the $\underline{\cal O}^{(1)}$ and
$\hbar \underline{\tilde{v}}^{\beta}$ contributions
of eq.\ (\ref{Hamil(1)}).
This rule also applies to higher-order terms.
The second-order term is transformed in the same way into
\begin{eqnarray}
&&\hspace{-9mm}
\underline{\Lambda}^{\!\prime(2)}({\mib k},{\mib q})
\nonumber \\
&&\hspace{-9mm}=
\frac{1}{2}\underline{{\cal O}}^{(2)}(\overline{{\mib k}\!+\!{\mib q}})
\underline{\Lambda}^{\!(0)}({\mib k},{\mib q})+\frac{1}{2}
\underline{\Lambda}^{\!(0)}({\mib k},{\mib q})
\underline{\cal O}^{(2)}({\mib k})
\nonumber \\
&&\hspace{-5mm}
+\frac{1}{2}h_{\alpha\beta}h_{\alpha'\beta'}\nabla_{\!{\mib k}}^{\beta}\biggl[
\bigl\{\underline{x}^{\alpha}(\overline{{\mib k}\!+\!{\mib q}}),
\underline{x}^{\alpha'}(\overline{{\mib k}\!+\!{\mib q}})\bigr\}
\nabla_{\!{\mib k}}^{\beta'}\!\underline{\Lambda}^{\!(0)}\!({\mib k},{\mib q})
\nonumber \\
&&\hspace{15mm}
+\bigl[
\nabla_{\!{\mib k}}^{\beta'}\!\underline{\Lambda}^{\!(0)}\!({\mib k},{\mib q})\bigr]
\bigl\{\underline{x}^{\alpha}({\mib k}),
\underline{x}^{\alpha'}({\mib k})\bigr\}\biggr]
\nonumber \\
&&\hspace{-5mm}
-\frac{1}{4}h_{\alpha\beta}h_{\alpha'\beta'}\biggl[
\bigl\{\underline{x}^{\alpha}(\overline{{\mib k}\!+\!{\mib q}}),
\underline{x}^{\alpha'}(\overline{{\mib k}\!+\!{\mib q}})\bigr\}
\nabla_{\!{\mib k}}^{\beta}\nabla_{\!{\mib k}}^{\beta'}
\!\underline{\Lambda}^{\!(0)}\!({\mib k},{\mib q})
\nonumber \\
&&\hspace{11mm}
+\bigl[\nabla_{\!{\mib k}}^{\beta}\nabla_{\!{\mib k}}^{\beta'}
\!\underline{\Lambda}^{\!(0)}\!({\mib k},{\mib q})
\bigr]
\bigl\{\underline{x}^{\alpha}({\mib k}),
\underline{x}^{\alpha'}({\mib k})\bigr\}\biggr]
\nonumber \\
&&\hspace{-5mm}
+\frac{i}{4}h_{\alpha\beta}h_{\alpha'\beta'}
\nabla_{\!{\mib k}}^{\beta}\nabla_{\!{\mib k}}^{\beta'}
\biggl[
\underline{x}^{\alpha'}(\overline{{\mib k}\!+\!{\mib q}})\nabla_{\!{\mib k}}^{\alpha}
\!\underline{\Lambda}^{\!(0)}\!({\mib k},{\mib q})
\nonumber \\
&&\hspace{20mm}
-\nabla_{\!{\mib k}}^{\alpha}
\!\underline{\Lambda}^{\!(0)}\!({\mib k},{\mib q})\,
\underline{x}^{\alpha'}({\mib k})\biggr] \, .
\label{Lambda'2}
\end{eqnarray}
\end{subequations}

Let us substitute eqs.\ (\ref{calSRk}) and (\ref{U'Exp}) into
eq.\ (\ref{U}) and carry out the procedure of eq.\ (\ref{MR}) repeatedly.
We thereby obtain eq.\ (\ref{UExp}).

\section{$\mbox{\boldmath $\Phi^{{\rm HF}(j)}$}$ and 
$\mbox{\boldmath $\Sigma^{{\rm HF}(1)}$}$}
\label{app:phi-HF}

In this Appendix we derive explicit expressions for 
$\Phi^{{\rm HF}(j)}$ $(j\!=\!1,2)$ and $\Sigma^{{\rm HF}(1)}$.
First of all, the Hartree contribution to $\Phi^{{\rm HF}(1)}$
is calculated from eq.\ (\ref{Phi-HF}) as
\begin{eqnarray}
&&\hspace{-10mm}\Phi^{{\rm H}(1)}=
\frac{1}{V} \sum_{{\mib k}{\mib k}'}\!\sum_{{\mib K}}
{\cal U}_{{\mib K}}
{\rm Tr}[\,\underline{n}({\mib k})\,
\underline{\Lambda}^{\!(1)}({\mib k},{\mib K})]
\nonumber \\
&& \hspace{20mm}\times
{\rm Tr}[
\,\underline{n}({\mib k}')\,
\underline{\Lambda}^{\!(0)}({\mib k}',-{\mib K})]
\nonumber \\
&&\hspace{-10mm} =-h_{\alpha\beta}
\sum_{{\mib k}}{\rm Tr}\,\underline{n}({\mib k})
\bigl\{ \underline{x}^{\alpha}({\mib k}),
\nabla_{\!{\mib k}}^{\beta}
\underline{\Sigma}^{{\rm H}(0)}({\mib k})\bigr\} \, ,
\label{Phi-H(1)}
\end{eqnarray}
where we have used eq.\ (\ref{Lambda1}) to obtain the second expression
with $\underline{\Sigma}^{{\rm H}(0)}({\mib k})$ given by
\begin{equation}
\underline{\Sigma}^{{\rm H}(0)}({\mib k})=
\frac{1}{V}\sum_{{\mib k}'{\mib K}}
{\cal U}_{{\mib K}} 
\underline{\Lambda}^{\!(0)}({\mib k},{\mib K})
{\rm Tr}[\,\underline{n}({\mib k}')\,
\underline{\Lambda}^{\!(0)}({\mib k}',-{\mib K})] \, .
\label{Sigma-H(0)}
\end{equation}
On the other hand, the Fock contribution has extra terms 
from the operator $\otimes$ as
\begin{eqnarray}
&&\hspace{-7mm}
\Phi^{{\rm F(1)}}
\nonumber \\
&&\hspace{-7mm}=
-\frac{1}{V}\!\sum_{{\mib k}{\mib q}}{\cal U}_{{\mib q}}{\rm Tr}
\,\underline{n}({\mib k})
\!\left\{\!
\underline{\Lambda}^{\!(1)}(\overline{{\mib k}\!+\!{\mib q}},-{\mib q})
\underline{n}(\overline{{\mib k}\!+\!{\mib q}})
\underline{\Lambda}^{\!(0)}({\mib k},{\mib q})
\right.
\nonumber \\
&&\hspace{-3mm}+\frac{1}{2}ih_{\alpha\beta}
[\nabla_{\!{\mib k}}^{\alpha}
\underline{\Lambda}^{\!(0)}(\overline{{\mib k}\!+\!{\mib q}},-{\mib q})]
\nabla_{\!{\mib k}}^{\beta}[
\underline{n}(\overline{{\mib k}\!+\!{\mib q}})
\underline{\Lambda}^{\!(0)}({\mib k},{\mib q})]
\nonumber \\
&&\hspace{-3mm}+\frac{1}{2}ih_{\alpha\beta}\!\left.
\underline{\Lambda}^{\!(0)}(\overline{{\mib k}\!+\!{\mib q}},-{\mib q})
[\nabla_{\!{\mib k}}^{\alpha}
\underline{n}(\overline{{\mib k}\!+\!{\mib q}})]
\nabla_{\!{\mib k}}^{\beta}
\underline{\Lambda}^{\!(0)}({\mib k},{\mib q})\right\} 
\nonumber \\
&&\hspace{-7mm}=
-h_{\alpha\beta}\sum_{{\mib k}}{\rm Tr}\left[\underline{n}({\mib k})
\bigl\{ \underline{x}^{\alpha}({\mib k}),
\nabla_{\!{\mib k}}^{\beta}
\underline{\Sigma}^{{\rm F}(0)}({\mib k})\bigr\}\right.
\nonumber \\
&&\hspace{8mm}
+[\nabla_{\!{\mib k}}^{\alpha}\underline{n}({\mib k})]
\bigl\{ \underline{x}^{\beta}({\mib k}),
\underline{\Sigma}^{{\rm F}(0)}({\mib k})\bigr\}
\nonumber \\
&&\hspace{8mm}
+i\!\left.\underline{x}^{\alpha}({\mib k})\underline{n}({\mib k})
\underline{x}^{\beta}({\mib k})
\underline{\Sigma}^{{\rm F}(0)}({\mib k})\right] .
\label{Phi-F(1)}
\end{eqnarray}
The last expression has been obtained through a calculation
of using the symmetry $h_{\alpha\beta}\!=\!-h_{\beta\alpha}$
and partial integrations with
\begin{equation}
\Sigma^{{\rm F(0)}}({\mib k})
=-\frac{1}{V}\!\sum_{{\mib q}}{\cal U}_{{\mib q}}
\underline{\Lambda}^{\!(0)}(\overline{{\mib k}\!+\!{\mib q}},-{\mib q})
\underline{n}(\overline{{\mib k}\!+\!{\mib q}})
\underline{\Lambda}^{\!(0)}({\mib k},{\mib q}) \, .
\label{Sigma-F(1)}
\end{equation}
The terms with $\nabla_{\!{\mib k}}^{\beta}
\underline{\Sigma}^{{\rm H}(0)}({\mib k})$ and 
$\nabla_{\!{\mib k}}^{\beta}
\underline{\Sigma}^{{\rm F}(0)}({\mib k})$ 
in eqs.\ (\ref{Phi-H(1)}) and (\ref{Phi-F(1)}) have the effect of renormalizing
the non-interacting velocity $\tilde{\underline{\mib v}}$ in eq.\ (\ref{barH(1)})
into $\underline{\mib v}\!\equiv\!
\mbox{\boldmath $\nabla$}_{\!{\mib k}}[\underline{\cal H}({\mib k})+
\underline{\Sigma}^{{\rm HF}}({\mib k})]/\hbar$.
However, there also appears extra contributions in $\Phi^{{\rm F(1)}}$
which are not directly connected with the renormalization effect.

The corresponding self-energy 
is obtained easily by $\underline{\Sigma}^{{\rm HF}(1)}({\mib k})=
\delta \Phi^{{\rm HF}(1)}/\delta \underline{n}({\mib k})$.
The Hartree part is calculated from eq.\ (\ref{Phi-H(1)}) as
\begin{eqnarray}
&&\hspace{-8mm}
\underline{\Sigma}^{{\rm H}(1)}({\mib k})
=-h_{\alpha\beta}
\bigl\{ \underline{x}^{\alpha}({\mib k}),
\nabla_{\!{\mib k}}^{\beta}
\underline{\Sigma}^{{\rm H}(0)}({\mib k})\bigr\}
\nonumber \\
&&\hspace{-4mm}
+\frac{1}{V}
\sum_{{\mib K}}{\cal U}_{{\mib K}}\underline{\Lambda}^{\!(0)}({\mib k},{\mib K})
\!\sum_{{\mib k}'}{\rm Tr}\,\underline{n}({\mib k}')
\underline{\Lambda}^{\!(1)}({\mib k}',-{\mib K}) \, .
\nonumber \\
\label{Sigma-H(1)}
\end{eqnarray}
Thus, the Hartree self-energy already has a term which can not
be expressed in terms of $\nabla_{\!{\mib k}}^{\gamma}
\underline{\Sigma}^{{\rm H}(0)}({\mib k})$.
The Fock self-energy may be obtained similarly from eq.\ (\ref{Phi-F(1)}).
It also has extra terms besides the one with 
$\nabla_{\!{\mib k}}^{\gamma}\underline{\Sigma}^{{\rm F}(0)}({\mib k})$.

The second-order functional $\Phi^{{\rm HF}(2)}$ 
may be calculated similarly.
The Hartree part is obtained from eqs.\ (\ref{Lambda0-2}) 
and (\ref{Phi-HF}) as
\begin{eqnarray}
&&\hspace{-8mm}
\Phi^{{\rm H}(2)}
\nonumber \\
&&\hspace{-8mm}=
\sum_{{\mib k}}{\rm Tr}\, 
\underline{n}({\mib k})\biggl[
h_{\alpha\beta}h_{\alpha'\beta'}\biggl(\!\nabla_{\!{\mib k}}^{\beta}\bigl\{\!
\{\underline{x}^{\alpha}\!,\underline{x}^{\alpha'}\},
\nabla_{\!{\mib k}}^{\beta'}\underline{\Sigma}^{{\rm H}(0)}\bigr\}
\nonumber \\
&&\hspace{28mm}
-\frac{1}{2}
\bigl\{\!\{\underline{x}^{\alpha},\underline{x}^{\alpha'}\},
\nabla_{\!{\mib k}}^{\beta}
\nabla_{\!{\mib k}}^{\beta'}\underline{\Sigma}^{{\rm H}(0)}
\bigr\}
\nonumber \\
&&\hspace{28mm}
+\frac{i}{4}\nabla_{\!{\mib k}}^{\beta}\nabla_{\!{\mib k}}^{\beta'}
\bigl[\underline{x}^{\alpha'},
\nabla_{\!{\mib k}}^{\alpha}\underline{\Sigma}^{{\rm H}(0)}\bigr]\!\biggr)
\nonumber \\
&&\hspace{14mm}
+h_{\alpha\beta}\bigl\{{\cal O}^{(1)},\bigl\{\underline{x}^{\alpha},
\nabla_{\!{\mib k}}^{\beta}\underline{\Sigma}^{{\rm H}(0)}\bigr\}\!\bigr\}\!
\nonumber \\
&&\hspace{14mm}
+\frac{1}{8}\bigl[\,{\cal O}^{(1)} ,
\bigl[\,{\cal O}^{(1)} ,
\underline{\Sigma}^{{\rm H}(0)}\,\bigr]\bigr] 
\nonumber \\
&&\hspace{14mm}
+\frac{i}{2}h_{\alpha\beta}
\nabla_{\!{\mib k}}^{\beta}
\bigl[\,{\cal O}^{(1)},
\nabla_{\!{\mib k}}^{\alpha}\underline{\Sigma}^{{\rm H}(0)}\,\bigr]\biggr]
\nonumber \\
&&\hspace{-5mm}
+\frac{1}{2V}\!\sum_{{\mib K}{\mib k}{\mib k'}}{\cal U}_{{\mib K}}
{\rm Tr}\, \underline{n}({\mib k})\underline{\Lambda}^{\!(1)}({\mib k},{\mib K})
{\rm Tr}\, \underline{n}({\mib k}')\underline{\Lambda}^{\!(1)}({\mib k}',-{\mib K})
\, .
\nonumber \\
\label{Phi-H(2)}
\end{eqnarray}
The terms with
$\underline{\Sigma}^{{\rm H}(0)}({\mib k})$ in the above expression
have the effect of
renormalizing the non-interacting energy $\underline{\cal H}^{(0)}$ and
the velocity $\hbar \tilde{\underline{\mib v}}$.
Indeed, we can find the correspondents to them in eq.\ (\ref{barH(2)}).
On the other hand, the last term cannot be expressed as the renormalization effect.

The Fock part may be calculated similarly.

\section{$\mbox{\boldmath $\Phi^{{\rm DFT}(j)}$}$ and 
$\mbox{\boldmath $\Sigma^{{\rm DFT}(1)}$}$}
\label{app:phi-DFT}

We here derive expressions of $\Phi^{{\rm DFT}(j)}$ for $j\!=\! 1,2$ and
$\Sigma^{{\rm DFT}(1)}$.
Among the two contributions in eq.\ (\ref{Phi-DFT}),
the Hartree part has already
been treated in Appendix B.
The quantities $\Phi^{{\rm H}(1)}$ and $\Phi^{{\rm H}(2)}$ are given by
eqs.\ (\ref{Phi-H(1)}) and (\ref{Phi-H(2)}), respectively,
and $\underline{\Sigma}^{{\rm H}(1)}$ is obtained as eq.\ (\ref{Sigma-H(1)})
with ${\cal U}_{\mib K}\!=\!4\pi e^{2}/K^{2}$ (${\mib K}\!\neq\!{\mib 0}$).

As for the exchange-correlation part, we here consider the cases
where $E_{\rm xc}$ is given by\cite{PW86}
\begin{subequations}
\label{Exc}
\begin{equation}
E_{\rm xc}[n_{\sigma}({\mib r}),{\mib \nabla} n_{\sigma}({\mib r}),
\triangle n_{\sigma}({\mib r})] \, .
\label{Exc1}
\end{equation}
Let us express $n_{\sigma}({\mib r})$ as eq.\ (\ref{n_sigma2})
and regard $E_{\rm xc}$ as a functional of $\underline{n}({\mib k})$
instead of $n_{\sigma}({\mib r})$.
We then expand $E_{\rm xc}$ with respect to the explicit ${\mib B}$ dependences 
through
$\underline{\rho}^{\pm}({\mib k},{\mib r},{\mib B})$ as
\begin{equation}
E_{\rm xc}=E_{\rm xc}^{(0)}+\sum_{j=1}^{\infty}E_{\rm xc}^{(j)} \, .
\label{Exc2}
\end{equation}
\end{subequations}
To obtain the expressions of $E_{\rm xc}^{(j)}$,
we expand $\underline{\rho}^{\pm}({\mib k},{\mib r},{\mib B})$
of eq.\ (\ref{rho})
in powers of ${\mib B}$ as
\begin{equation}
\underline{\rho}^{\pm}({\mib k},{\mib r},{\mib B})= 
\underline{\rho}^{\pm(0)}({\mib k},{\mib r})+\sum_{j=1}^{\infty}
\underline{\rho}^{\pm(j)}({\mib k},{\mib r},{\mib B}) \, .
\label{rho-E}
\end{equation}
The expansion coefficients are found easily as
\begin{subequations}
\label{rho0-2}
\begin{equation}
{\rho}_{bb'}^{\pm(0)}({\mib k},{\mib r})\equiv 
\vec{u}_{b{\mib k}}^{\dagger}({\mib r})
\frac{N_{\rm c}}{2}
(\underline{1}\pm\underline{\tau}_{3})\,
\vec{u}_{b'{\mib k}}({\mib r}) \, ,
\label{rho(0)}
\end{equation}
\begin{equation}
\underline{\rho}^{\pm(1)}
=\underline{\rho}^{\prime\, \pm(1)} +2
\bigl\{ \underline{\cal S}^{(1)},\underline{\rho}^{\pm(0)}\bigr\}\, ,
\label{rho(1)}
\end{equation}
\begin{eqnarray}
&&\hspace{-14mm}
\underline{\rho}^{\pm(2)}=
\underline{\rho}^{\prime\, \pm(2)}+
2\bigl\{ \underline{\cal S}^{(2)},
\underline{\rho}^{\pm(0)}\bigr\}
+
2\bigl\{ \underline{\cal S}^{(1)},
\underline{\rho}^{\prime\, \pm(1)}\bigr\}
\nonumber \\
&&\hspace{-9mm}
+\,
\underline{\cal S}^{(1)}
\underline{\rho}^{\prime\, \pm(0)}\,\underline{\cal S}^{(1)}\!
-ih_{\alpha\beta} 
\bigl[\nabla_{{\mib k}}^{\beta} \underline{\cal S}^{(1)},
\nabla_{{\mib k}}^{\alpha}\underline{\rho}^{\pm(0)}\bigr] ,
\label{rho(2)}
\end{eqnarray}
\end{subequations}
with $\rho^{\prime\,\pm(j)}$ obtained from eq.\ (\ref{rho'}) as
\begin{subequations}
\label{rho'1-2}
\begin{equation}
{\rho}_{bb'}^{\prime\, \pm(1)}({\mib k},{\mib r},{\mib B})\equiv 
ih_{\alpha\beta}
[\nabla_{{\mib k}}^{\alpha} 
\vec{u}_{b{\mib k}}^{\dagger}({\mib r})]\,
\frac{N_{\rm c}}{2}
(\underline{1}\pm\underline{\tau}_{3})
\nabla_{{\mib k}}^{\beta}
\vec{u}_{b'{\mib k}}({\mib r}) \, ,
\label{rho'(1)}
\end{equation}
\begin{eqnarray}
&&\hspace{-9mm}
{\rho}_{bb'}^{\prime\, \pm(2)}({\mib k},{\mib r},{\mib B})\equiv
-\frac{1}{2}h_{\alpha\beta}h_{\alpha'\beta'}
[\nabla_{{\mib k}}^{\alpha}\nabla_{{\mib k}}^{\alpha'} 
\vec{u}_{b{\mib k}}^{\dagger}({\mib r})]\,\frac{N_{\rm c}}{2}
(\underline{1}\pm\underline{\tau}_{3})
\nonumber \\
&&\hspace{18mm}
\times 
\nabla_{{\mib k}}^{\beta}\nabla_{{\mib k}}^{\beta'}
\vec{u}_{b'{\mib k}}({\mib r}) \, .
\label{rho'(2)}
\end{eqnarray}
\end{subequations}

It follows from eq.\ (\ref{Exc}) with eqs.\ (\ref{n_sigma2}) and (\ref{rho-E}) 
that $E_{\rm xc}^{(j)}$ for $j\!=\!1,2$ are given by
\begin{eqnarray}
&&\hspace{-12mm}
E_{\rm xc}^{(j)}
=
\!\int \! {\rm d}{\mib r}\! \left[\,\sum_{\sigma=\pm}
F^{(1)}_{\sigma}({\mib r}) n_{\sigma}^{(j)}({\mib r},{\mib B})\right.
\nonumber \\
&&\hspace{-1mm}
\left. \! +\delta_{j2}\sum_{\sigma\sigma'}
n_{\sigma}^{(1)}({\mib r},{\mib B})F^{(2)}_{\sigma\sigma'}({\mib r})
n_{\sigma'}^{(1)}({\mib r},{\mib B})\right] ,
\label{Exc(j)}
\end{eqnarray}
where $n_{\sigma}^{(j)}({\mib r},{\mib B})$, $F^{(1)}_{\sigma}({\mib r})$, and
$F^{(2)}_{\sigma\sigma'}({\mib r})$ are defined by
\begin{equation}
n_{\sigma}^{(j)}({\mib r},{\mib B})\equiv\frac{1}{N_{\rm c}}\sum_{{\mib k}}
{\rm Tr} \,\underline{n}({\mib k}) \,
\underline{\rho}^{\sigma (j)}({\mib k},{\mib r},{\mib B}) \, ,
\label{eta-def}
\end{equation}
\begin{subequations}
\label{F1-2}
\begin{equation}
F^{(1)}_{\sigma}({\mib r})
\equiv\sum_{\nu=0}^{4}
(-1)^{p_{\nu}}
D_{\nu}\frac{\delta E_{\rm xc}}{\delta [D_{\nu}n_{\sigma}({\mib r})]} \, ,
\label{F(1)}
\end{equation}
\begin{equation}
F^{(2)}_{\sigma\sigma'}({\mib r})\!
\equiv\! \sum_{\nu\nu'} 
(-1)^{p_{\nu}+p_{\nu'}}
D_{\nu}D_{\nu'}\frac{\delta^2 E_{\rm xc}}{\delta [D_{\nu}n_{\sigma}({\mib r})]
\delta [D_{\nu'}n_{\sigma'}({\mib r})]} 
\, ,
\label{F(2)}
\end{equation}
\end{subequations}
with $(D_{\nu},p_{\nu})\!=\!(1,1)$, $(\nabla_{x},-1)$, $(\nabla_{y},-1)$,
$(\nabla_{z},-1)$, and $(\triangle,1)$ for $\nu\!=\!0,\cdots,4$, respectively.

Noting
$F_{\sigma}^{(1)}({\mib r}+{\mib R})\!=\!F_{\sigma}^{(1)}({\mib r})$,
we can further simplify the first term in the square bracket of
eq.\ (\ref{Exc(j)}) by using the completeness of 
$\{\vec{u}_{b''{\mib k}}({\mib r})\}$ 
over the unit cell.
To be more specific,
we expand
$\sum_{\sigma}F_{\sigma}^{(1)}({\mib r})\frac{N_{\rm c}}{2}(\underline{1}\!+\!\sigma
\underline{\tau}_{3})\vec{u}_{b'{\mib k}}({\mib r})$ 
in $\sum_{\sigma}F_{\sigma}^{(1)}({\mib r})
\underline{\rho}^{\sigma (j)}({\mib k},{\mib r},{\mib B})$ as
\begin{equation}
\sum_{\sigma}F_{\sigma}^{(1)}({\mib r})
\frac{N_{\rm c}}{2}(\underline{1}\!+\!\sigma\underline{\tau}_{3})
\vec{u}_{b'{\mib k}}({\mib r})=\sum_{b''}
\vec{u}_{b''{\mib k}}({\mib r})\Sigma^{{\rm xc}(0)}_{b''b'}({\mib k}) \, ,
\end{equation}
where
\begin{equation}
\Sigma^{{\rm xc}(0)}_{bb'}({\mib k})\equiv
\frac{\delta E_{\rm xc}^{(0)}}{\delta n_{b'b}({\mib k})} 
=\frac{1}{N_{\rm c}} \sum_{\sigma}\int\!
F^{(1)}_{\sigma}({\mib r})\rho^{\sigma(0)}_{bb'}({\mib k},{\mib r}) \, {\rm d}{\mib r}
\label{Sigma-xc(0)}
\end{equation}
denotes the exchange-correlation self-energy in zero field.
With the same procedure as that of deriving eqs.\ (\ref{Hamil(1)}) and (\ref{barH(1)}),
eq.\ (\ref{Exc(j)}) for $j\!=\! 1$ is then transformed into
\begin{subequations}
\label{Exc1-2}
\begin{equation}
E_{\rm xc}^{(1)}= -h_{\alpha\beta} 
\sum_{{\mib k}}{\rm Tr} \,\underline{n}({\mib k})
\bigl\{\underline{x}^{\alpha}({\mib k}), 
{\nabla}_{\!{\mib k}}^{\beta}\underline{\Sigma}^{{\rm xc}(0)}({\mib k}) \bigr\} 
\, .
\label{Exc(1)}
\end{equation}
Equations (\ref{Phi-H(1)}) and (\ref{Exc(1)}) have the effect of
turning $\tilde{\underline{\mib v}}$ in eq.\ (\ref{barH(1)}) into
the renormalized velocity $\underline{\mib v}$.

The second-order term $E_{\rm xc}^{(2)}$ can be transformed similarly
with the procedure of deriving eqs.\ (\ref{Hamil(2)}) and (\ref{barH(2)}).
We thereby obtain an expression of $E_{\rm xc}^{(2)}$ as
\begin{eqnarray}
&&\hspace{-9mm}
E_{xc}^{(2)}
=
\sum_{{\mib k}}{\rm Tr}\, 
\underline{n}({\mib k})\biggl[
h_{\alpha\beta}h_{\alpha'\beta'}\biggl(\!\nabla_{\!{\mib k}}^{\beta}\bigl\{\!
\{\underline{x}^{\alpha}\!,\underline{x}^{\alpha'}\},
\nabla_{\!{\mib k}}^{\beta'}\underline{\Sigma}^{{\rm xc}(0)}\bigr\}
\nonumber \\
&&\hspace{19mm}
-\frac{1}{2}
\bigl\{\!\{\underline{x}^{\alpha},\underline{x}^{\alpha'}\},
\nabla_{\!{\mib k}}^{\beta}
\nabla_{\!{\mib k}}^{\beta'}\underline{\Sigma}^{{\rm xc}(0)}
\bigr\}
\nonumber \\
&&\hspace{19mm}
+\frac{i}{4}\nabla_{\!{\mib k}}^{\beta}\nabla_{\!{\mib k}}^{\beta'}
\bigl[\underline{x}^{\alpha'},
\nabla_{\!{\mib k}}^{\alpha}\underline{\Sigma}^{{\rm xc}(0)}\bigr]\!\biggr)
\nonumber \\
&&\hspace{9mm}
+h_{\alpha\beta}\bigl\{{\cal O}^{(1)},\bigl\{\underline{x}^{\alpha},
\nabla_{\!{\mib k}}^{\beta}\underline{\Sigma}^{{\rm xc}(0)}\bigr\}\!\bigr\}\!
\nonumber \\
&&\hspace{9mm}
+\frac{1}{8}\bigl[\,{\cal O}^{(1)} ,
\bigl[\,{\cal O}^{(1)} ,
\underline{\Sigma}^{{\rm xc}(0)}\,\bigr]\bigr] 
\nonumber \\
&&\hspace{9mm}
+\frac{i}{2}h_{\alpha\beta}
\nabla_{\!{\mib k}}^{\beta}
\bigl[\,{\cal O}^{(1)},
\nabla_{\!{\mib k}}^{\alpha}\underline{\Sigma}^{{\rm xc}(0)}\,\bigr]\biggr]
\nonumber \\
&&\hspace{2mm}
+\sum_{\sigma\sigma'}\int\!
n_{\sigma}^{(1)}({\mib r},{\mib B})F^{(2)}_{\sigma\sigma'}({\mib r})
n_{\sigma'}^{(1)}({\mib r},{\mib B}) \, {\rm d}{\mib r}\, ,
\label{Exc(2)}
\end{eqnarray}
\end{subequations}
which is analogous with eq.\ (\ref{Phi-H(2)}).
The terms with
$\underline{\Sigma}^{{\rm xc}(0)}({\mib k})$ in the above expression
have the effect of
renormalizing the non-interacting energy $\underline{\cal H}^{(0)}$ and
velocity $\hbar \tilde{\underline{\mib v}}$.
Indeed, we can find the correspondents in eq.\ (\ref{barH(2)}).
On the other hand, the last term cannot be expressed as the renormalization effect.

The first-order self-energy is obtained from
$\Sigma^{{\rm xc}(1)}_{bb'}({\mib k})\!=\!
{\delta E_{\rm xc}^{(1)}}/{\delta n_{b'b}({\mib k})}$ as
\begin{eqnarray}
&&\hspace{-15mm}
\underline{\Sigma}^{{\rm xc}(1)}({\mib k})=
-h_{\alpha\beta} 
\bigl\{\underline{x}^{\alpha}({\mib k}), 
{\nabla}_{\!{\mib k}}^{\beta}\underline{\Sigma}^{{\rm xc}(0)}({\mib k}) \bigr\} 
\nonumber \\
&&\hspace{-10mm}
+\frac{1}{N_{\rm c}}\sum_{\sigma\sigma'}\int\!
\underline{\rho}^{\sigma(0)}({\mib k},{\mib r})F^{(2)}_{\sigma\sigma'}({\mib r})
n_{\sigma'}^{(1)}({\mib r},{\mib B}) \, {\rm d}{\mib r}\, .
\label{Sigma-xc(1)}
\end{eqnarray}
Thus, $\underline{\Sigma}^{{\rm xc}(1)}({\mib k})$ also has a term which can not
be expressed in terms of $\nabla_{\!{\mib k}}^{\gamma}
\underline{\Sigma}^{{\rm xc}(0)}({\mib k})$.

\section{Derivation of eq.\ (\ref{chi-LP})}
\label{app:chi-LP}

We here derive eq.\ (\ref{chi-LP}) valid for ${\mib B}\!\rightarrow\! {\mib 0}$
by expanding the logarithmic term in eq.\ (\ref{Omega-2}) with respect to
the ${\mib B}$ dependence in ${\mib \kappa}$
up to the second order.
Our method is a slight modification of the one developed 
by Sondheimer and Wilson\cite{SW51} and refined by
Roth.\cite{Roth62} 
It enables us to treat the correlation effects exactly, as seen below.

Let us define the matrix $\underline{H}(\varepsilon_{n},{\mib k})$ by
\begin{equation}
\underline{H}(\varepsilon_{n},{\mib k})\equiv
\underline{\cal H}({\mib k})+
\underline{\Sigma}(\varepsilon_{n},{\mib k})- i\varepsilon_{n}\underline{1}\, .
\end{equation}
It is connected with the Green's
function as $\underline{H}\!=\!-\underline{G}^{-1}$
and appears in the logarithm of eq.\ (\ref{Omega-2}) as 
$\underline{H}(\varepsilon_{n},{\mib\kappa})$.
Let us denote the eigenvalues of $\underline{H}(\varepsilon_{n},{\mib k})$ and
$\underline{H}(\varepsilon_{n},{\mib \kappa})$ 
as $E_{\ell}$ and $E_{\lambda}$, respectively.
Specifically, $\ell$ is given by $\ell\!=\! b{\mib k}$, 
and $\lambda$ is composed of the band-spin index $b$, 
the Landau level $N$ accompanied by
an additional quantum number distinguishing its degenerate states,
and the wave vector $k_{z}$ perpendicular to the magnetic field.
Both
$E_{\ell}\!\equiv\! E_{\ell}(\varepsilon_{n})$ and 
$E_{\lambda}\!\equiv\! E_{\lambda}(\varepsilon_{n})$ are complex
in general and can be written as
\begin{equation}
E_{\lambda}(\varepsilon_{n})=E_{\lambda}'(\varepsilon_{n})
-i E_{\lambda}''(\varepsilon_{n}) \, ,
\label{ReE}
\end{equation}
for example, where $E_{\lambda}'$ and $E_{\lambda}''$ are some real numbers.
We will proceed by assuming that $E_{\lambda}''$ has the same sign as 
$\varepsilon_{n}$, i.e.,
$E_{\lambda}''(\varepsilon_{n})
\!=\!|E_{\lambda}''(\varepsilon_{n})|{\rm sgn}(\varepsilon_{n})$,
which is expected from the analytic properties of Green's function.

Let us concentrate on the case $\varepsilon_{n}\!>\!0$ where $E_{\lambda}''\!>\!0$.
Then the trace of the logarithmic term in eq.\ (\ref{Omega-2})
is transformed as
\begin{eqnarray}
&&\hspace{-5mm} 
S\equiv {\rm Tr}\,{\rm Tr}_{{\mib k}}\,
\ln [\,
\underline{H}(\varepsilon_{n},{\mib \kappa})]=
\sum_{\lambda} 
\ln E_{\lambda} 
\nonumber \\
&&\hspace{-5mm} =
-\sum_{\lambda}
\int_{0}^{\infty} 
\frac{{\rm e}^{-iE_{\lambda}t}-{\rm e}^{-i1_{-}t}}{t} \, {\rm d}t
\nonumber \\
&&\hspace{-5mm} =
-{\rm Tr}\,{\rm Tr}_{{\mib k}} \int_{0}^{\infty} 
\frac{{\rm e}^{-i\underline{H}(\varepsilon_{n},{\mib \kappa})t}-
{\rm e}^{-i\underline{1}_{-}t}}{t} 
\, {\rm d}t
\, ,
\label{f-def}
\end{eqnarray}
with $1_{-}\!\equiv\!1\!-\!i0_{+}$.
The final expression enables us to expand ${\rm e}^{-i\underline{H}({\mib \kappa})t}$ 
in powers of ${\mib B}$ in ${\mib \kappa}$.
To this end, we adopt the procedure of the perturbation expansion
in the field theory
and define $\underline{U}({\mib k},t)$ through
\begin{equation}
{\rm e}^{-i\underline{H}({\mib \kappa})t}\equiv 
{\rm e}^{-i\underline{H}({\mib k})t}\,\underline{U}({\mib k},t)
\, .
\label{U-def}
\end{equation}
It can be written explicitly as
\begin{equation}
\underline{U}({\mib k},t)= 
T\exp\left[-i\int_{0}^{t}\underline{H}'({\mib k},t')\, {\rm d}t'\right] ,
\label{U-exp}
\end{equation}
where $T$ is the time-ordering operator and  $\underline{H}'({\mib k},t)\!\equiv\!
{\rm e}^{i\underline{H}({\mib k})t}\underline{H}'({\mib k})
{\rm e}^{-i\underline{H}({\mib k})t}$
with
\begin{eqnarray}
&&\hspace{-7mm}
\underline{H}'({\mib k})\equiv\sum_{{\mib k}'}\delta_{{\mib k}'{\mib k}}
\left({\rm e}^{ih_{\alpha\beta}\nabla_{\!{\mib k}'}^{\alpha}\nabla_{\!{\mib k}}^{\beta}}-1\right)
\underline{H}({\mib k}')
\nonumber \\
&&\hspace{-7mm}
= ih_{\alpha\beta}(\nabla_{\!{\mib k}}^{\alpha}\underline{H})
\nabla_{\!{\mib k}}^{\beta} -\frac{1}{2}h_{\alpha\beta}h_{\alpha'\beta'}
(\nabla_{\!{\mib k}}^{\alpha}\nabla_{\!{\mib k}}^{\alpha'}\underline{H})
\nabla_{\!{\mib k}}^{\beta}\nabla_{\!{\mib k}}^{\beta'}+\cdots
\nonumber \\
&&\hspace{-7mm}
=\underline{H}^{\prime(1)}({\mib k})+\underline{H}^{\prime(2)}({\mib k})+\cdots  \, .
\label{H'}
\end{eqnarray}
We perform the expansion of eq.\ (\ref{U-exp}) up to the second order in ${\mib B}$.
As noted by Roth,\cite{Roth62} the term $\underline{H}^{\prime(1)}$ of eq.\ (\ref{H'})
does not contribute since it necessarily yields
$h_{\alpha\beta}(\nabla_{\!{\mib k}}^{\alpha}H)(\nabla_{\!{\mib k}}^{\beta}H)\!=\!0$.
Thus, the ${\mib B}$ dependence
through ${\mib \kappa}$ yields no terms first-order in ${\mib B}$. It hence follows that 
this dependence of the first category can be treated independently from the others
in obtaining the expression for the zero-field susceptibility.
Thus, the expansion of $\underline{U}({\mib k},t)$ up to 
the second order is given by
\begin{eqnarray}
&&\hspace{-10mm}
\underline{U}({\mib k},t)=1-i\int_{0}^{t}{\rm e}^{i\underline{H}({\mib k})t'}
\underline{H}^{\prime(2)}({\mib k})
{\rm e}^{-i\underline{H}({\mib k})t'}{\rm d}t'
\nonumber \\
&&\hspace{-10mm}
=1-\frac{1}{2}h_{\alpha\beta}h_{\alpha'\beta'}
(\nabla_{\!{\mib k}}^{\alpha}\nabla_{\!{\mib k}}^{\alpha'}\!\underline{H})
\left[\frac{(-it)^{2}}{2}
(\nabla_{\!{\mib k}}^{\beta}\nabla_{\!{\mib k}}^{\beta'}\!\underline{H})\right.
\nonumber \\
&&\hspace{10mm}
\left.
+ \frac{(-it)^{3}}{3}(\nabla_{\!{\mib k}}^{\beta}\underline{H})
(\nabla_{\!{\mib k}}^{\beta'}\!\underline{H})
\right] .
\label{U-2}
\end{eqnarray}
Let us substitute eqs.\ (\ref{U-def}) and (\ref{U-2}) into eq.\ (\ref{f-def}).
We then carry out a similarity transformation in the trace
to express $S$ in terms of $E_{\ell}$.
We also use the identity ($p\!=\!0,1,2,\cdots$):
\begin{eqnarray}
&&\hspace{-5mm}
-\int_{0}^{\infty} 
\frac{(-it)^{p}{\rm e}^{-iE_{\ell}t}-{\rm e}^{-i1_{-}t}}{t} \, {\rm d}t 
= \frac{\partial^{p}}{\partial E_{\ell}^{p}}\ln  E_{\ell}
\nonumber \\
&&\hspace{-5mm}
= \delta_{p0}\ln E_{\ell} - (1-\delta_{p0})(p\!-\!1)! \, G_{\ell}^{p}
\, ,
\end{eqnarray}
where $G_{\ell}\!\equiv\! -E_{\ell}^{-1}$ is Green's function.
We finally perform an integration by parts
to the contribution from the second term
in the square bracket of eq.\ (\ref{U-2}) as
$\sum_{{\mib k}}(\nabla_{\!{\mib k}}^{\beta}\! E_{\ell})
(\nabla_{\!{\mib k}}^{\beta'}\! E_{\ell})2G_{\ell}^{3}
\!=\!\sum_{{\mib k}}(\nabla_{\!{\mib k}}^{\beta}\! E_{\ell})
\nabla_{\!{\mib k}}^{\beta'}\!G_{\ell}^{2}
\!=\!-\sum_{{\mib k}}(\nabla_{\!{\mib k}}^{\beta}\nabla_{\!{\mib k}}^{\beta'}\! E_{\ell})
G_{\ell}^{2}$.
We thereby obtain a simple expression for eq.\ (\ref{f-def})
as
\begin{equation}
S= \sum_{\ell}\!\left[
\ln E_{\ell}+\frac{1}{12}h_{\alpha\beta}h_{\alpha'\beta'}
(\nabla_{\!{\mib k}}^{\alpha}\nabla_{\!{\mib k}}^{\alpha'}\! E_{\ell})
(\nabla_{\!{\mib k}}^{\beta}\nabla_{\!{\mib k}}^{\beta'}\! E_{\ell})\,
G_{\ell}^{2}\,
\right] .
\label{f-def2}
\end{equation}

The case $\varepsilon_{n}\!<\!0$ can be treated similarly. Indeed,
we only need to change the range of integration in eq.\ (\ref{f-def})
into $(-\infty,0)$, replace $1_{-}$ by $1_{+}$, and 
remove the minus signs in front of the integrals.
We thereby arrive at the same expression as eq.\ (\ref{f-def2}).
Differentiating the second contribution of
eq.\ (\ref{f-def2}) twice with respect to ${\mib B}$ 
and going back to the original ${\mib k}$ representation,
we arrive at eq.\ (\ref{chi-LP}).



\begin{thebibliography}{99}
\bibitem{comment1}We here drop the band index for simplicity,
which will be recovered appropriately in Sec.\ II.
\bibitem{comment}The two terms with ${\cal O}^{\alpha}$ on the right-hand side
of eq.\ (\ref{chi-DFT2}) are apparently absent in the expression by Roth.\cite{Roth62}
They were absorbed into eq.\ (\ref{chi-DFT-PvV}) by using 
different expressions for ${\cal H}^{{\rm DFT}(\alpha')}$ and ${\cal H}^{{\rm DFT}(\alpha)}$
in the present paper.
\end{thebibliography}
\end{document}